\documentclass[preprint,aps,prd,floatfix,nofootinbib,11pt]{revtex4-2}
\pdfoutput=1
\usepackage{graphicx}
\usepackage{nicefrac}
\usepackage{natbib}
\usepackage{braket}

\usepackage{hyperref}

\usepackage{xcolor}
\usepackage[normalem]{ulem}

\usepackage{bm}
\usepackage{amsfonts}
\usepackage{latexsym}
\usepackage{graphicx}
\usepackage{amsmath}
\newenvironment{psmallmatrix}
{\left(\begin{smallmatrix}}
	{\end{smallmatrix}\right)}
\usepackage{palatino}
\usepackage{bigints}
\usepackage{mathpazo}
\usepackage{textcomp}
\usepackage{braket}
\usepackage{float}
\usepackage{booktabs}
\usepackage{dcolumn}
\usepackage{booktabs}
\usepackage{multirow}
\usepackage{hyperref}
\hypersetup{colorlinks,citecolor=blue}
\usepackage{amsmath}
\usepackage{xcolor}
\usepackage{orcidlink}
\usepackage{commath}
\usepackage{subcaption}

\def\jnl@style{\it}
\def\aaref@jnl#1{{\jnl@style#1}}

\def\aaref@jnl#1{{\jnl@style#1}}

\def\aj{\aaref@jnl{AJ}}                   
\def\apj{\aaref@jnl{ApJ}}                 
\def\apjl{\aaref@jnl{ApJ}}                
\def\apjs{\aaref@jnl{ApJS}}               
\def\apss{\aaref@jnl{Ap\&SS}}             
\def\aap{\aaref@jnl{A\&A}}                
\def\aapr{\aaref@jnl{A\&A~Rev.}}          
\def\aaps{\aaref@jnl{A\&AS}}              
\def\mnras{\aaref@jnl{Mon.~Not.~Roy.~Astron.~Soc.}}             
\def\prd{\aaref@jnl{Phys.~Rev.~D}}        
\def\prc{\aaref@jnl{Phys.~Rev.~C}}  
\def\prl{\aaref@jnl{Phys.~Rev.~Lett.}}    
\def\qjras{\aaref@jnl{QJRAS}}             
\def\skytel{\aaref@jnl{S\&T}}             
\def\ssr{\aaref@jnl{Space~Sci.~Rev.}}     
\def\zap{\aaref@jnl{ZAp}}                 
\def\nat{\aaref@jnl{Nature}}              
\def\aplett{\aaref@jnl{Astrophys.~Lett.}} 
\def\apspr{\aaref@jnl{Astrophys.~Space~Phys.~Res.}} 
\def\physrep{\aaref@jnl{Phys.~Rep.}}      
\def\physscr{\aaref@jnl{Phys.~Scr}}       
\def\commat{\aaref@jnl{Comm.~Math.~Phys.}}              
\def\science{\aaref@jnl{Science}}               
\def\cqg{\aaref@jnl{Classical Quant.~Grav.}}            
\def\jpcs{\aaref@jnl{JPCS}}                                     
\def\ijmpd{\aaref@jnl{Int.~J.~Mod.~Phys.~D}}                    
\def\grg{\aaref@jnl{Gen.~Relat.~Gravit.}}               
\def\rpp{\aaref@jnl{Rep.~Prog.~Phys.}}          
\def\npa{\aaref@jnl{Nucl.~Phys.~A}}        
\def\lrr{\aaref@jnl{Living Rev.~Rel.}}                   
\def\jcap{\aaref@jnl{J.~Cosmology Astropart.~Phys.}}    
\def\rmp{\aaref@jnl{Rev.~Mod.~Phys.}}   
\def\epjc{\aaref@jnl{Eur.~Phys.~J.~C}}

\allowdisplaybreaks[1]

\begin{document}

\title{The $D$-dimensional charged AdS black holes solutions in polytropic dark energy from Barrow entropy}
	
	
		\author{Y. Sekhmani}
	\email{sekhmaniyassine@gmail.com}
	\affiliation{Center for Theoretical Physics, Khazar University, 41 Mehseti Street, Baku, AZ1096, Azerbaijan.}
	\affiliation{Department of Mathematical and Physical Sciences, College of Arts and Sciences, University of Nizwa, Nizwa 616, Sultanate of Oman}
	
	\author{B. Hazarika}
	\email{rs\_bidyuthazarika@dibru.ac.in}
	\affiliation{Department of Physics, Dibrugarh University, Dibrugarh, Assam,786004}

	\author{P. Phukon}
	\email{prabwal@dibru.ac.in}
	\affiliation{Department of Physics, Dibrugarh University, Dibrugarh, Assam,786004}
	\affiliation{Theoretical Physics Division, Centre for Atmospheric Studies, Dibrugarh University, Dibrugarh, Assam,786004}

	\author{A. Landry}
	\email{a.landry@dal.ca}
	\affiliation{Department of Mathematics and Statistics, Dalhousie University, Halifax, Nova Scotia, Canada, B3H 3J5}

	\author{S. K. Maurya}
	\email{sunil@unizwa.edu.om}
	\affiliation{Department of Mathematical and Physical Sciences, College of Arts and Sciences, University of Nizwa, Nizwa 616, Sultanate of Oman}

	\author{J. Rayimbaev}
	\email{javlon@astrin.uz}
	\affiliation{National Research University, 100174 Tashkent, Uzbekistan}
	\affiliation{University of Tashkent for Applied Sciences, Gavhar Str. 1, 100149 Tashkent, Uzbekistan }
	\affiliation{Urgench State University, Kh. Alimdjan str. 14, 220100 Urgench, Uzbekistan}
	\affiliation{Tashkent State Technical University, 100095 Tashkent, Uzbekistan}
	\affiliation{Department of Nuclear Physics and Astronomy, Samarkand State University, 140104 Samarkand, Uzbekistan}

	\begin{abstract}
		
		\vspace*{0.5cm}
	{This paper mainly aims to solve the Anti-deSitter Black Holes (AdS BH) under the Barrow entropy under polytropic gas fluid, especially the Chaplygin Gas. First, we develop this last polytropic model in detail to then obtain the possible solutions and thermodynamic conditions on the energy momentum and the Barrow entropy for spacetimes of spatial dimension $D>3$. Then, we focus on thermidynamic solutions and the different impacts on the Barrow entropy of the black hole, the temperature profile, the mass, and the various physical quantities involved. Afterwards, we focus on the specific cases of the solutions of dimensions $D=4$ and $5$ to concretely test the models, especially from the point of view of the thermodynamic topology. Finally, we generalize everything by elaborating and testing the stability of the models to arrive at the thermal geometry of the AdS BH.}
	
	\end{abstract}
	
	
	
	

\maketitle


\section{Introduction}

The spherically symmetric spacetime physical problem is a well-interesting challenge, especially concerning a black hole (BH) as a gravitational source in General Relativity (GR) and some more generalized and complex gravity theories. Not only the BH horizon solutions are in the most interest for physicists in gravity, but there are also neutron stars (NS) and white dwarf (WD) solutions as interesting physical processes using the same formalism. However, there has been an increasing interest since a while toward the thermodynamic aspects of physical processes implying {  BHs} and any remnants. This interest leads to the emission of thermal radiation from spherically symmetric mass processes such as Hawking radiation \cite{hawking1,hawking2,bharea1,areatheorem,bharea2,bhevap1,BHdynamics1,BHdynamics2,BHdynamics3,BHdynamics4,nonsymbekersteinhawkBH}. This type of thermal radiation leads to the Hawking temperature, which is essentially the evaporating temperature of BH can be measured from the BH horizon surface as in infinity. A BH will evaporate its thermal energy in some conditions as a black body radiation. There are several works in the literature on the various physical processes leading to Hawking BH radiation and may also imply {  BHs} and accretion systems \cite{BHdynamics1,BHdynamics2,BHdynamics3,BHdynamics4,nonsymbekersteinhawkBH}. We should note that there is an equivalent for linear accelerated mass thermal radiation called the Unruh radiation working a bit on the same principles \cite{unruh1,unruh2,landryhammad1}.

From this point, because the thermal radiation from spherically symmetric astrophysical objects leads to thermodynamic process, we need to get interested on the entropy related to this thermal radiation. Because a BH has an intrinsic temperature, the Hawking temperature $T_H=T_H(r\rightarrow\infty)=\frac{1}{8\pi\,M}$, this consideration implies that the first thermodynamic law needs to be satisfied as $dE=T\,dS-P\,dV$ where $T$ can be expressed in terms of the Hawking temperature $T_H$ and $dS$ will be the related instant entropy of this same BH. We can also express this in a macroscopic manner as $\left(S-S_0\right) \approx 8\pi M\,{  (E-E_0)}\approx 8\pi M\,dE$ {  (or $dS \approx 8\pi M\,dE$ infinitesimally by assuming $S(E+\Delta E)-S(E)\approx\frac{dS}{dE}\,\Delta E$ at the first order)} and the BH entropy is proportional to its surface area \cite{bharea1,areatheorem,bharea2,bhevap1,BHdynamics1,BHdynamics2,BHdynamics3,BHdynamics4}. The Hawking BH evaporation temperature leads to the Hawking-Bekerstein entropy described by $S \sim A_{H}$ where $A_{H}$ is the surface area of the horizon. Depending on the type of BH, there are several types of more developed entropy processes and specific relations \cite{Bardeen1973}.

Before going into the Barrow entropy with a more in-depth discussion and precision, we need to discuss more about the {  appropriate cosmological fluids. In the current paper, we will work with the polytropic fluids as described by the equation of state (EoS)} \cite{polytropic1,polytropic2,polytropic3,polytropic4,polytropic5,polytropic6,polytropic7,polytropic8}:
\begin{equation}\label{polytropicEoS}
	P( \rho) =-\gamma \rho^{1+\frac{1}{n}} ,
\end{equation}
where $n$ is the polytropic index and the possible values are $1<n< \infty$. {  Some parallels with the conventional cosmological perfect fluid can easily be made as follows. Firstly, {  Eq.} \eqref{polytropicEoS} establishes the fundamental polytropic EoS used to model the dark energy (DE)} fluids. Setting $\frac{1}{3}<\gamma<1$ and $n\rightarrow\infty$ in {  Eq.} \eqref{polytropicEoS}, we obtain {  from the polytropic EoS} the solutions of quintessence {  DE} fluids. This state constitutes a fifth matter form in the universe, and this quintessence process is related to an associated scalar field. The cosmological constant $\Lambda_0$ is obtained by setting $\gamma=1$ and $n\rightarrow\infty$, but this also constitutes the {  most elementary} {  DE} form \cite{Arun:2017dm,Lambiase2009}. However, quintessence is a second form of {  DE} as stated in P. Steinhardt's works predicting the accelerating universe expansion \cite{steinhardt1,steinhardt2,caldwell1}. The quintessence {  DE} form is in the meantime a specific subcase of polytropic gas according to {  Eq.} \eqref{polytropicEoS}. This EoS as stated by {  Eq.} \eqref{polytropicEoS} can be generalized by using scalar field EoS with $P_{\phi}$ and $\rho_{\phi}$ as pressure and fluid density function \cite{darkenergy2,coleylandrygholami} as:
\begin{align}
	\rho_{\phi} =& \frac{\dot{\phi}^2}{2}+V(\phi) ,
	\\
	P_{\phi} =&\frac{\dot{\phi}^2}{2}-V(\phi),
\end{align}
where $\dot{\phi}$ is the time-derivative of the scalar field $\phi$ and $V(\phi)$ is the scalar potential. The quintessence ratio is defined as $w_Q=-\gamma=\frac{P_{\phi}}{\rho_{ \phi}}$.

In addition, there are some possible generalizations of this {  Eq.} \eqref{polytropicEoS} EoS such as logatropic and the double polytropic fluids with an EoS as \cite{murnaghan1,murnaghan2}:
\begin{equation}\label{doublepolytropicEoS}
	P( \rho) =-\gamma_1 \rho^{1+\frac{1}{n_1}}-\gamma_2 \rho^{1+\frac{1}{n_2}} ,
\end{equation}
where $n_1$ and $n_2$ are polytropic indexes. One special double polytropic fluid case is the Murnaghan fluids where {  Eq.} \eqref{doublepolytropicEoS} will simplify as \cite{murnaghan1,murnaghan2}: 
\begin{equation}\label{murnaghanEoS}
	P( \rho) =-\gamma \rho^{1+\frac{1}{n}}+P_2 ,
\end{equation}
where the second part of fluid is an isobar fluid (constant pressure). From all of these previous EoS, we can easily find several types of BH and cosmological solutions.

There is some literature on polytropic cosmological structure models explaining BHs and cosmological solutions with {  DE} models \cite{polytropic1,polytropic2,polytropic3,polytropic4,polytropic5,polytropic6,polytropic7,polytropic8}. From this consideration, there is the polytropic EoS leading to the polytropic cosmological $\gamma$-models. One of the important studies on polytropic gas systems was on the {  DE} system with cold matter scenario \cite{polytropic7}. This study was done with some specific polytropic gases, especially with Chaplygin gas. This last kind of gas was originally described by $P(\rho)=-\gamma \rho^{-1}$, where $n=-\frac{1}{2}$ as a polytropic index inside {  Eq.} \eqref{polytropicEoS} \cite{Kamenshchik2001,Bento2002}. This original Chaplygin gas EoS leads to the Quartessence phenomena, a generalization and more complete process in comparison to quintessence models \cite{Bilic:2002chg,Makler:2003iw,zhu2004}. This physical process can be described by two different manifestations of a single dark fluid leading to two thermodynamic states: Dark Matter (DM) and {  DE}. This explanation was also confirmed by cosmological background observations \cite{Makler:2003iw,zhu2004}. Both states of the Quartessence process will have some topological consequences and especially will allow the Anti-de Sitter BH (AdS BH) solutions by this thermodynamic and topological approach. 

Another good example of the application of polytropic gases is the accretion of matter by a charged BH inside such a gas \cite{polytropic8}. In this case, the authors found some critical points of possible accretion and the temperature profile and limits close to the event horizon for a Maxwell-Boltzmann gas. These previous solutions are still using the Hawking thermal radiation formalism, even if these papers resolve the physical problem for an EoS of some polytropic gases. However, the most important consequence for the thermodynamic process is that the thermal process of BH can easily be expressed by the Barrow entropy.  

{ Barrow~\cite {Barrow:2020tzx} recently introduced an innovative statistical framework that moves away from Gaussian assumptions by attributing a fractal geometry to the BH horizon; this feature arises from quantum gravitational phenomena. Indeed, the Barrow entropy, being a state function, assigns a unique entropy value to each equilibrium state of the black hole system. It generalizes the standard area law by incorporating a fractional deformation parameter, $\Delta$, motivated by spacetime microstructure effects. In this approach, the conventional BH entropy is modified to 
\begin{equation}
S_{\Delta} = \left(\frac{A_g}{A_{pl}}\right)^{1+\frac{\Delta}{2}},
\label{eq:BarrowEntropy}
\end{equation}
where \(S_{\Delta}\) represents the Barrow entropy. Here, \(A_g = 4\pi R_g^2\) is the event horizon area of a Schwarzschild BH, and \(A_{pl}\approx 4G\) denotes the Planck area. The parameter \(\Delta\), which satisfies \(0 \leq \Delta \leq 1\), quantifies the impact of quantum gravity: \(\Delta=0\) corresponds to the standard smooth BH while \(\Delta=1\) represents the case with maximal fractal deformation. Although this parameter introduces quantum gravitational corrections into the model, it does not incorporate any additional quantum constants. This power-law correction is applicable to static, spherically symmetric Schwarzschild {  BHs} under conditions that ensure a finite volume, even when the surface area diverges as the structure becomes increasingly intricate at smaller scales. The parameter \(\Delta\) effectively encodes the corresponding quantum gravity effects. Drawing on the gravity-thermodynamics correspondence, various cosmological phenomena have been examined using modified dynamical equations derived within this framework~\cite{Saridakis:2020lrg}. In particular, the holographic principle~\cite{tHooft:1993dmi} has been revisited from the perspective of Barrow entropy to model the late-time acceleration of the universe through the Barrow holographic dark energy (BHDE) models~\cite{Saridakis:2020zol,Dabrowski:2020atl,Sheyki2,Barrowholononflat,Huang:2021zgj,Nojiri:2021jxf,Luciano:2022viz,Chanda:2022tpk,barrow2,Ghaffari:2022skp,Maity:2022gdy}. Furthermore, the early inflationary epoch has been analyzed by integrating these quantum gravity effects into cosmological frameworks~\cite{Ghaffari:2022skp,Maity:2022gdy,Luciano:2023roh,Saha:2024dhr}. Since the inception of Barrow statistics in BH physics, additional studies have focused on elucidating various physical properties of BHs within this expanded paradigm~\cite{Cappozielo2025,Abreu:2020dyu,Abreu:2020wbz,Abreu:2021kwu,Abreu:2022pil,Jawad:2022lww,Wang:2022hun,adscftbarrow1,upperbound1,Hayward1994, AshtekarKrishnan2004,Cai2009}.
}

There is a direct generalization of Barrow with a similar definition called the Tsallis entropy. The exact definition is \cite{tsallis2,tsallisbarrow1,tsallissekh1}:
\begin{equation}\label{TsallisEntropy}
S_{Ts} = \gamma\,A^{\beta},
\end{equation}
where $\beta$ is the Tsallis parameter which is in principle $0<\beta <\infty$. Barrow entropy becomes, with this generalization, a specific subcase of Tsallis entropy where $\beta=1+\frac{\Delta}{2}$, and then $1 \leq \beta \leq \frac{3}{2}$. Another possible entropy process case is the Kaniadakis entropy defined as \cite{kaniadakis1,kaniadakis2,kaniadakis3}:
\begin{equation}\label{KaniadakisEntropy}
	S_{K} = -k_B\,\sum_{i}^{N}\,n_i\,\frac{n_i^K-n_i^{-K}}{2K} ,
\end{equation}
where $k_B$ is the Boltzmann constant, $K$ is the Kaniadakis parameter with $-1<K<1$ and $n_i$ is the statistical distribution function as defined in ref. \cite{kaniadakis3}. For a simple BH, {  Eq.} \eqref{KaniadakisEntropy} will simplify as $S_{K}=\frac{\sinh(K\,S_{BH})}{K}$, where $S_{BH}\sim A_H$ as defined for the Hawking-Bekenstein entropy. There are some other possible investigations to be done on these more complex entropy cases, but they are going further than the aims of this current paper.

We will now focus on the aims of this current paper, the thermodynamic { topology} of high-dimensional AdS BH solutions under Barrow entropy defined by {  Eq.} \eqref{eq:BarrowEntropy} for a polytropic gas EoS defined by {  Eq.} \eqref{polytropicEoS}. High-dimensional spacetimes are generally defined when the total number of spatial dimensions is $D>{ 4}$. { The introduction of a five-dimensional $(5D)$ spacetime framework allows for a comparative study with the four-dimensional $(4D)$ case, focusing on the thermodynamic topology of the system using Duan's method. This approach enhances our understanding of how thermodynamic topology is influenced by different spacetime dimensions. In this way,} we will see in particular how the thermodynamics of the polytropic gas and the Barrow entropy will influence the parameters of {  $4$D and $5$D BH solutions. We analyze the thermodynamic phase space topology using Duan’s topological current theory \cite{Duan:1979}. By constructing a two-dimensional vector field (e.g., ($T$,\,$\partial_T F$)) and examining its zero points, we compute the winding number about critical points. This topological charge characterizes stability and phase structure, following the method established by Wei and Liu \cite{Wei:2022}.}

The manuscript is organized as follows: in Sec.~\ref{sec2} we introduce field equations (FEs) for a spherically symmetric metric in the presence of polytropic structure, featured by EoS {  Eq.} \eqref{polytropicEoS} and deliberate on the primary characteristics of the resulting equations \cite{Kamenshchik:2001cp}. In Sec.~\ref{ex}, we first solve our main equations for the general exact BH solutions with the most important conditions to satisfy. We will then be interested in Sec.~\ref{secther} by the general thermodynamic effects of the main solutions under the Barrow entropy. We will enchain and go further in Sec.~\ref{topology} by working on the topological effects of the BH Barrow thermodynamics. In Sections \ref{ThermalBH5D} to \ref{Sectthermalgeometry}, we will study more in detail some thermodynamic concerns, with specific cases such as $4D$ and $5D$ AdS BH solutions, on stability, critical points and some other important physical aspects and solutions on the higher dimensional BH Barrow solutions. We will finally conclude on all the previous parts in Sec.~\ref{conc}.


\section{Polytropic Gas Model of Dark Energy with EoS: $p =-\gamma \rho^{1+\frac{1}{n}}$}
\label{sec2}

To set up a corresponding physical set for charged scalar gas fields with a polytropic EoS in the GR framework, an appropriate action ought to be considered. To this end, a higher-dimensional action featuring the Einstein-Hilbert term, the Maxwell term, and a polytropic background, expressed by \cite{Witten:1998qj,Maldacena:1997re}
    \begin{equation}\label{action}
    \mathcal{I}=\int d^Dx\sqrt{-g}\left[\frac{1}{2\alpha^2}(\mathcal{R}+6L^{-2} )-\frac{1}{2}F_{\mu\nu}F^{\mu\nu} \right]+\mathcal{I}_{Poly},
\end{equation}
where the gravity is ruled out by the scalar curvature of the Ricci scalar $\mathcal{R}$, the parameter ${  L^2=-\frac{(D-1)(D-2)}{2\Lambda}}$ encodes information on the length of the AdS space {  from which $\Lambda$ acts as a negative cosmological constant}, $g=\det (g_{\mu\nu})$ is the determinant of the metric 2-rank tensor $g_{\mu\nu}$, $F_{\mu\nu}$ is the electromagnetic 2-rank tensor associated with abelian gauge symmetry $\textsc{U}(1)$ by the gauge potential $\varphi_\mu$ via $F_{\mu\nu}=\partial_\mu \varphi_\nu-\partial_\nu \varphi_\mu$, and $\mathcal{I}_{Poly}$ stands for the polytropic action \footnote{In the following, we assume that $\alpha=8\pi G=c=1$, where $G$ and $c$ are, respectively, the Newtonian gravitational constant and the speed of light.}.

From that point on, the variation of the action $(\ref{action})$ results in the following FEs:
\small
 \begin{eqnarray}
 \mathcal{I}_{\mu\nu}=\mathcal{R}_{\mu\nu}-\frac{1}{2}g_{\mu\nu}(\mathcal{R}+6L^{-2})-\alpha(\mathcal{T}_{\mu\nu}^{\text{Poly}}+\mathcal{T}_{\mu\nu}^{\text{em}})&=&0,\hspace{0.6cm}\label{eom}\\
  \partial_\mu(\sqrt{-g}F^{\mu\nu})&=&0,\hspace{0.8cm}\label{em}
 \end{eqnarray}
 \normalsize
where the Maxwell FEs are identified by  \begin{align}
    \mathcal{T}_{\mu\nu}^{\text{em}}&=-\frac{1}{2}g_{\mu\nu}{  F^2}+2F_{\mu\sigma}F^{\sigma}_{\nu},\\
    {  (\mathrm{d}\varphi)_a} &=\varphi(r)(\mathrm{d}t)_a,
\end{align}
with ${  F^2}=F_{\mu\nu}F^{\mu\nu}$ is an invariant,  $\mathcal{T}_{\mu\nu}^{\text{Poly}} $ is the energy-momentum tensor for polytropic structure, {  $\mathrm{d}\varphi_a$ is the exterior derivative of the electric potential one-form, $\varphi(r)$ a scalar function describing the radial electric potential profile, and $(\mathrm{d}t)_a$ is the time-coordinate one-form \cite{Bardeen1973,Wald1984}.}

In order to model an exact BH solution, the processing has to incorporate a static, spherically symmetric $D$-dimensional spacetime metric ansatz with $g_{tt}\,g_{rr}=-1$ involving the unknown metric function ${  A(r)}$, such that
 \begin{align}
    \mathrm{d}s^2=-{  A(r)}\,\mathrm{d}t^2+{  A(r)}^{-1}\,\mathrm{d}r^2+r^2\,\mathrm{d}\Omega_{D-2}^2,\label{met}
 \end{align}
where $\mathrm{d}\Omega_{D-2}^2$ identifies the line element of a $(D-2)$-dimensional hypersurface $\Sigma$ with constant curvature $(D-2)(D-3)$. It is accordingly described by
    \begin{eqnarray}
        \mathrm{d}\Omega_{D-2}^2 =\mathrm{d}\theta^2+\sin^2\theta\left[\mathrm{d}\phi_1+\sum_{i =2}^{D-3}\prod_{j= 1}^{i-1}\sin^2\phi_j\mathrm{d}\phi_i^2\right],
        \nonumber\\
    \end{eqnarray}
with $\theta \in [0, \frac{\pi}{2}]$. It is then handy to specify the coordinates $x_i$ for a spherical hypersurface as follows
    \begin{eqnarray}
        x_1&=& r\cos\theta    ,\\
x_i&=&r\sin\theta\cos\phi_{D-i-1} 
\prod_{j= 1}^{D-i-2}\sin\phi_j,\,\, i= 2,\cdots,D-2 \\
        x_{D-1}&=&r\sin\theta\prod_{j= 1}^{D-4}\sin\phi_j,
    \end{eqnarray}
where $r$ is the radius of the hypersphere and the angles are of the standard ranges.
    
In this study, it is worth considering the perfect fluid state, which is featured in terms of the stress-energy tensor
\begin{equation}
    \mathcal{T}_{\mu\nu}=\left(\rho+p\right)u_\mu u_\nu+ p g_{\mu\nu}\,.
\end{equation}
Here, $\rho$ and $p$ are, respectively, the energy density and isotropic pressure, which are gauged by an observer moving with the fluid, and $u_\mu$ denotes the velocity vector $D$. Numerous studies are currently being carried out in the field of GR, involving static spherically symmetric solutions with the surrounding part of typical perfect fluid (dust, radiation, {  DE}, or ghost energy) with an EoS $p = \omega \rho$ ($\omega$ is a constant) \cite{Kiselev:2002dx,Li:2014ixn,Setare:2007jw,Benaoum:2012uk,Bilic:2002chg,Chen:2005qh,Arun:2017dm,Kubiznak:2015bh}. Multiple indicators suggest that the ideal cosmological fluid around the BH may be considered anisotropic because of gravitational influence. By way of example, the polytropic gas with a nonlinear EoS has been widely deployed in the literature to probe the features of compact objects. As polytropes are self-gravitating gaseous spheres, they can serve as a \textit{fuzzy} approximation to more refined stellar models. Furthermore, it helps in describing the internal structure of {  NS}, inclusive of their maximum mass, surface temperature, pulsar glitches, and other features, due to its efficient fit to the EoS of NS. On the other hand, certain evidence invokes a scalar tachyon field $\phi$ which is held to be the source of {  DE}, and a potential tachyon field $V(\phi)$, with Born-Infeld type Dirac Lagrangian $\mathcal{L}_\phi=-V(\phi)\sqrt{1-g^{\mu\nu}{ \partial_{\mu}\partial_{\nu}}\phi}$ \cite{Garousi:2000tr}. Further, drawing on ideas in terms of the $k$-essence scalar field concepts, the polytropic structure can be reconstructed as a scalar field gas. This reconstruction is generated in fact by considering the $K$-essence scalar field action, $S=\int \mathrm{d}^4x\sqrt{-g}\, p(\phi,\chi)$ with $p(\phi,\chi)$ being the Lagrangian density \cite{MagalhaesBatista:2009cus,Raposo:2018rjn}. The foregoing implies that the polytropic structure is properly emulated such that its radial pressure is distinct from the tangential pressure. This is in accordance with anisotropic fluids, entailing a covariant form of the stress-energy tensor for the polytropic structure, given by 
\begin{equation}\label{T1}
    \mathcal{T}_{\mu\nu}=\left(\rho+p_t\right)u_\mu u_\nu-p_t g_{\mu\nu}+\left(p_r-p_t\right)\chi_\mu \chi_\nu\,,
\end{equation}
in which $p_r$ is the radial pressure along the direction of $\chi_\mu$, $p_t$ is the tangential pressure perpendicular to $\chi_\mu$, and $\chi_\mu$ is the unit spacelike vector perpendicular to the velocity $u_\mu$. In addition, $u_\mu$ and $\chi_\mu$ comply with the constraint $u_\mu u^\mu=-\chi_\mu \chi^\mu=1$.

The present examination as regards the polytropic structure is performed by considering the frame in comoving with the fluid, allowing us to have $u^a=\sqrt{{  A(r)}}\,\delta_0^a$ and $\chi^a=1/\sqrt{{  A(r)}}\,\delta_1^a$. Accordingly, the stress-energy tensor of {  Eq.} (\ref{T1}) can be restated as follows
\begin{equation}\label{T2}
    \mathcal{T}_\mu^\nu=-\left(\rho+p_t\right)\delta_\mu^0 \delta^\nu_0+p_t\delta_\mu^\nu+\left(p_r-p_t\right)\delta_\mu^1\delta_1^\nu\,,
\end{equation}
where the term $p_r-p_t$ is termed the anisotropic factor, and for $p_r=p_t$, the case will be subject to the standard isotropic background.

For the purpose of exhaustiveness, we assume that the scalar field gas is in a state on either side of an event horizon identified by the stress energy \eqref{T2}. To be more precise, it should be noted that inside the horizon, where $g_{rr} <0$ and $g_{tt} >0$, the behavior of the spatial coordinate $r$ is similar to that of the temporal coordinate $t$. The energy density is thus $\mathcal{T}_r^r= p_r$, while the pressure along the spatial direction $t$ is defined as $\mathcal{T}_t^t=-\rho$. Based on this connection, the energy density and the pressure remain continuous, provided that the criterion $p_r=-\rho$ is verified. However, in the cases of $p_r\neq-\rho$ and $\rho(r_h)\neq 0$, the pressure at the horizon is discontinuous, and the relative phase of the solution evolves dynamically. 

Going forward, we will address just the case where $p_r=-\rho$ \cite{Kiselev:2002dx}, where the polytropic structure is static and, barring restrictions on the solution, the energy density is continuous across the horizon. Specifically, borrowing ideas from anisotropic fluids, the tangential pressure $p_t$ is constrained by adopting isotropic averaging over the angles and asserting that $\braket{\mathcal{T}_i^{(D)j}}=p(r) \delta_i^j$. Thus, it is possible to get
\begin{equation}\label{ptt}
   p(r) =p_t+\frac{1}{D-1}\left(p_r-p_t\right),
\end{equation}
where the standard formula $\braket{\delta_i^1 \delta_1^j}\equiv\frac{1}{D-1}$ is taken into consideration. Pragmatically, in view of analogous concepts, the standard tangential pressure formulation for quintessential {  DE} is given taking into account Eq. (\ref{ptt}) as $p_t = \frac{1}{D-2}\left((D-1)\omega+1\right)\rho$, which is in line with the radial pressure $p_r = -\rho$.

The polytropic structure is characterized to exhibit a non-linear EoS as $p = -\gamma \rho^{1+\frac{1}{n}}$, where $\xi$ is a positive parameter. In considering the rule $p_r = -\rho$, the tangential pressure of the polytropic structure is $p_t =\frac{1}{D-2}\rho-\frac{(D-1)}{(D-2)}\gamma \rho^{1+\frac{1}{n}}$. Therefore, the elements of the stress-energy tensor of the polytropic structure can be specified in terms of the expressions
\begin{align}
    \mathcal{T}_t^t&=\mathcal{T}_r^r=-\rho\label{m1},\\
\mathcal{T}_{\theta_{1}}^{\theta_{1}}&=\mathcal{T}_{\theta_{i}}^{\theta_{i}}=\frac{1}{D-2}\rho-\frac{(D-1)}{(D-2)}\gamma \rho^{1+\frac{1}{n}}\label{m2}.
\end{align}
We later show that the anisotropy of the polytropic structure diminishes and that the EoS $p = -\gamma \rho^{1+\frac{1}{n}}$ is retained on the cosmological scale.

\section{Exact solutions}
\label{ex}
Due to the static and spherically symmetric characteristics of spacetime, the requirement $\mathcal{T}_t^t =\mathcal{T}_r^r$ is fully fulfilled. Thus, the elements of the gravitational FE \eqref{eom} in $D$-dimensions are given by 
    \begin{eqnarray}
\mathcal{I}_t^t=\mathcal{I}_r^r&=&\frac{1}{2r^2}(D-2)(D-3)({  A(r)}-1)
+\frac{1}{2r}(D-2){  A'(r)}-\frac{3}{L^2},\label{g1}\\
     \mathcal{I}_{\theta_i}^{\theta_i}= \mathcal{I}_{\theta_1}^{\theta_1}&=&\frac{{  A''(r)}}{2}+\frac{({  A(r)}-1)(D-3)(D-4)}{2r^2} 
     +\frac{(D-3){  A'(r)}}{r}-\frac{3}{L^2}.\label{g2}
    \end{eqnarray}

On the other hand, insights into the structure of Maxwell's charged source are thus explored by considering the time component of Eq. \eqref{em} in conjunction with the space-time metric \eqref{met}, giving the following differential equation
\begin{equation}
    (D-2)\varphi'(r)+r\,\varphi''(r)=0 ,
\end{equation}
which admits an exact electric field given by
\begin{equation}
   \varphi(r) = \begin{cases}\label{EEMM}
    -\frac{q}{(D-3)r^{D-3}}+\Phi_0\hspace{0.5cm} D>3, 
     \vspace{3mm}\\q\, \text{ln} r+\Phi_0 \hspace{1.6cm}  D=3.
    \end{cases}\,
\end{equation}
where $q$ and $\Phi_0$ are integration constants.

As a rule of thumb, the integration constant $q$ is correlated to the physical parameter of the charge $Q$ on a two-dimensional sphere of infinite radius by means of Gauss's law
\begin{align}
    Q&=\frac{1}{4\pi}\int_{S^2_{\infty}}\star \mathbf{F}=\frac{1}{8\pi}\int_{S^2_{\infty}}F^{ab}\,\epsilon_{abc_1c_2\cdots c_{D-2} }\nonumber\\
    &=\frac{1}{4\pi}\int_{S^2_{\infty}}\varphi'(r)\sqrt{-g}\,\mathrm{d}\theta_1\,\mathrm{d}\theta_2\cdots \mathrm{d}\theta_{D-2} \nonumber\\
    &=\frac{q}{4\pi}\omega_{D-2},
\end{align}
where $\omega_{D-2}$ is the volume of the unit $(D-2)$-dimensional spherical topology\footnote{$\omega_{D-2}=\frac{2\pi^{\frac{D-1}{2}}}{\Gamma(\frac{D-1}{2})}$}.

Seeking an analytically exact solution for the surrounding polytropic structure in the context of GR must take into account the dual equations of the gravitational field and the matter field (\ref{eom}-\ref{em}). Therefore, a specific analysis of the polytropic structure side using the Bianchi identity $(T^{\mu\nu}_{;\nu}=0)$ could supply a first-order differential equation in terms of $\rho(r)$ such that 
\begin{equation}\label{ut}
    r\,\rho'(r) +(D-1)\,\rho(r)-\gamma(D-1) \rho(r)^{1+\frac{1}{n}} =0\,,
\end{equation}
where the first prime refers to the first derivative with respect to the radial variable $r$. An attempt to analytically solve Eq. \eqref{ut} results in an exact solution for the polytropic structure energy density in the form of
\begin{equation}
\rho(r) =\left(\xi^2\, r^{\frac{D-1}{n}}+\gamma\right)^{-n},\label{ho}
\end{equation}
where $\xi$ is a normalisation parameter specifying the intensity of the polytropic scalar field gas. In closer scrutiny, Eq. (\ref{ho}) revolves around the result of the stress energy tensor conservation law $\partial_\mu T^{\mu\nu}=0$.

To gain a deeper insight into the behavior of the polytropic structure, it is worth examining its criticality at certain limits. In this respect, at large radial coordinates (i.e. $\xi^2 r^{\frac{D-1}{n}}\gg \gamma)$, it gives rise to
\begin{equation}
    \rho(r)\sim \xi^{-2n}\big/ r^{D-1}\,,
\end{equation}
which states that the polytropic structure behaves like a matter content whose energy density varies with $r^{D-1}$. On the other hand, at small radial coordinates (i.e. $\xi^2 r^{\frac{3}{n}}\ll \gamma$), we obtain 
\begin{equation}
     \rho(r)\sim \gamma^{-n}   \,,
\end{equation}
which in turn means that the polytropic structure resembles a positive cosmological constant in a small-scale scenario, and that the closer it gets to the BH, the more it gravitationally clumps up. 

Alternatively, picking $r\rightarrow \infty$ results in properties 
\begin{align}
    p_r&\rightarrow-\left(\xi^2 r^{\frac{D-1}{n}}\right)^{-n} ,\\ 
    p_{\theta,\phi}& \rightarrow\frac{1}{D-2}\left(\frac{\xi^{-2n}}{r^{D-1} }\right)\left(1-\gamma (D-1)\xi^{-2}r^{-\frac{D-1}{n}}\right),
\end{align}
which means that the anisotropic nature is not guaranteed at large distances. In Table \ref{Tab1}, we work out the large distance bounds on the pressure components for the polytropic structure, revealing that the anisotropic factor asymptotically fails to tend to zero. This is unlike the effects observed in the quintessential fluid \cite{Kiselev:2002dx} and CDF \cite{Sekhmani:2024udl}, where the anisotropy decreases with distance; for the polytropic structure, the anisotropy factor reaches zero in the case of $\gamma=2(D-2)\left(r^{(D-1)/n}\xi^2\right)\big/(D-1)$. This lingering anisotropy raises a number of implications for the evolution of the universe, such as Hubble expansion anisotropies and the fluxes of mass phenomenon. These insights are supported by an array of observational surveys, such as those focused on galaxy clusters \cite{Migkas:2017vir,Migkas:2020fza,Migkas:2021zdo} and type Ia supernovae (SN Ia) within the framework of the Lematre-Tolman-Bondi (LTB) patterns \cite{DelCampo:2012wkb}. Our preliminary findings pave the way for more in-depth investigations into the polytropic structure, drawing on the studies referred to herein and others.

For a slightly clearer insight into the behavior of our solution, we shall focus on the classical ECs, i.e., the null energy condition (NEC), the dominant energy condition (DEC), the weak energy condition (WEC), and the strong energy condition (SEC), defined as follows~\cite{Kontou:2020bta}:
 \begin{figure*}[tbh!]
    \centering
    \begin{subfigure}[b]{0.5\textwidth}
        \centering
        \includegraphics[scale=0.88]{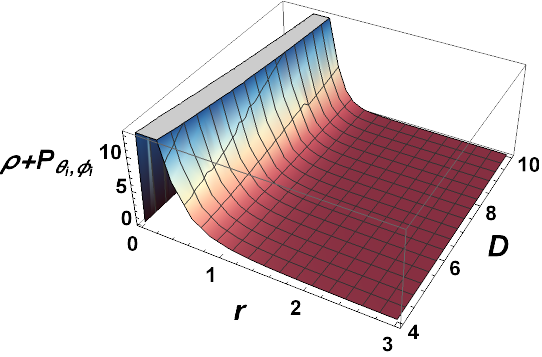}
    \end{subfigure}%
    \hfill
    \begin{subfigure}[b]{0.5\textwidth}
        \centering
        \includegraphics[scale=0.9]{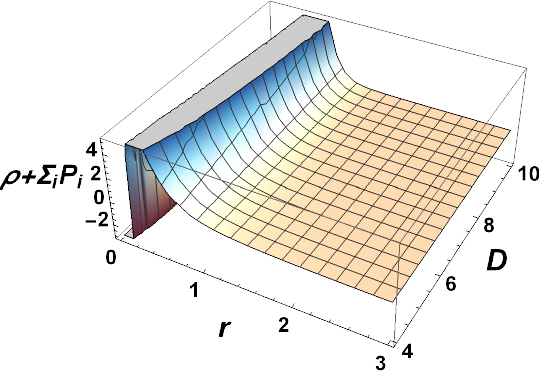}
    \end{subfigure}%
    \\
    \begin{subfigure}[b]{0.5\textwidth}
        \centering
        \includegraphics[scale=0.9]{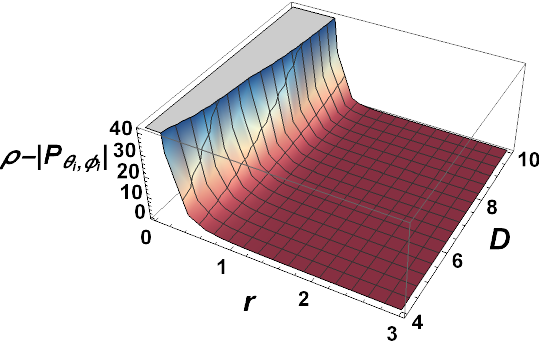}
    \end{subfigure}
    \caption{The variation of $\rho+\sum_i P_i$ (strong energy condition), $\rho+P_{\theta_i,\phi_i} $ (null energy condition), and $\rho-\mid P_{\theta_i,\phi_i}\mid $ (dominant energy condition) against $r$ for multiple values of the parameter dimension $D$ and for the fixed set, i.e., $\gamma=0.095$, $\xi=1$, and $n=2$.}
    \label{fig2}
\end{figure*}
\begin{eqnarray}\label{18}
\textbf{WEC}&:& \rho\geq 0,\; \rho+P_i\geq 0,\nonumber\\[2mm]
\textbf{SEC}&:& \rho+\sum_{i}P_i\geq 0,\, \;\rho+P_i\geq 0,\nonumber \\[2mm]
\textbf{NEC}&:& \rho+P_i\geq 0,\nonumber \\[2mm]
\label{19}
\textbf{DEC}&:& \rho\geq 0, \;|P_i|\leq \rho.
\label{20}
\end{eqnarray}
Accordingly, full expressions may be furnished as follows:
\begin{align}
   \rho + P_r =0,\,\, \rho+ P_{\theta_{i},\phi_{i}}&=\frac{D-1}{D-2}\bigg(\rho-\frac{\gamma}{\rho}\bigg),\nonumber\\
   \rho+ \sum_{i}P_i&=\rho-\left(\frac{D-1}{D-2}\right)\frac{\gamma}{\rho}\,,\\
   \rho-|P_r|=0,\,\,  \rho-|P_{\theta_{i},\phi_{i}}|&=\rho-\Bigl|\frac{1}{D-2}\rho-\left(\frac{D-1}{D-2}\right)\frac{\gamma}{\rho}\Bigr|\,.\nonumber
\end{align}
where explicitly, in terms of the BH parameter space, one has
    \begin{align}
\rho+ P_{\theta_{i},\phi_{i}}&=\frac{(D-1) }{D-2}\Bigg(\left(\gamma +\xi ^2 r^{\frac{D-1}{n}}\right)^{-n} 
-\gamma  \left(\left(\gamma +\xi ^2
   r^{\frac{D-1}{n}}\right)^{-n}\right)^{\frac{1}{n}+1}\Bigg),\nonumber\\
   \rho+ \sum_{i}P_i&=\left(\gamma +\xi ^2 r^{\frac{D-1}{n}}\right)^{-n} 
   -\frac{\gamma  (D-1) \left(\left(\gamma +\xi ^2
   r^{\frac{D-1}{n}}\right)^{-n}\right)^{\frac{1}{n}+1}}{D-2},\\
\rho-|P_{\theta_{i},\phi_{i}}|&=\frac{\left(\gamma +\xi ^2 r^{\frac{D-1}{n}}\right)^{-n}}{D-2}\Bigg(D-3+\gamma  (D-1) 
\left(\gamma +\xi ^2
   r^{\frac{D-1}{n}}\right)^{-1}\Bigg)\,.\nonumber
\end{align}
    
\begin{figure*}[ht!]
      	\centering{
       \includegraphics[scale=0.71]{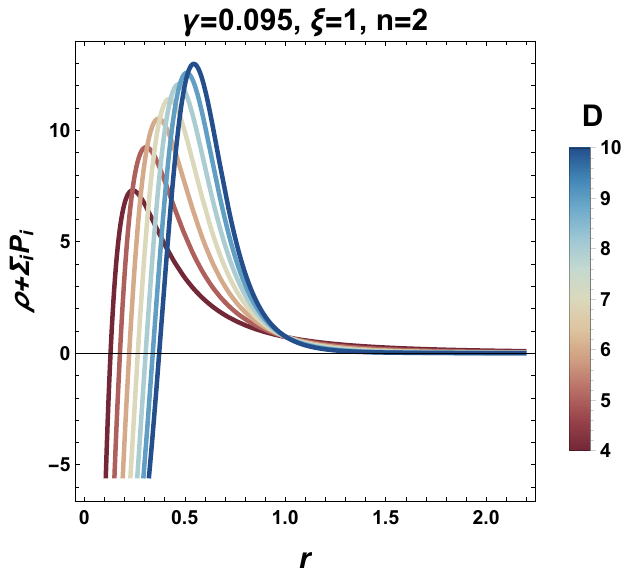} \hspace{2mm}
      	\includegraphics[scale=0.72]{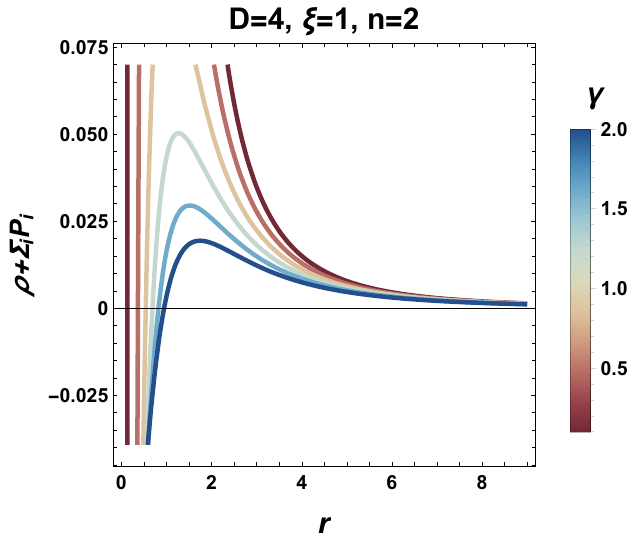} \hspace{2mm}
      	}
      	\centering{
       \includegraphics[scale=0.69]{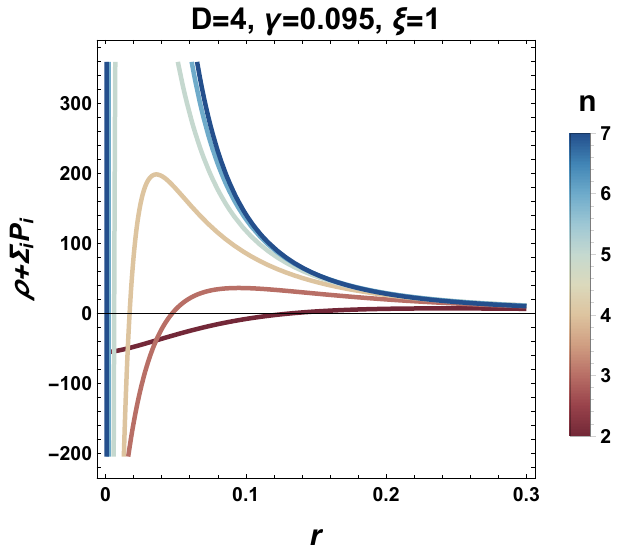} \hspace{2mm}
      	\includegraphics[scale=0.69]{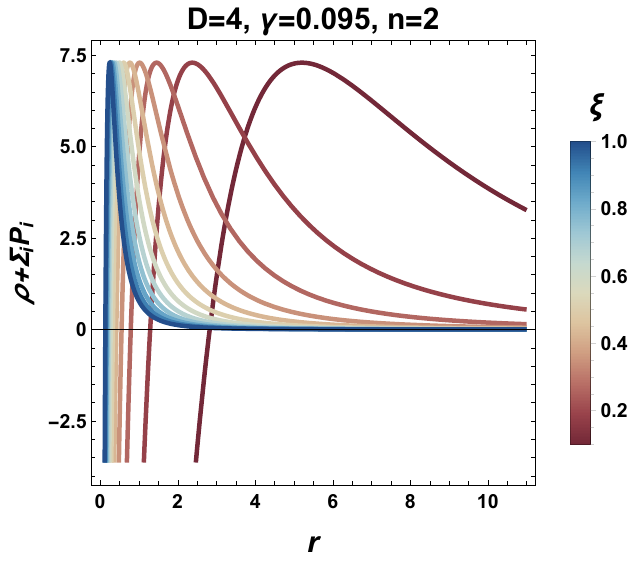} \hspace{2mm}
      }

      	\caption{Strong $(\rho+ \sum_{i= 1}^3P_i $) energy condition for several fixed parameters}
      	\label{fig3}
      \end{figure*}

To highlight the inspection of the violation/satisfaction of the constraints of ECs, Fig.~\ref{fig2} represents, for a multiple value of the BH parameter system, three types of ECs such as $\rho+\sum_{i}P_i$, $\rho+ P_{\theta_{i},\phi_{i}}$ and $\rho-|P_{\theta_{i},\phi_{i}}|$ against the variable $r$. Upon detailed scrutiny, $\rho+ P_{\theta_{i},\phi_{i}}$ and $\rho-|P_{\theta_{i},\phi_{i}}|$ are all definite positively, while $\rho+\sum_{i}P_i$ undergoes a sign change at $r_0=\left(\gamma/(D-2)\xi^2\right)^{\frac{n}{-1+D}}$, entailing a negative-to-positively switch between small $r$ and large one (see Fig. \ref{fig3}). Of interest is the fact that the root $r_0$ is effectively the transition point in the sign of $P_{\theta_{i},\phi_{i}}$, i.e. the point at which the tangential pressure switches from a repulsive to an attractive phase. In a nutshell, the concluding assessment and compliance reveals that the polytropic structure satisfies the NEC, WEC, and DEC criteria, though it is in violation of the SEC. Interestingly, this scenario is exactly the similar to the quintessential model for {  DE}. Roughly speaking, any violation of the SEC in GR has been demonstrated to be a violation of the attractive nature of gravity, as manifest in the case of {  DE}. However, this scenario is not necessarily viable in the context of extended severity, as shown for the extended gravity model of the model $f(R)$~\cite{Santos:2016vjg}.

It is similarly worthwhile approaching the EC set-up from a thermodynamic standpoint. Notice that the thermodynamic description of the BH system in the extended phase space is substantially given by the radius $r$, which is related to the thermodynamic volume by Eq.~\eqref{VVV}. 

These considerations are worth contrasting with current findings in the literature. For example, logotropic fluid-type BHs can violate the SEC for high radii, as argued in~\cite{Capozziello:2022ygp}. In turn, the study of~\cite{Fan:2016rih} discloses that, while the WEC is satisfied, the SEC, in fact, remains violated for regular Hayward-AdS BHs. According to~\cite{Rodrigues:2020pem}, such new solutions are suggested for regular BHs with multihorizons, inferring that the SEC is not satisfied at all inside the event horizon for all solutions, while the other ECs rely on the ratio of extreme charges of the isolated solutions. Here, in~\cite{Rodrigues:2020pem} further solutions for regular BHs featuring multihorizons are offered, concluding by stating that the SEC is never satisfied across the event horizon in all solutions, while the remaining ECs rely on the ratio of extreme charges of the isolated solutions. For non-rotating or spinning BHs in conformal gravity, analogous conclusions are offered in~\cite{Toshmatov:2017kmw}, where the SEC has to be satisfied for particular BH sizes that rely on the theory's ultimate scaling.

\begin{center}
\begin{table*}[t]
    \centering
    \caption{Quintessence, CDF and polytropic structures in \textit{higher}-dimensional spacetime formulation.}
    \scalebox{0.7}{\begin{tabular}{@{}c||c|c|c|c|c@{}}
    \hline
         anisotropic fluid  &  EoS  & $p_r$ & $p_t$ & $\rho$ & asymptotic behavior \\
         \hline\hline
       \text{Quintess. ({  DE})} \cite{Kiselev:2002dx} \,&\, $p=\rho \omega$ $\left(-1<\omega<-1/3\right)$&\, $-\rho$ \,&\, $\frac{1}{D-2}\left((D-1)\omega+1\right)\rho$ \,&\, $\frac{c\,\omega(D-1)(D-2)}{4r^{(D-1)(\omega+1)}}$ \,&\, $\begin{array}{lcl}
 \rho&\rightarrow&0\\
 p_r&\rightarrow&0\\
 p_t&\rightarrow&0
 \end{array}$   \\
 \hline
       \text{CDF} \cite{Sekhmani:2024udl} \,&\, $p=-\gamma/\rho$ $\left(\gamma>0\right)$ \,&\,$-\rho$ \,&\, $\frac{1}{D-2}\left(\rho-\frac{(D-1)\gamma}{\rho}\right)$ \,&\, $\sqrt{\gamma+\frac{Q^2}{r^{2(D-1)}}}$ \,&\, $\begin{array}{lcl}
 \rho&\rightarrow&\sqrt{\gamma}\\
 p_r&\rightarrow&-\sqrt{\gamma}\\
 p_t&\rightarrow&-\sqrt{\gamma}
 \end{array}$   \\
       \hline
       \text{Polytropic} \,&\, $p=-\gamma \rho^{1+\frac{1}{n}}$ $\left(\gamma>0\right)$ \,&\,$-\rho$ \,&\, $\frac{1}{D-2}\left(\rho-\frac{(D-1)}{\gamma\rho^{1+\frac{1}{n}}}\right)$ \,&\, $\left(\xi^2\, r^{\frac{D-1}{n}}+\gamma\right)^{-n}$ \,&\, 
       $\begin{array}{lcl}
 \rho&\rightarrow&\frac{\xi^{-2n}}{r^{(D-1)}}\\
 p_r&\rightarrow&-\frac{\xi^{-2n}}{r^{(D-1)}}\\
 p_t&\rightarrow&-p_r\big(\frac{1}{(D-2)}\\
 &+&\gamma(D-2)p_r^{\frac{1}{n}}\big)
 \end{array}$ 
 \\ \hline
          \end{tabular}}
    \label{Tab1}
\end{table*}
\end{center}

   	By taking into account the exact analytical solutions for the Maxwell electric field \eqref{EEMM} with respect to Einstein's equations \eqref{eom}-\eqref{em}, we are able to obtain two sets of second differential equations in terms of the metric function ${  A(r)}$, such that 
 \small
    \begin{eqnarray}
\mathcal{I}_t^t=\mathcal{I}_r^r&=&\frac{1}{2r^2}(D-2)(D-3)({  A(r)}-1)+\frac{(D-2)}{2r}{  A'(r)}
-\frac{3}{L^2}+\frac{q^2}{4r^{2(D-2)} }-\rho(r),\hspace{0.75cm}\label{g1a}\\
     \mathcal{I}_{\theta_i}^{\theta_i}= \mathcal{I}_{\theta_1}^{\theta_1}&=&\frac{{  A''(r)}}{2}+\frac{({  A(r)}-1)(D-3)(D-4)}{2r^2} 
     +\frac{(D-3){  A'(r)}}{r} -\frac{3}{L^2}
    -   \frac{q^2}{4r^{2(D-2)}}+\frac{1}{D-2}\rho(r)
    \nonumber\\
    & & \,-\frac{D-1}{D-2}\gamma\,\rho^{1+\frac{1}{n}}.\hspace{1cm}\label{g2a}
    \end{eqnarray}
\normalsize
While considering the energy density of the polytropic structure \eqref{ho}, and proceeding with the $\begin{psmallmatrix}t  \\t \end{psmallmatrix}$ components of the FEs allows us to obtain the following differential equation

    \begin{align}
     &(D-2) \left[(3-D) (1-{  A(r)})+r\, {  A'(r)}\right] 
     +2 r^2 \left(\left(\xi^2\, r^{\frac{D-1}{n}}+\gamma\right)^{-n}-3L^{-2} +\frac{q^2}{4r^{2(D-2)} } \right)=0 ,
    \end{align}
where an exact analytical solution for ${  A(r)}$ can be represented in the following form
    \begin{align}\label{metric}
  {  A(r)} =1-&\frac{2M}{r^{D-3}}+\frac{q^2}{r^{2(D-3)}} +\frac{ r^2}{L^2} 
  -
  \frac{2 r^2 \gamma ^{-n} \, _2F_1\left(n,n;n+1;-\frac{\xi^2 r^{\frac{D-1}{n}}}{\gamma }\right)}{(D-2) (D-1)}\,,
\end{align}
where $M$ represents the mass of the BH, in this case we are treating the BH as a point mass, which means that $M$ is a constant. Here, ${_2}F_1[a_1, a_2; a_3; a_4]$ is a hypergeometric function which describes the regular solution of the hypergeometric differential equation, is held to hold for $\lvert a_4\rvert<1$ and is specified by a series of powers of the form
\begin{equation}
{_2}F_1\left(a_1, a_2; a_3; a_4\right)=\sum_{p=0}^{\infty}\bigg((a_1)_p(a_2)_p\big/(a_3)_p\bigg)a_4^p/p! \,,
\end{equation}
with $(k)_p$ is the (rising) Pochhammer symbol~\cite{Abra}.
      
For the sake of disclosure of the asymptotic behavior of BH solution, such a large distance analysis is taken into account with respect to the metric function ${  A(r)}$, resulting in
\begin{equation}
    {  A(r)}\rightarrow 1+\frac{r^2}{L^2},
\end{equation}
which demonstrates that in the asymptotic limit, our BH solution is uniquely expressed by the length of AdS $L$. Furthermore, a thorough analysis showed that BH solution modeling of the background can be brought back to the higher-dimensional Einstein-Maxwell theory as \cite{Chamblin:1999tk}
 \begin{align}
  {  A(r)}\mid_{\gamma\rightarrow\infty} =1-&\frac{2M}{r^{D-3}}+\frac{q^2}{r^{2(D-3)}} +\frac{ r^2}{L^2}.
\end{align}

\begin{figure*}[!htp]
    \centering
    \begin{subfigure}[b]{0.5\textwidth}
        \centering
        \includegraphics[scale=0.9]{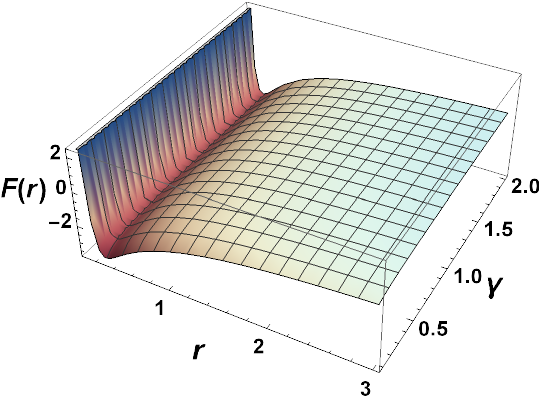}
        \label{ch1}
    \end{subfigure}%
    \hfill
    \begin{subfigure}[b]{0.5\textwidth}
        \centering
        \includegraphics[scale=0.9]{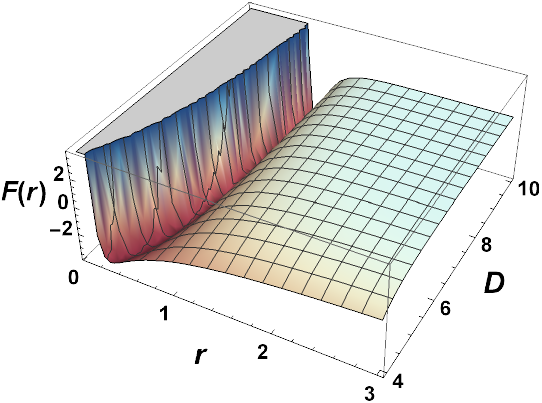}
        \label{ch2}
    \end{subfigure}%
    \caption{Left panel: The behavior of ${  A(r)}$ \eqref{metric} by varying $\gamma$ and setting $D = 4$. Right panel: The behavior of ${  A(r)}$ with varying $D$ and setting $\gamma = 0.095$. For the two panels, set $M = 1,\, L=1,\, q=0.5,\, \xi=1$ and $n =2$.} 
    \label{figb}
\end{figure*}

A concrete visualization of the behavior of the metric function ${  A(r)}$ is shown in a two-dimensional representation for different values of the parameters ($D, \gamma$) with a fixed set of rest parameters in Fig. \ref{figb}. In broad terms, it can be observed that the metric function exhibits two possible horizon radii, namely the Cauchy horizon (inner horizon) and the event horizon radius (outer horizon).

Next, we investigate in terms of curvature singularity tools the relevant features of BH solution. Thus, the verification of the uniqueness and singularity of BH solution is carried out by analysing the Ricci scalar $(\mathcal{R})$ and Kretschmann scalar ($\mathcal{R}_{\alpha\beta\mu\nu}\mathcal{R}^{\alpha\beta\mu\nu}$) invariants given, respectively, in the essence of the metric function \eqref{metric} by 
    \begin{align}\label{R}
    \mathcal{R}&=-\frac{D(D-1)}{L^2}+\frac{2 \left(\gamma +\xi ^2
   r^{\frac{D-1}{n}}\right)^{-n} \left(D\gamma  r^{\frac{1}{n}}+\xi ^2
   r^{D/n}\right)}{(D-2) \left(\xi ^2 r^{D/n}+\gamma 
   r^{\frac{1}{n}}\right)} 
   -(D-4) (D-3) q^2 r^{4-2 D},
\end{align}
and
\small
    \begin{align}\label{RRR}
&\mathcal{R}_{\alpha\beta\mu\nu}\mathcal{R}^{\alpha\beta\mu\nu} =-\Bigg(\frac{ \left(\xi ^2 r^{\frac{D-1}{n}}+\gamma\right)^{-n} \left(2 (2 D-5) \xi ^2 r^{D/n}+2 \gamma  (D-4) r^{\frac{1}{n}}\right)}{(D-2) \left(\xi ^2 r^{D/n}+\gamma 
   r^{\frac{1}{n}}\right)}
   \nonumber\\
   &-\frac{2 (D-3) \gamma ^{-n} \, _2F_1\left(n,n;n+1;-\frac{r^{\frac{D-1}{n}} \xi ^2}{\gamma }\right)}{D-1}
   -2 (D-3) r^{1-2 D} \left((D-2) M r^D+(5-2 D) q^2
   r^3\right)+\frac{2}{L^2}\Bigg)^2
   \nonumber\\
   &+2 (D-3) (D-2) \bigg(-\frac{2 \gamma ^{-n} \, _2F_1\left(n,n;n+1;-\frac{r^{\frac{D-1}{n}} \xi ^2}{\gamma }\right)}{(D-2) (D-1)}+r^{1-2 D} \left(q^2 r^3-2 M
   r^D\right)+\frac{1}{L^2}\bigg)^2
   \nonumber\\
   &+8 (D-2) \Bigg(\frac{(D-3) \gamma ^{-n} \, _2F_1\left(n,n;n+1;-\frac{r^{\frac{D-1}{n}} \xi ^2}{\gamma }\right)}{(D-2) (D-1)}+(D-3) r^{1-2 D} \left(M r^D-q^2
   r^3\right)\nonumber\\
   &-\frac{\gamma ^{-n} \left(\frac{\xi ^2 r^{\frac{D-1}{n}}}{\gamma }+1\right)^{-n}}{D-2}+\frac{1}{L^2}\Bigg)^2\,.
\end{align}
\normalsize
By carefully examining the expressions \eqref{R} and \eqref{RRR}, we can confirm that the BH solution by means of the metric function \eqref{metric} is singular. This holds for any given set of parameter spaces with the given constraint $D>3$. For practical considerations, the occurrence of the singularity is caused by the mass, the charge, and polytropic structure terms in the BH metric. However, opposing alternative situations can be thought of to abolish the singularity, such as the choice of $D=3$ leading to elimination of the singularity. Notably, to implement such removal of singular behavior, numerous studies invented such practical mechanisms for this purpose (see, for example, ~\cite{Balart:2014cga}). Throughout the sequel, we will consider no such situation and rely instead on the metric function \eqref{metric}. In a more concrete inspection, the singularity is evidenced on the basis of the Ricci scalar and the Kretschmann scalar at the centre $r=0$, which yields the results
\begin{align}
    \lim\limits_{r\to 0} \mathcal{R} &\approx\infty\,,\\
    \lim\limits_{r\to 0}\mathcal{R}_{\alpha\beta\mu\nu}\mathcal{R}^{\alpha\beta\mu\nu} &\approx\infty\,.
\end{align}
On the other hand, looking at large distance behavior is a worthwhile follow-up feature, since
\begin{align}
    \lim\limits_{r\to \infty} \mathcal{R} &\approx-\frac{D(D-1)}{L^2}\,,\\
    \lim\limits_{r\to \infty}\mathcal{R}_{\alpha\beta\mu\nu}\mathcal{R}^{\alpha\beta\mu\nu} &\approx\frac{2(D^2-D-4)}{L^2}\,,
\end{align}
which point out that the Kretschmann and Ricci scalars exhibit a finite term at large distances. In summary, the scalar tools demonstrate that the BH solution is unique and significantly altered by the AdS backgrounds.

\section{Thermodynamics and Smarr relation}
\label{secther}
In the context of gauge/gravity duality, strongly coupled gauge theories can be interrelated with the closely associated weakly coupled string theories. From the holographic perspective, bulk string theory can in turn be apt to inform boundary gauge theory. By means of the AdS/CFT correspondence \cite{Maldacena:1997re,Witten:1998qj}, in which the conformal field theory maps onto the asymptotically AdS spacetime in a higher dimension, the thermodynamic characteristics of a BH can be made to disclose the characteristics of the dual physical state. Analogously, the BH horizon in the asymptotically AdS spacetime provides insight into the finite temperature of its dual field theory.

In the remainder of this part, we shall explore the thermodynamic quantities of charged AdS BHs solution with a surrounding polytropic structure and ascertain the first law of thermodynamics as well as the Smarr relation. So as to establish the Hawking temperature, it needs initially to be considered the surface gravity, given by~\cite{Kubiznak:2016qmn}
\begin{equation}\label{r30}
    \kappa=\left(-\frac{1}{2}\nabla_\mu\xi_\nu\nabla^\mu\xi^\nu\right)^{1/2}=\frac{1}{2}F'\left(r_h\right),
\end{equation}
with $\xi^\mu=\partial/\partial t$ is a Killing vector for the metric. Thus, the formula $T=\kappa/2\pi$ gives the Hawking temperature expressed in terms of the parameters of the BH system, as follows,
    \begin{align}\label{43}
        T&=\frac{1}{4 \pi  r_h}\Biggl\{r_h^2 \Bigg(-\frac{2 \left(\gamma +\xi^2 r_h^{\frac{D-1}{n}}\right)^{-n}}{D-2}+3 q^2 r_h^{4-2 D}-\frac{1}{L^2}\Bigg)
        +\left(-q^2 r_h^{6-2 D}+\frac{r^2}{L^2}+1\right)D-3\Biggr\},
    \end{align}
where the horizon constraint $F\left(r_h\right)=0$ has been taken into account. 

On the other hand, entropy might arise from the fact that it is computed in accordance with the so-called area law \cite{Gibbons:1977mu,Bekenstein:1973ur}, which states that $S_{BH}$ is one quarter of area of the BH event horizon given by
\begin{equation}\label{mass1}
   S_{BH} = \frac{\omega_{D-2}}{4}\,r_h^{D-2}.
\end{equation}

In the limit as $r$ approaches infinity, the mass of the AdS BH solution with an encompassing polytropic structure is expressed as
\begin{equation}\label{ms}
    M_{ADM}=\frac{(D-2) \omega_{D-2}}{16\pi}M\,,
\end{equation}
where $M$ is defined by the formula $F(r_h) = 0$ in Eq. \eqref{metric}. Thus, the Arnowitt-Deser-Misner equation may be understood in terms of the parameter BH spacetime as  
(ADM) \cite{Ashtekar:1984zz,Ashtekar:1999jx} mass is given by
\small
    \begin{align}
  M_{ADM}&=\frac{(D-2) \omega_{d-2}}{8\pi}r_h^{D-3} \Biggl\{4 \pi ^{3-D} Q^2 r_h^{6-2 D} \Gamma \left(\frac{D-1}{2}\right)^2 
  +\frac{r^2}{L^2}+1-\frac{2 r^2 \gamma ^{-n} \, _2F_1\left(n,n;n+1;-\frac{\xi^2 r_h^{\frac{D-1}{n}}}{\gamma }\right)}{(D-2) (D-1)}\Biggr\}.\label{mass}
\end{align}
\normalsize

To achieve a proper derivation of the relevant thermodynamic quantities, a useful insight is the implementation of the first law of BH thermodynamics. Concretely, the first law can be written as~\cite{Cai:1998vy}
\begin{equation}\label{r32}
    \mathrm{d} M=T\mathrm{d}S+\sum_i \mu_i \,\mathrm{d}\mathcal{N}_i\,,
\end{equation}
where $\mu_i$ represent the chemical potentials associated with the conserved charges $\mathcal{N}_i$. Consequently, the first law of BH thermodynamics is fulfilled by the physical properties of BHs with a surrounding polytropic configuration defined by 
\begin{equation}\label{r40}
\mathrm{d}M=T\mathrm{d}S+\mathcal{L}\mathrm{d}L+U \mathrm{d}Q+\mathcal{A}\mathrm{d}\gamma.
\end{equation}
Here, $T$ corresponds to the Hawking temperature, which is correlated with the horizon's gravity, and $\mathcal{L}$ refers to the thermodynamic variable dual to $L$, $U$ is the electric potential, and $\mathcal{A}$ represents the thermodynamic variable associated with $\gamma$. To guarantee that the fundamental equation depends only on extensive variables, the variable $L$, which is related to the cosmological constant, should be considered.

Based on the specific volume $v =\frac{4r_h}{D-2}$ and making use of {  Eq.} \eqref{ms}, the conjugate quantity of the pressure results in the following amount:
\begin{equation}\label{VVV}
 V=\frac{\omega_{D-2}}{D-1}r_h^{D-1}\,.
\end{equation}
In parallel, to properly set up the thermodynamic phase space environment, the parameter function $F(r_h, M, L, Q, \gamma)$ ought to vanish under any given parameter transformation. Related notes offer analogous arguments for recognising the constraints $F(r_h, M, L, Q, \gamma)= 0$ and $\delta F(r_h, M, L, Q, \gamma)= 0$ on the evolution along the parameter space. However, an alternative solution is envisaged by treating the mass parameter, $M$, as a function of the parameters $M(r_h,L, Q, \gamma)$ as well.

More specifically, the thermodynamic variables in question are entropy $S$, pressure $P$, physical charge $Q$ and tye variable $\mathcal{A}$, which is the conjugate potential of the parameter $\xi$. It is then fairly straightforward to redefine $M=M(S,Q,L, \gamma)$ as a 1-form differential such that 
\begin{eqnarray}\label{r41}
\mathrm{d}M&=&\bigg(\frac{\partial M}{\partial S}\bigg)_{Q,L,\gamma}\mathrm{d}S+\bigg(\frac{\partial M}{\partial Q}\bigg)_{S,L,\gamma}\mathrm{d}Q
+\bigg(\frac{\partial M}{\partial L}\bigg)_{S,Q,\gamma}\mathrm{d}L 
+\bigg(\frac{\partial M}{\partial \gamma}\bigg)_{S,Q,L}\mathrm{d}\gamma, ,\hspace{1cm}
\end{eqnarray}
where the components stand for temperature $T$, thermodynamic volume $V$,the related electric potential, and the conjugate potential of the parameter $\xi$ which are giving as
\begin{align}\label{r42}
T&=\bigg(\frac{\partial M}{\partial S}\bigg)_{Q,L,\gamma},\,
U=\bigg(\frac{\partial M}{\partial Q}\bigg)_{S,L,\gamma},
\mathcal{L}=\bigg(\frac{\partial M}{\partial L}\bigg)_{S,Q,\gamma},\, \mathcal{A}=\bigg(\frac{\partial M}{\partial \gamma}\bigg)_{S,L,Q}.\, 
\end{align}

The above findings can otherwise be expressed as a function of the variation in parameter space of the condition identified by $F(r_h, M, Q, L,\gamma)$ as
\begin{equation}\label{r43}\nonumber
0=\mathrm{d}F(r_h, M, Q,L,\gamma)=\frac{\partial F}{\partial r_h}\mathrm{d}r_h+\frac{\partial F}{\partial M}\mathrm{d}M+\frac{\partial F}{\partial Q}\mathrm{d}Q+\frac{\partial F}{\partial L}\mathrm{d}L+\frac{\partial F}{\partial \gamma}\mathrm{d}\gamma,
\end{equation}
which can be recast into another term for $\mathrm{d}M$, whichever gives the following:
\begin{align}\label{r44}
\mathrm{d}M&=\bigg(\frac{1}{4\pi}\frac{\partial F}{\partial r_h}\bigg)\bigg(-\frac{1}{4\pi}\frac{\partial F}{\partial M}\bigg)^{-1}\mathrm{d}r_h
+\bigg(-\frac{\partial F}{\partial M}\bigg)^{-1}\bigg(\frac{\partial F}{\partial Q}\bigg)\mathrm{d}Q\nonumber\\
&+\bigg(-\frac{\partial F}{\partial M}\bigg)^{-1}\bigg(\frac{\partial F}{\partial L}\bigg)\mathrm{d}L 
+\bigg(-\frac{\partial F}{\partial M}\bigg)^{-1}\bigg(\frac{\partial F}{\partial \gamma}\bigg)\mathrm{d}\gamma.
\end{align}
To be more specific, Eq. (\ref{r40}) would be consistent with Eq. (\ref{r44}) which involves the presence of temperature, geometrically defined as follows
\begin{equation}\label{r45}
T=\frac{1}{4\pi}\frac{\partial F}{\partial r_h},
\end{equation}
and then, the entropy can be defined by
\begin{eqnarray}\label{r46}
\mathrm{d}S=\bigg(-\frac{1}{4\pi}\frac{\partial F}{\partial M}\bigg)^{-1}\,\mathrm{d}r_h.
\end{eqnarray}
This can also be derived using Wald's formalism; in other words, $\delta S = \int\frac{\partial \mathcal{L}}{\partial R}$ as long as $\mathrm{d}F = 0$ is satisfied.

Furthermore, the thermodynamic volume and the conjugate potential are defined according to the formula:
\begin{align}\label{r47}
U&=\bigg(\frac{\partial M}{\partial Q}\bigg)_{S,L,\gamma}=\bigg(-\frac{\partial F}{\partial M}\bigg)^{-1}\bigg(\frac{\partial F}{\partial Q}\bigg),\\
\mathcal{L}&=\bigg(\frac{\partial M}{\partial L}\bigg)_{S,Q,\gamma}=\bigg(-\frac{\partial F}{\partial M}\bigg)^{-1}\bigg(\frac{\partial F}{\partial L}\bigg),\\
\mathcal{A}&=\bigg(\frac{\partial M}{\partial \gamma}\bigg)_{S,Q,L}=\bigg(-\frac{\partial F}{\partial M}\bigg)^{-1}\bigg(\frac{\partial F}{\partial \gamma}\bigg).
\end{align}
As far as the BH system is concerned, the enthalpy is represented by the total mass of the system. Depending on the parameters of the BH system and in the extended phase space environment, the thermodynamic volume, electric potential, and $\mathcal{A}$ can be expressed as follows:
\begin{align}\label{r52}
U&=\left(\frac{\partial M}{\partial Q}\right)_{S,L,\gamma}=\frac{2\pi^{\frac{-D+3}{2}}\Gamma(\frac{D-1}{2})}{r_h^{-3+D}}Q, \\
\mathcal{L}&=\left(\frac{\partial M}{\partial L}\right)_{S,Q,\gamma}=-\frac{r_h^{D-1}}{L^3},\\
\mathcal{A}&=\left(\frac{\partial M}{\partial \xi}\right)_{S,Q,L}=\frac{n r_h^{D-1} \left(\gamma +\xi^2 r_h^{\frac{D-1}{n}}\right)^{-n}}{\gamma  (D-2) (D-1)} ,
\end{align}
and the Hawking temperature is found to be in the extended phase space as follows:
    \begin{eqnarray}\label{Tbar}
      T=\bigg(\frac{\partial M}{\partial S}\bigg)_{P,Q}
      &=&\frac{1}{4 \pi  r_h}\Biggl\{r_h^2 \left(-\frac{2 \left(\gamma +Q^2 r_h^{\frac{D-1}{n}}\right)^{-n}}{D-2}-\frac{1}{L^2}\right)
        +\frac{r_h^2D}{L^2}
      \nonumber\\
       & &\, -4 (D-3) \pi ^{3-D} Q^2 r_h^{6-2 D} \Gamma \left(\frac{D-1}{2}\right)^2+D-3\Biggr\}.\hspace{0.6cm}
    \end{eqnarray}

    Inspecting the related Smarr relation is useful. For that reason, the application of Euler's theorem~\cite{Kastor:2009wy,Altamirano:2014tva}  can provide the dimensional analysis for the implication of the polytropic structure such that $[M]=(D-3)$, $[L]=1$, $[S]=(D-2)$, $[Q]=(D-3)$, $[\gamma]=2/n$ and $[\xi]=-(D-3)/2n$. We find
    \begin{equation}\label{r48}
(D-3) M=(D-2) TS+\mathcal{L}L+(D-3)U Q+\frac{2}{n}\mathcal{A}\gamma.
\end{equation}
It is worth mentioning that for a particular case of the parameter $n$, the Smarr formula relation brings to a reduced one, such that for $n=1$, 
\begin{equation}\label{r48a}
(D-3) M=(D-2) TS+\mathcal{L}L+(D-3)U Q+2\mathcal{A}\gamma\,,
\end{equation}
and for $n=2$, one has
\begin{equation}\label{r48b}
(D-3) M=(D-2) TS+\mathcal{L}L+(D-3)U Q+\mathcal{A}\gamma\,,
\end{equation}
where all the thermodynamic variables emerge. These quantities in the classical limit where $(Q=0$ $\&$ $\xi=0)$ are consistent with the corresponding quantities for the AdS Schwarzchild BH~\cite{Kumar:2020cve}.

Barrow introduced a fractal shape for the BH horizon, which enhances its surface area. According to Barrow's modifications, the modified entropy relation is represented by \cite{Barrow:2020tzx,barrow2,Luciano:2023roh,barrow3,leon1}
\begin{equation}
    S=\left(\frac{A}{A_0}\right)^{1+\frac{\Delta}{2}}.
\end{equation}
So, the next stage consists of drawing up a thermodynamic structure based on Barrow entropy. We might begin by identifying the link between the horizon radius and Barrow entropy by means of $S_{BH}$ as
\begin{equation}
   S = \left(\frac{\omega_{D-2}}{4}\,r_h^{D-2}\right)^{1+\frac{\Delta}{2}}\,, 
\end{equation}
where a reciprocal link can be illustrated in terms of the horizon radius as
\begin{equation}\label{rbar}
    r_h=\left(\frac{4}{\omega_{D-2}}S^{\frac{2}{1+\Delta}}\right)^{\frac{1}{D-2}}.
\end{equation}
\section{Thermodynamic topology}
\label{topology}
Recent developments in the study of BH thermodynamics have highlighted the importance of thermodynamic topology as a tool for exploring the complex phase structure of BHs. Initially, topological methods were applied to investigate phenomena such as light rings and time-like circular orbits in BH spacetimes \cite{PRL119-251102, PRL124-181101, PRD102-064039, PRD103-104031, PRD105-024049, PRD108-104041, 2401.05495, PRD107-064006, JCAP0723049, 2406.13270}. The application of topological techniques in BH thermodynamics was first introduced in \cite{28}, drawing inspiration from the earlier work of Duan \cite{d1, d2} in relativistic particle systems. This approach revolves around the concept of topological defects represented by zero points of a vector field, which correspond to the critical points of the system. These zero points serve as indicators of phase transitions and can be classified by their winding numbers, offering a way to categorize BH systems into distinct topological classes based on shared thermodynamic properties.

In particular, the classification into different topological classes is determined by the structure and nature of these critical points. This methodology has been widely employed in various BH models across the literature \cite{PRD105-104053, PLB835-137591, PRD107-046013, PRD107-106009, JHEP0623115, 2305.05595, 2305.05916, 2305.15674, 2305.15910, 2306.16117, PRD106-064059, PRD107-044026, PRD107-064015, 2212.04341, 2302.06201, 2304.14988, 64, 2309.00224, 2312.12784, 2402.18791, 2403.14730, 2404.02526}. 

In this paper, we utilize the topological framework discussed in \cite{29}, which is particularly well-suited for studying BH thermodynamics. By adopting the off-shell free energy method, BHs are treated as topological defects within their thermodynamic structure. This method provides insights into both the local and global topological aspects of BHs, using winding numbers to describe their topological charge and stability. Crucially, the stability of a BH can be inferred from the sign of its winding number, underscoring the connection between thermodynamic topology and BH stability. This method has been successfully applied to numerous BH systems in a variety of gravitational frameworks \cite{PRD107-064023, PRD107-024024, PRD107-084002, PRD107-084053, 2303.06814, 2303.13105, 2304.02889, 2306.13286, 2304.05695, 2306.05692, 2306.11212, EPJC83-365, 2306.02324, PRD108-084041, 2307.12873, 2309.14069, AP458-169486, 2310.09602, 2310.09907, 2310.15182, 2311.04050, 2311.11606, 2312.04325, 2312.06324, 2312.13577, 2312.12814, PS99-025003, 2401.16756, AP463-169617, PDU44-101437, 686, 2404.08243, 2405.02328, 685, 2405.20022, 2406.08793, AC48-100853, 682, 683}. 

The general form of generalized off-shell free energy was introduced in \cite{29,york}, defined as:
\begin{equation}
\mathcal{F} = E - \frac{S}{\tau} ,
\label{offshell}
\end{equation}
where $E$ refers to the BH's energy (or mass $M$), and $S$ denotes its entropy. The variable $\tau$ represents a varying time scale, which can be interpreted as the inverse of the equilibrium temperature at the shell surrounding the BH. Using this generalized free energy, a vector field $\phi$ is formulated as \cite{29}:
\begin{equation}
\phi = \left(\phi^r, \phi^\Theta \right) = \left(\frac{\partial \mathcal{F}}{\partial S}, -\cot \Theta \, \csc \Theta \right).
\label{phi}
\end{equation}

The zero points of this vector field are essential, since they denote the critical points of the BH solutions.  Specifically, these zero points are located at $(\tau, \Theta) = \left(\frac{1}{T}, \frac{\pi}{2}\right)$, where $T$ is the BH’s equilibrium temperature within the surrounding cavity. Each of these defects, or zero points, carries a topological charge, which can be calculated using Duan’s $\phi$-mapping method. To determine this charge, the unit vector $n$ is derived from the field $\phi$. The unit vector $n^a$ must satisfy the following conditions:
\begin{equation}
n^a n^a = 1 \quad \text{and} \quad n^a \partial_\nu n^a = 0.
\end{equation}

A topological current $j^\mu$ can then be defined in the coordinate space $x^\nu = \{t, S, \Theta\}$ as:
\begin{equation}
j^\mu = \frac{1}{2\pi} \epsilon^{\mu \nu \rho} \epsilon_{ab} \partial_\nu n^a \partial_\rho n^b,
\end{equation}
where $\epsilon^{\mu \nu \rho}$ is the Levi-Civita symbol, and $\partial_\nu = \frac{\partial}{\partial x^\nu}$. This current is conserved, implying:
\begin{equation}
\partial_\mu j^\mu = 0.
\end{equation}

The current $j^\mu$ can be re-expressed in terms of the topological density as:
\begin{equation}
j^\mu = \delta^2(\phi) J^\mu \left( \frac{\phi}{x} \right),
\end{equation}
where $J^\mu$ relates to the Jacobi tensor $\epsilon^{ab} J^\mu \left( \frac{\phi}{x} \right) = \epsilon^{\mu \nu \rho} \partial_\nu \phi^a \partial_\rho \phi^b$. Utilizing the Laplacian Green function $\triangle_{\phi^a} \ln ||\phi|| = 2 \pi \delta^2(\phi)$, the previous expression is derived.

The topological charge $W$ is determined by integrating the zeroth component of the current density:
\begin{equation}
W = \int_\Sigma j^0 \, d^2x = \sum_{i=1}^N w_i,
\end{equation}
where $w_i$ represents the winding number associated with each zero point of the vector field $\phi$, and $\Sigma$ denotes the region over which the winding numbers are calculated. The contours used to define this region are constructed as:
\begin{equation}
\begin{cases}
S = S_1 \cos \nu + S_0, \\
\Theta = S_2 \sin \nu + \frac{\pi}{2},
\end{cases}
\label{contour}
\end{equation}
where $\nu \in (0, 2\pi)$, and $S_1$, $S_2$ determine the size of the contour, while $S_0$ marks its center. The relationship between the winding number and the deflection angle $\Omega$ is given by:
\begin{equation}
w = \frac{\Omega (2\pi)}{2\pi},
\label{winding}
\end{equation}
where $\Omega$ is computed as:
\begin{equation}
\Omega (\nu) = \int_{0}^{\nu} \epsilon_{12} n^1 \partial_\nu n^2 \, d\nu.
\label{deflection}
\end{equation}

The sum of all winding numbers provides the total topological charge:
\begin{equation}
W = \sum_i w_i.
\end{equation}

This topological charge $W$ characterizes the structural features of BH solutions within the framework of thermodynamic topology. Importantly, $j^\mu$ is nonzero only at the zero points of the vector field $\phi$; if no such points exist in the parameter space, the total topological charge remains zero.
In this section, we will conduct an in-depth investigation of the thermodynamic topology of these BHs within the Barrow entropy framework, with a focus on both 4-dimensional and 5-dimensional solutions. Our study will consider two distinct ensembles: the canonical ensemble and the grand canonical ensemble. We will begin with the analysis of 5-dimensional BHs.
	\section{Thermodynamics of $D=5$ charged AdS Black Holes immersed in polytropic dark energy  }\label{ThermalBH5D}
 Before starting our analysis, we would like to draw attention towards an important assumption.
 The integration constant $q$ is related to charge $Q$ as:
 $$Q=\frac{q}{4\pi}\omega_{D-2}.$$
 From the area law of entropy in eq. \eqref{mass1}, we draw the expression of $\omega_{D-2}$ as
$$\omega_{D-2}=\frac{4 S_{BH}}{r_h^{d-2}}.$$
	Again, the general form of entropy for D-dimensional BHs is given by 
 \begin{equation}
     S_{BH}=\int\frac{dM}{T}=\frac{2 \pi  r^{D-2}}{D-2}.
     \label{genentropy}
 \end{equation}
 Using Eq. \eqref{genentropy}, we can establish a relation between integration q and charge Q as, 
 $$Q=\frac{2 q}{D-2} ,$$
 hence Q can be considered equivalent to $q$ as both are mere constants. So, during the topological analysis, we will consider $q$ as a charge.\\
 
We start with evaluating mass of the 5D BHs by solving the equation $F(r=r_h)=0$ , which comes out to be
	\begin{align}
		M=\frac{1}{2} r_h^2 \biggl(r_h^2 \biggl(\frac{1}{L^2}-\frac{1}{6} \gamma ^{-n} \, _2F_1\biggl(n,n;n+1;-\frac{\xi ^2 r_h^{4/n}}{\gamma }\biggr)\biggr) +\frac{q^2}{r_h^4}+1\biggr) .
		\label{mass5}
	\end{align}
	The temperature of these BHs is evaluated to be
	 \begin{equation}
		T=\frac{r_h^6 \left(\frac{6}{L^2}-\gamma ^{-n} \left(\frac{\xi ^2 r_h^{4/n}}{\gamma }+1\right){}^{-n}\right)+3 r_h^4-3 q^2}{6 \pi  r_h^5},
		\label{temp5}
	\end{equation}
	for $D=5$ we found $S_{BH}=\frac{2 \pi  r^{3}}{3}$
	The BH mass in terms of BH entropy $S_{BH}$ is calculated as
	\begin{align}\label{mmass4}
	M=-\frac{\sqrt[3]{\frac{3}{2}} S_{\text{BH}}^{4/3} \gamma ^{-n} \, _2F_1\left(n,n;n+1;-\frac{\left(\frac{3}{2 \pi }\right)^{\frac{4}{3 n}} \xi ^2 S_{\text{BH}}^{\frac{4}{3 n}}}{\gamma }\right)}{8 \pi ^{4/3}}+\frac{3 \sqrt[3]{\frac{3}{2}} S_{\text{BH}}^{4/3}}{4 \pi ^{4/3} L^2}
 +\frac{\pi ^{2/3} q^2}{\sqrt[3]{2} 3^{2/3} S_{\text{BH}}^{2/3}}+\frac{1}{2} \left(\frac{3}{2 \pi }\right)^{2/3} S_{\text{BH}}^{2/3}.
	\end{align}
	The relation between GB entropy, $S_{BH}$ and Barrow entropy $S$ is given by 
	$$S_{BH}=S^{\frac{2}{\Delta +2}}.$$
	Replacing the entropy $S_{BH}$ by Barrow entropy $S$ in the expression for mass will give us the new mass as :
 \small
	\begin{align}
		M=& 
  \frac{-\sqrt[3]{\frac{3}{2}} \gamma ^{-n} \left(S^{\frac{2}{\Delta +2}}\right)^{4/3} \, _2F_1\left(n,n;n+1;-\frac{\left(\frac{3}{2 \pi }\right)^{\frac{4}{3 n}} \left(S^{\frac{2}{\Delta +2}}\right)^{\frac{4}{3 n}} \xi ^2}{\gamma }\right)}{8 \pi ^{4/3}} 
  +\frac{3 \sqrt[3]{\frac{3}{2}} \left(S^{\frac{2}{\Delta +2}}\right)^{4/3}}{4 \pi ^{4/3} L^2}
  \nonumber\\
&\, +\frac{\left(\frac{\pi }{3}\right)^{2/3} q^2}{\sqrt[3]{2} \left(S^{\frac{2}{\Delta +2}}\right)^{2/3}}
  +\frac{1}{2} \left(\frac{3S^{\frac{2}{\Delta +2}}}{2 \pi }\right)^{2/3}.
		\label{massB5}
	\end{align}
 \normalsize
	At $\Delta=0$, the expression of mass becomes equivalent to the mass in GB statistics written in Eq. \eqref{mmass4}.
	The temperature is now calculated again for Barrow entropy as
\begin{eqnarray}
		T&=& -\frac{\gamma ^{-n} S^{\frac{2}{\Delta +2}-1} \sqrt[3]{S^{\frac{2}{\Delta +2}}} \left(\frac{\left(\frac{3}{2 \pi }\right)^{\frac{4}{3 n}} \xi ^2 \left(S^{\frac{2}{\Delta +2}}\right)^{\frac{4}{3 n}}}{\gamma }+1\right)^{-n}}{\sqrt[3]{2} 3^{2/3} \pi ^{4/3} (\Delta +2)} 
  -\frac{2 \left(\frac{2 \pi }{3}\right)^{2/3} q^2 S^{-\frac{2}{\Delta +2}-1} \sqrt[3]{S^{\frac{2}{\Delta +2}}}}{3 (\Delta +2)}\nonumber\\
  & &\,+ \frac{\sqrt[3]{\frac{2}{3}} \left(S^{\frac{2}{\Delta +2}}\right)^{2/3}}{\pi ^{2/3} (\Delta +2) S}+\frac{2^{2/3} \sqrt[3]{3} S^{\frac{2}{\Delta +2}-1} \sqrt[3]{S^{\frac{2}{\Delta +2}}}}{\pi ^{4/3} (\Delta +2) L^2}.
		\label{tempB5}
	\end{eqnarray}

\begin{widetext}
	\begin{figure*}[!htp]	
		\centering
		\begin{subfigure}{0.43\textwidth}
			\includegraphics[width=\linewidth]{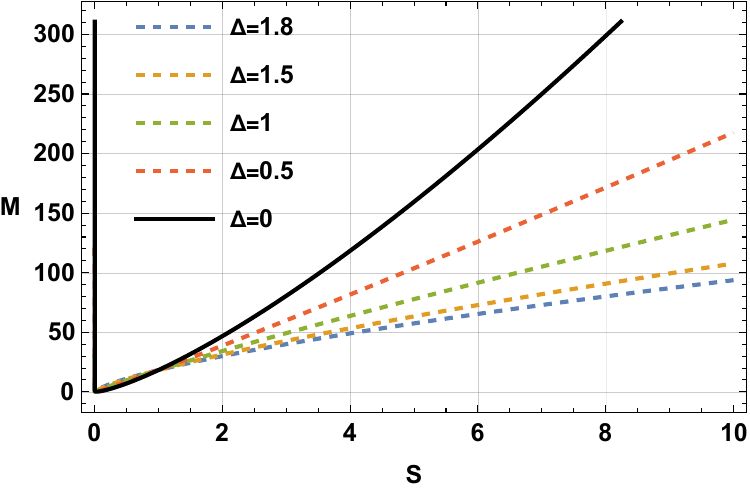}
			\caption{}
			\label{b3a}
		\end{subfigure}
		\begin{subfigure}{0.43\textwidth}
			\includegraphics[width=\linewidth]{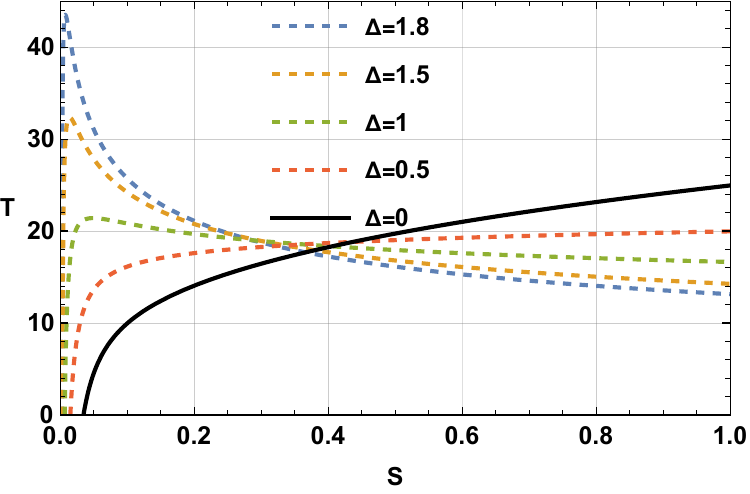}
			\caption{}
			\label{b3b}
		\end{subfigure}
		\begin{subfigure}{0.43\textwidth}
		\includegraphics[width=\linewidth]{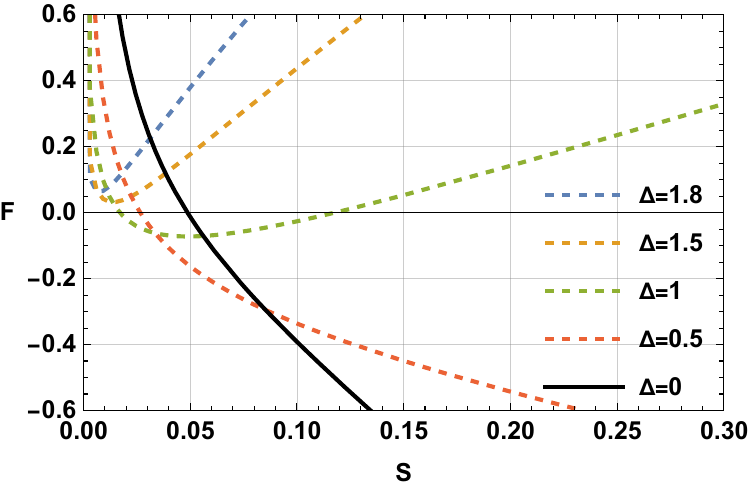}
		\caption{}
		\label{b3c}
	\end{subfigure}
	\begin{subfigure}{0.43\textwidth}
		\includegraphics[width=\linewidth]{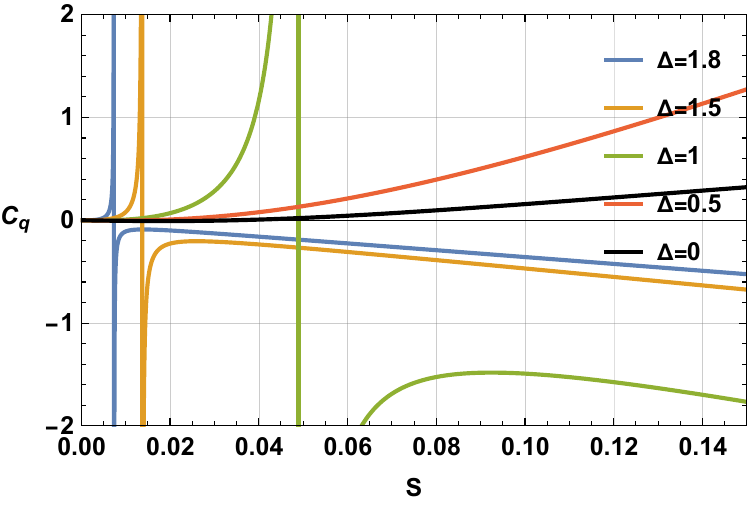}
		\caption{}
		\label{b3d}
	\end{subfigure}
	\begin{subfigure}{0.43\textwidth}
		\includegraphics[width=\linewidth]{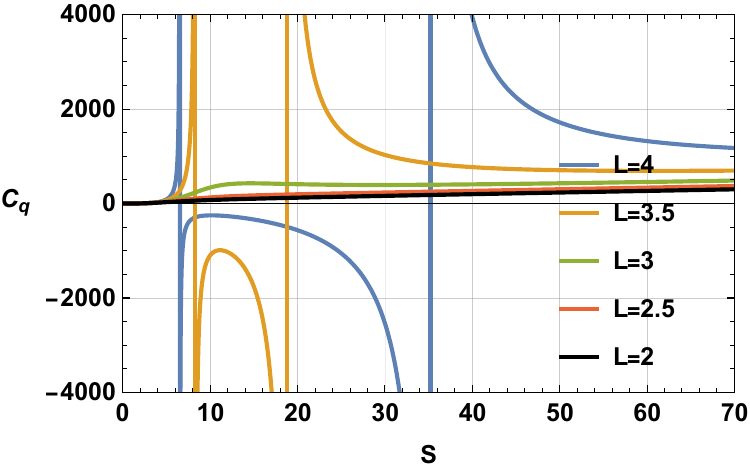}
		\caption{}
		\label{b3e}
	\end{subfigure}
	\begin{subfigure}{0.43\textwidth}
		\includegraphics[width=\linewidth]{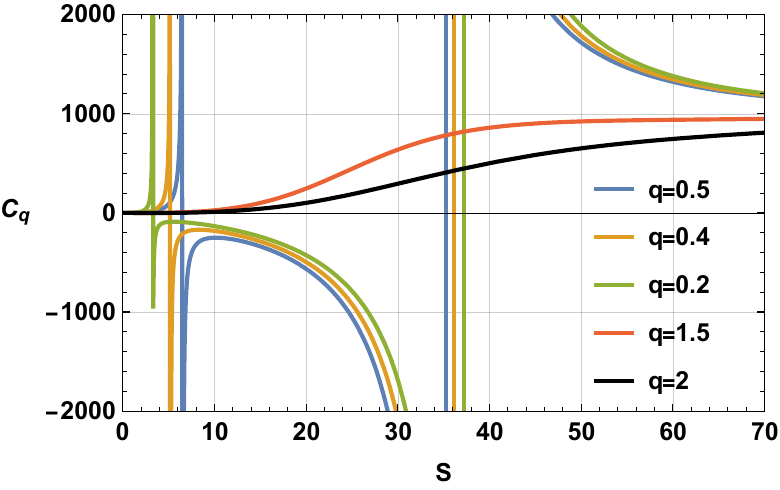}
		\caption{}
		\label{b3f}
	\end{subfigure}
		\caption{Thermodynamic property of $D=5$ dimensional  charged AdS BHs immersed in polytropic dark energy (DE) in the framework of Barrow entropy
		}
		\label{b3}
	\end{figure*}
	\end{widetext}

The influence of the parameter $\Delta$ on {BH} mass and temperature becomes more significant at higher values of Barrow entropy $S$. Figure \ref{b3a} shows the {BH} mass in the context of Barrow entropy for different values of $\Delta$, with the parameters fixed at $q = 0.2$, $\gamma = 0.0095$, $\xi = 1$, and $L = 0.1$. Figure \ref{b3b} displays the temperature $T$ as a function of entropy $S$, illustrating small-to-large {BH} phase transitions. For larger values of $\Delta$, two distinct {BH} phases are evident in the figure. The solid black line indicates that no phase transition occurs in GB statistics for the same set of thermodynamic parameters. Figure \ref{b3c} offers a detailed view of the phase transition phenomena in the free energy vs. entropy plot, where two distinct phases are clearly visible for higher values of $\Delta$, while no phase transition occurs for values of $\Delta$ near zero. In Figure \ref{b3d}, the same behaviour is depicted in the heat capacity $C_q$ vs. entropy $S$ plot.
For larger values of $\Delta$, two BH branches are observed: the small BH branch is stable, while the large BH branch is unstable. For values $\Delta=0$ or near zero, either three BH branches or a single branch exist, depending on the values of $L$ and $q$. The variation of the BH branches with $L$ and $q$ is shown in Figs.~\ref{b3e} and \ref{b3f}, respectively. 

In Figure \ref{b3e}, with $\Delta=0$, $q = 0.5$, $\gamma = 0.0095$, and $\xi = 1$ fixed, we observe that when three BH branches exist, the small branch is stable, the intermediate branch is unstable, and the large branch is stable. Conversely, when only a single branch exists, it is found to be stable. Figure \ref{b3f} illustrates the variation of BH branches with $q$, while keeping $\Delta=0$,$L = 4$, $\gamma = 0.0095$, and $\xi = 1$ fixed.\\

\begin{figure*}[!htp]	
		\centering
		\begin{subfigure}{0.43\textwidth}
			\includegraphics[width=\linewidth]{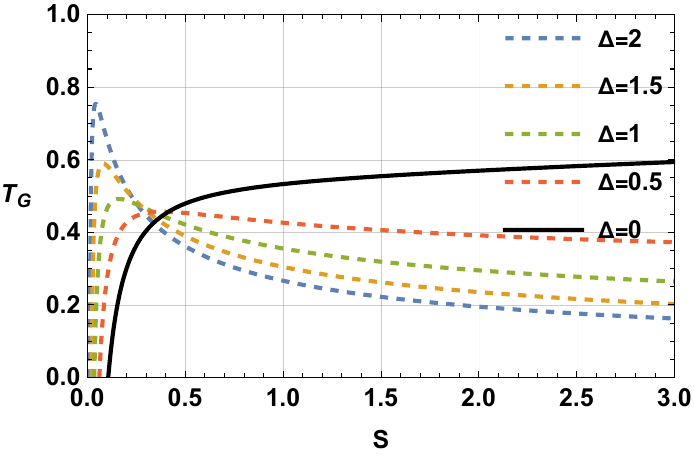}
			\caption{}
			\label{bh4a}
		\end{subfigure}
		\begin{subfigure}{0.43\textwidth}
			\includegraphics[width=\linewidth]{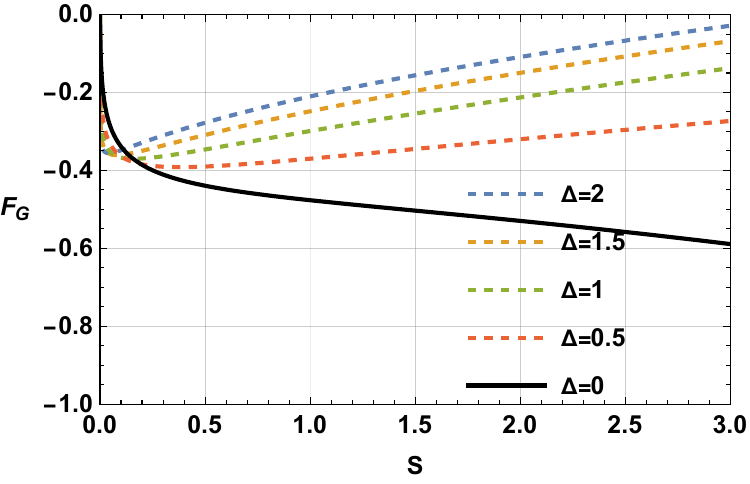}
			\caption{}
			\label{bh4b}
		\end{subfigure}
		\begin{subfigure}{0.43\textwidth}
		\includegraphics[width=\linewidth]{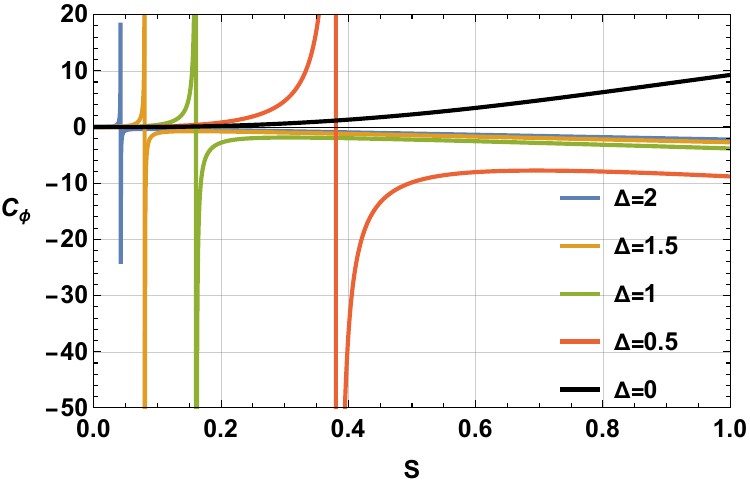}
		\caption{}
		\label{bh4c}
	\end{subfigure}
	\caption{Thermodynamic property of $D=5$ dimensional  charged AdS BHs in grand canonical ensemble
		}
		\label{bh4}
	\end{figure*}

Next, we discuss the thermodynamic properties of these BHs in the grand canonical ensemble, where the  potential $\phi$ is kept fixed instead of its conjugate parameter $q.$ $\phi$ can be calculated as :
	\begin{equation}
	\phi =\frac{\partial M}{\partial q}=\frac{\left(\frac{2 \pi }{3}\right)^{2/3} q}{\left(S^{\frac{2}{\Delta +2}}\right)^{2/3}} ,
	\label{phi5GC}
	\end{equation}
	solving Eq. \eqref{phi5GC}, we can obtain expression of $q$ in terms of $\phi$. Next $q$ is substituted in the equation given below and the grand canonical mass $M_G$ is obtained
 \small
	\begin{align}
	M_G=M-q \phi=&\left(\frac{3}{2 \pi }\right)^{2/3} \left(S^{\frac{2}{\Delta +2}}\right)^{2/3}-\frac{\sqrt[3]{\frac{3}{2}} \gamma ^{-n} \left(S^{\frac{2}{\Delta +2}}\right)^{4/3} \, _2F_1\left(n,n;n+1;-\frac{\left(\frac{3}{2 \pi }\right)^{\frac{4}{3 n}} \left(S^{\frac{2}{\Delta +2}}\right)^{\frac{4}{3 n}} \xi ^2}{\gamma }\right)}{8 \pi ^{4/3}}
    \nonumber\\
    &+\frac{3 \sqrt[3]{\frac{3}{2}} \left(S^{\frac{2}{\Delta +2}}\right)^{4/3}}{4 \pi ^{4/3} L^2}-\left(\frac{3}{2 \pi }\right)^{2/3} \phi ^2 \left(S^{\frac{2}{\Delta +2}}\right)^{2/3}.
	\end{align}
 \normalsize
Similarly, the expression for temperature in the grand canonical ensemble is obtained as :
\small
\begin{align}
T_G&=\frac{2^{2/3} \sqrt[3]{3} S^{-\frac{\Delta }{\Delta +2}} \sqrt[3]{S^{\frac{2}{\Delta +2}}}}{\pi ^{4/3} (\Delta +2) L^2}+\frac{2 \sqrt[3]{\frac{2}{3}} S^{-\frac{\Delta }{\Delta +2}-\frac{2}{\Delta +2}} \left(S^{\frac{2}{\Delta +2}}\right)^{2/3}}{\pi ^{2/3} (\Delta +2)}-\frac{\gamma ^{-n} S^{-\frac{\Delta }{\Delta +2}} \sqrt[3]{S^{\frac{2}{\Delta +2}}} \left(\frac{\left(\frac{3}{2 \pi }\right)^{\frac{4}{3 n}} \xi ^2 \left(S^{\frac{2}{\Delta +2}}\right)^{\frac{4}{3 n}}}{\gamma }+1\right)^{-n}}{\sqrt[3]{2} 3^{2/3} \pi ^{4/3} (\Delta +2)}
\nonumber\\
&-\frac{2 \sqrt[3]{\frac{2}{3}} \phi ^2 S^{-\frac{\Delta }{\Delta +2}-\frac{2}{\Delta +2}} \left(S^{\frac{2}{\Delta +2}}\right)^{2/3}}{\pi ^{2/3} (\Delta +2)}.
\end{align}
\normalsize
In the $T_G$ vs $S$ graph (Fig. \ref{bh4a}), we observe that the system exhibits either one or two branches for different values of $\Delta$, with parameters fixed at $\phi = 0.2$, $\gamma = 0.0095$, $\xi = 1$, and $L = 1$. The semi-positive nature of the temperature is found to depend on the parameter $\gamma$. Notably, we observe a Davis-type phase transition, rather than a van der Waals-like phase transition, as seen in the canonical ensemble. This observation is further supported by the $F_G$ vs $S$ plot in Fig. \ref{bh4b}. 

The stability of these branches is described by the $C_\phi$ vs $S$ graph in Fig. \ref{bh4c}. When two {BH} branches are present, the small {BH} branch is stable, while the large {BH} branch is unstable. In cases where only a single branch is observed, its stability depends on the values of $L$ and $\phi$. To better understand the differences in stability and the dependence of these temperature-driven phase transitions on various thermodynamic parameters, a topological analysis is more appropriate. This analysis is discussed in the next section.

	\subsection{Thermodynamic Topology of $5D$ Black Holes}

\begin{widetext}
    \begin{figure*}[!htp]	
		\centering
		\begin{subfigure}{0.43\textwidth}
			\includegraphics[width=\linewidth]{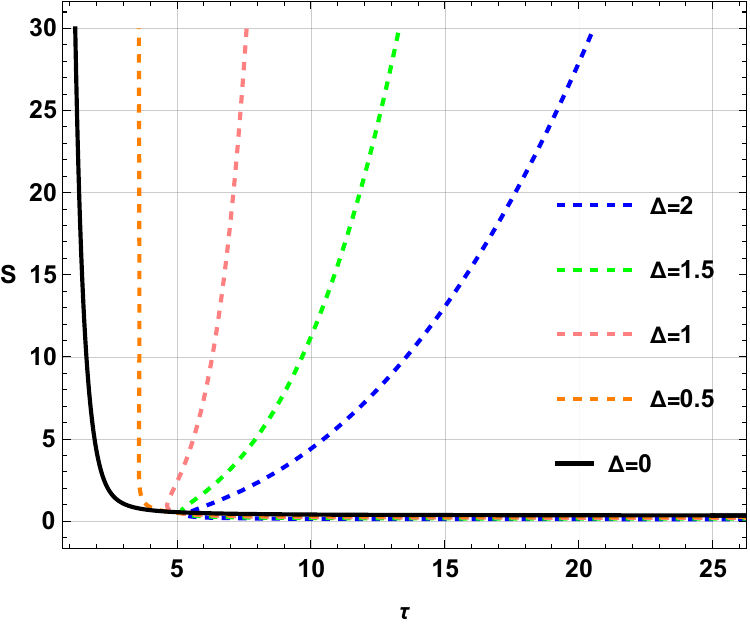}
			\caption{}
			\label{bh5a}
		\end{subfigure}
		\begin{subfigure}{0.43\textwidth}
			\includegraphics[width=\linewidth]{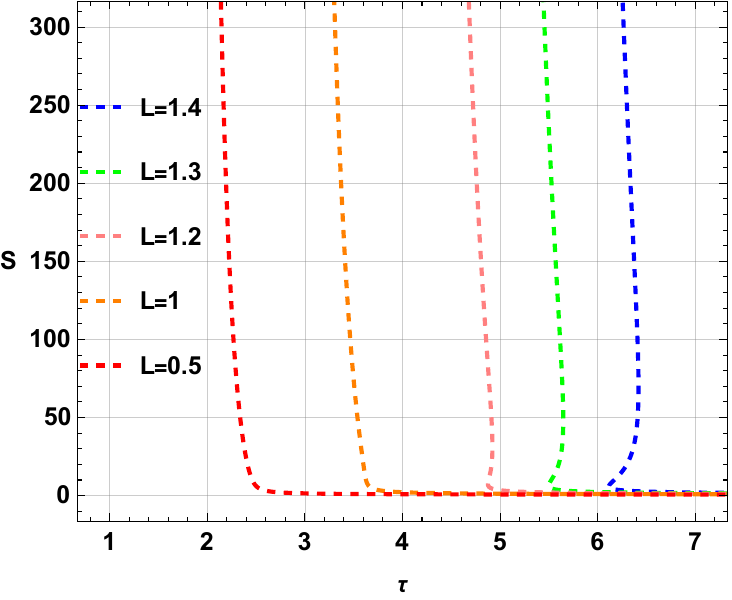}
			\caption{}
			\label{bh5b}
		\end{subfigure}
		\begin{subfigure}{0.43\textwidth}
		\includegraphics[width=\linewidth]{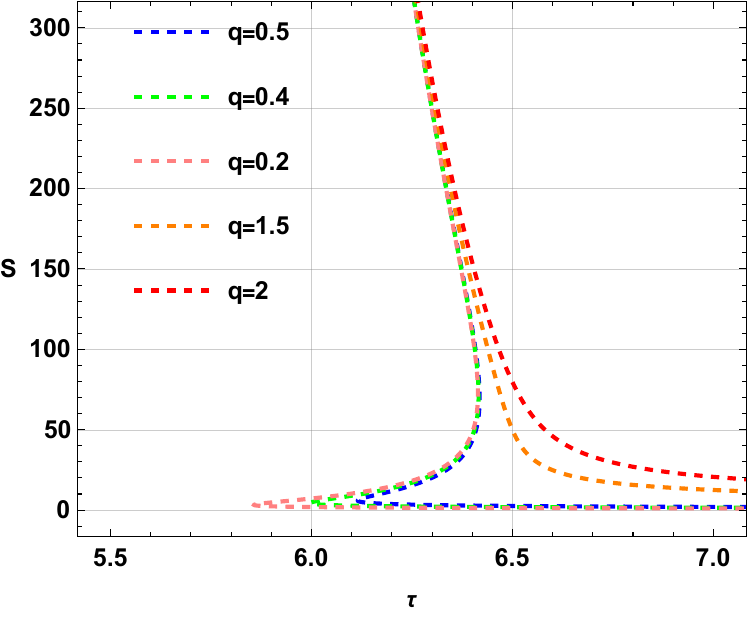}
		\caption{}
		\label{bh5c}
	\end{subfigure}
	\caption{Variation of defect curve of $D=5$ dimensional  charged AdS BHs in canonical ensemble
		}
		\label{bh5}
	\end{figure*}
\end{widetext}

First we conduct our analysis in canonical ensemble. Using Eqs. \eqref{offshell} and \eqref{massB5}, we write the off shell free energy as :
	\begin{align}
		\mathcal{F}=M-\frac{S}{\tau }=&-\frac{\sqrt[3]{\frac{3}{2}} \gamma ^{-n} \left(S^{\frac{2}{\Delta +2}}\right)^{4/3} }{8 \pi ^{4/3}} 
  \, _2F_1\left(n,n;n+1;-\frac{\left(\frac{3}{2 \pi }\right)^{\frac{4}{3 n}} \left(S^{\frac{2}{\Delta +2}}\right)^{\frac{4}{3 n}} \xi ^2}{\gamma }\right)
  \nonumber\\
  &+\frac{3 \sqrt[3]{\frac{3}{2}} \left(S^{\frac{2}{\Delta +2}}\right)^{4/3}}{4 \pi ^{4/3} L^2}+\frac{\left(\frac{\pi }{3}\right)^{2/3} q^2}{\sqrt[3]{2} \left(S^{\frac{2}{\Delta +2}}\right)^{2/3}}+\frac{1}{2} \left(\frac{3}{2 \pi }\right)^{2/3} \left(S^{\frac{2}{\Delta +2}}\right)^{2/3}-\frac{S}{\tau } .
	\end{align}
	Using Eq. \eqref{phi}, components of the vector $\phi $ are found to be :
	\begin{align}
		\phi ^{S} =& -\frac{1}{ \tau }+\frac{2^{2/3} \sqrt[3]{3} S^{\frac{2}{\Delta +2}-1} \sqrt[3]{S^{\frac{2}{\Delta +2}}}}{\pi ^{4/3} (\Delta +2) L^2}-\frac{\gamma ^{-n} S^{\frac{2}{\Delta +2}-1} \sqrt[3]{S^{\frac{2}{\Delta +2}}} \left(\frac{\left(\frac{3}{2 \pi }\right)^{\frac{4}{3 n}} \xi ^2 \left(S^{\frac{2}{\Delta +2}}\right)^{\frac{4}{3 n}}}{\gamma }+1\right)^{-n}}{\sqrt[3]{2} 3^{2/3} \pi ^{4/3} (\Delta +2)}
        \nonumber\\
        &-\frac{2 \left(\frac{2 \pi }{3}\right)^{2/3} q^2 S^{-\frac{2}{\Delta +2}-1} \sqrt[3]{S^{\frac{2}{\Delta +2}}}}{3 (\Delta +2)}+\frac{\sqrt[3]{\frac{2}{3}} \left(S^{\frac{2}{\Delta +2}}\right)^{2/3}}{\pi ^{2/3} (\Delta +2) S},
  \end{align}
  and
	\begin{equation}
		\phi ^{\Theta }=-\cot \Theta ~\csc \Theta.
\end{equation}
	Subsequently, we identify the zero points or singularities of the vector field. One zero point of the vector field consistently occurs at $\Theta=\frac{\pi}{2}$ as a result of the careful selection of the $\Theta$-component of the field. To determine the additional zero points, we get an equation for $\tau$ via solving $\phi^{S} = 0$, which is formulated as: 
	\begin{equation}
	\tau=\frac{\alpha}{\beta},
	\end{equation}
	where 
	\begin{align}
	\alpha= 3\ 6^{2/3} \pi ^{4/3} (\Delta +2) L^2 S \gamma ^n \left(S^{\frac{2}{\Delta +2}}\right)^{2/3} 
 \left(\frac{\left(\frac{3}{2 \pi }\right)^{\frac{4}{3 n}} \xi ^2 \left(S^{\frac{2}{\Delta +2}}\right)^{\frac{4}{3 n}}}{\gamma }+1\right)^n
	\end{align}
	and
	\begin{align}
	\beta=& L^2 \biggl( 2 \pi ^{2/3} \gamma ^n \biggl( 3 \sqrt[3]{3} \left(S^{\frac{2}{\Delta +2}}\right)^{4/3} 
- 2 \sqrt[3]{2} \pi ^{4/3} q^2 \biggr) \biggl( \frac{\biggl( \frac{3}{2 \pi }\biggr)^{\frac{4}{3 n}} \xi ^2 
\left(S^{\frac{2}{\Delta +2}}\right)^{\frac{4}{3 n}}}{\gamma } + 1 \biggr)^n 
- 3 \sqrt[3]{2} S^{\frac{4}{\Delta +2}} \biggr) 
\nonumber\\
&+ 18 \sqrt[3]{2} \gamma ^n S^{\frac{4}{\Delta +2}} \biggl( \frac{\biggl( \frac{3}{2 \pi }\biggr)^{\frac{4}{3 n}} \xi ^2 
\left(S^{\frac{2}{\Delta +2}}\right)^{\frac{4}{3 n}}}{\gamma } + 1 \biggr)^n
\biggr).
	\end{align}
	By plotting the defect curve in Fig. \ref{bh5}, we observe three distinct behaviours depending on the values of the thermodynamic parameters. To determine the topological class in the canonical ensemble, we analyze each pattern and calculate the corresponding topological charge. Fig. \ref{bh5a} illustrates the variation of the defect curve with different values of $\Delta$, while keeping $q = 0.2$, $\gamma = 0.0095$, $\xi = 1$, and $L = 1$ fixed. The solid black line represents the behaviour in the case of GB statistics. For higher values of $\Delta$, two BH branches are present, while for values of $\Delta$ near zero, we observe only a single branch for the same set of thermodynamic parameters.
Next, we examine the effect of varying the parameters $L$,$q$ and $\gamma$ on the defect curve. The impact is not significant for larger $\Delta$ values; however, when $\Delta$ is fixed at lower values, the phase transition behavior changes noticeably. Fig. \ref{bh5b} and Fig. \ref{bh5c} demonstrate the effects of varying $L$ and $q$, respectively. It is clear that as the values of $L$ and $q$ change, the number of BH branches transitions from one to three.Similar observations were made when we vary the $\gamma$ parameter.
 
\begin{figure}[!htp]	
		\centering
		\begin{subfigure}{0.43\textwidth}
			\includegraphics[width=\linewidth]{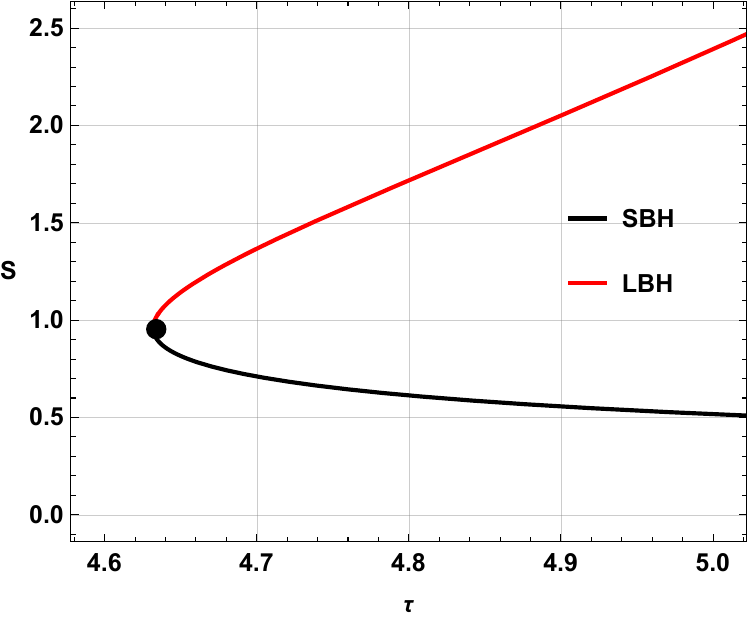}
			\caption{}
			\label{b6a}
		\end{subfigure}
		\begin{subfigure}{0.43\textwidth}
			\includegraphics[width=\linewidth]{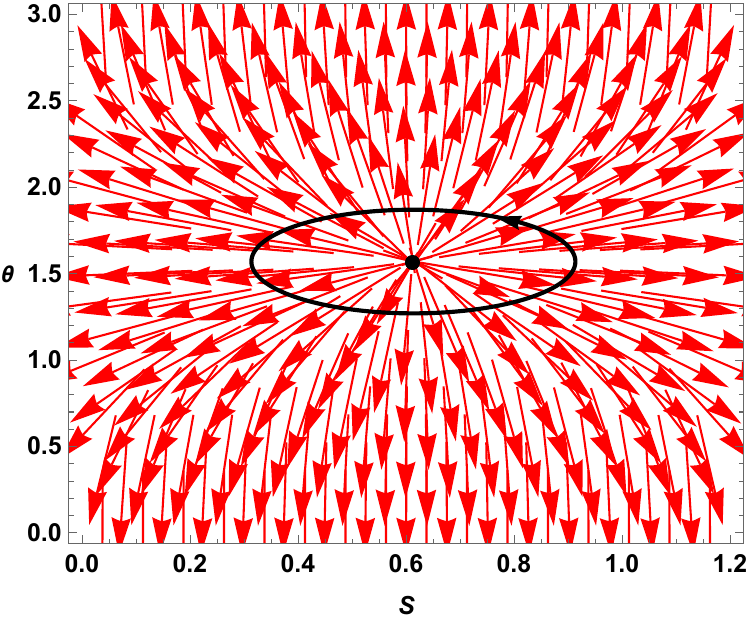}
			\caption{}
			\label{b6b}
		\end{subfigure}
		\begin{subfigure}{0.43\textwidth}
		\includegraphics[width=\linewidth]{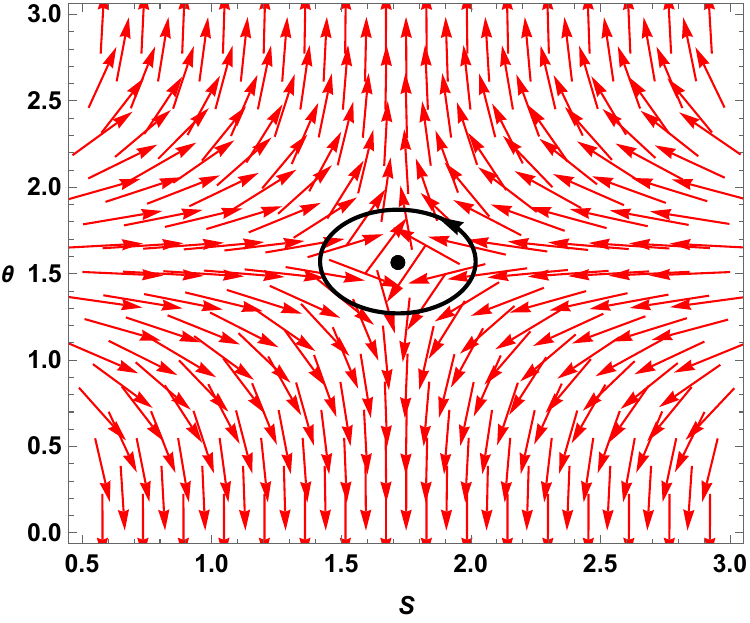}
		\caption{}
		\label{b6c}
	\end{subfigure}
	\begin{subfigure}{0.43\textwidth}
		\includegraphics[width=\linewidth]{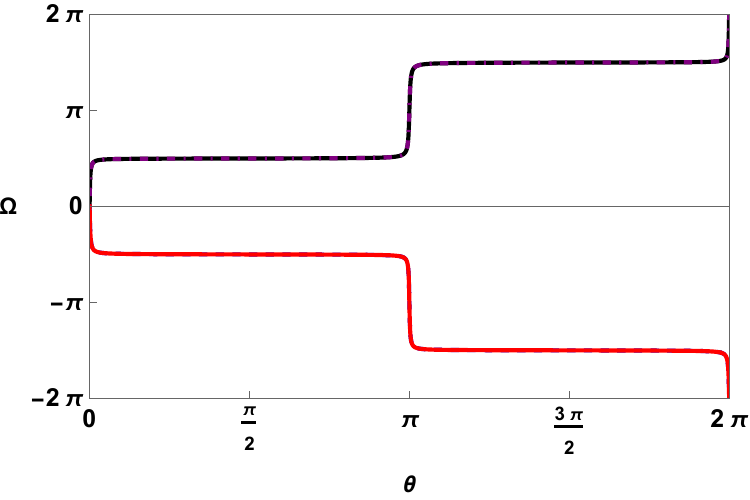}
		\caption{}
		\label{b6d}
	\end{subfigure}
	\caption{Variation of defect curve of $D=5$ dimensional  charged AdS BHs in canonical ensemble
		}
		\label{b6}
	\end{figure}

In Barrow entropy framework, when considering the parameters $q = 0.2$, $\gamma = 0.0095$, $\xi = 1$, $\Delta = 1$, and $L = 1$, two BH branches are observed, as shown in Fig. \ref{b6a}. The black solid line represents the smaller BH branch $(S<0.95862)$, while the red solid line represents the larger BH branch$(S>0.95862)$. The winding number calculation is illustrated in Fig. \ref{b6d}, where we focus on the zero points for $\tau =4.8$. Two zero points are identified at $S = 0.613272$ and $S = 1.71772$, corresponding to the smaller BH (SBH) and the larger BH (LBH), respectively. 
Figs.~\ref{b6b} and \ref{b6c} display the vector plots of the normalized vector field, with the zero points located at $\theta = \pi/2$. The winding number for the SBH is calculated as $+1$, indicating its stability, while the LBH has a winding number of $-1$, signifying instability. The overall topological charge is determined by summing the winding numbers, resulting in a total topological charge of $-1 + 1 = 0$ for this particular defect curve. The point at which phase transition occurs, i.e $S=0.95862$ the point is identified to be annihilation point as the stable phase ends here and a unstable phase starts. \\

\begin{figure}[!htp]
\includegraphics[width=8cm, height=6cm]{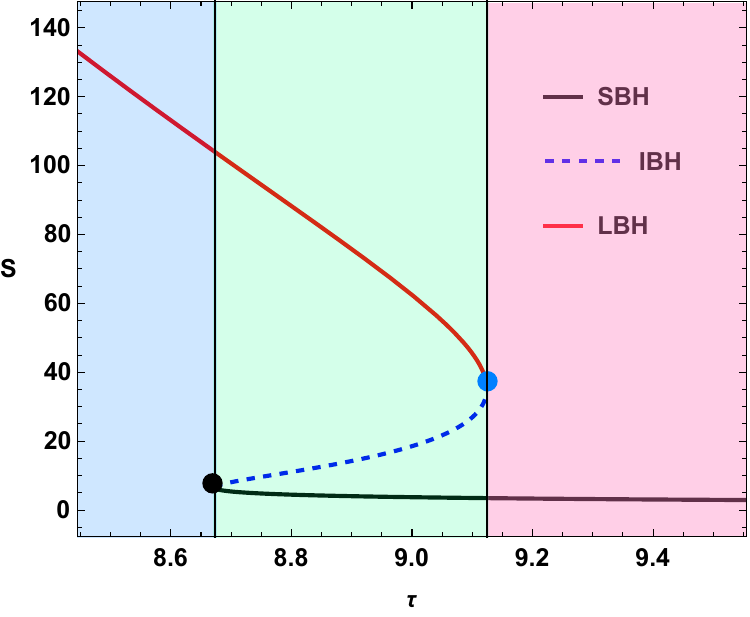}
\caption{Defect curve of $D=5$ dimensional  charged AdS BHs in canonical ensemble
		}
		\label{b7}
\end{figure}

Now we move to GB statistics framework where we consider small value of $\Delta.$ Considering the parameters $q = 0.5$, $\gamma = 0.0095$, $\xi = 1$, $\Delta = 0.5$, and $L = 4$, we observe the presence of three distinct BH branches, as depicted in Fig. \ref{b7}. The black solid line represents the SBH branch, the blue dashed line corresponds to the intermediate BH branch (IBH), and the red solid line represents the LBH branch. For $\tau = 42$, three zero points are identified at $S = 1.1016$, $S = 18.5452$, and $S =62.4727$, corresponding to the SBH, IBH, and LBH, respectively.We observe a generation point $S=35.2148$ and a annihilation point at $S=6.5074$
Upon varying the values of $L$, $q$, and $\gamma$, these three branches converge into a single BH branch, exhibiting behavior similar to that observed in GB statistics. Thus, for  values  $\Delta$ near zero  the thermodynamic topological properties of the BH solution demonstrate a pattern consistent with that seen in GB statistics.\\

\begin{figure}[!htp]	
		\centering
		\begin{subfigure}{0.43\textwidth}
			\includegraphics[width=\linewidth]{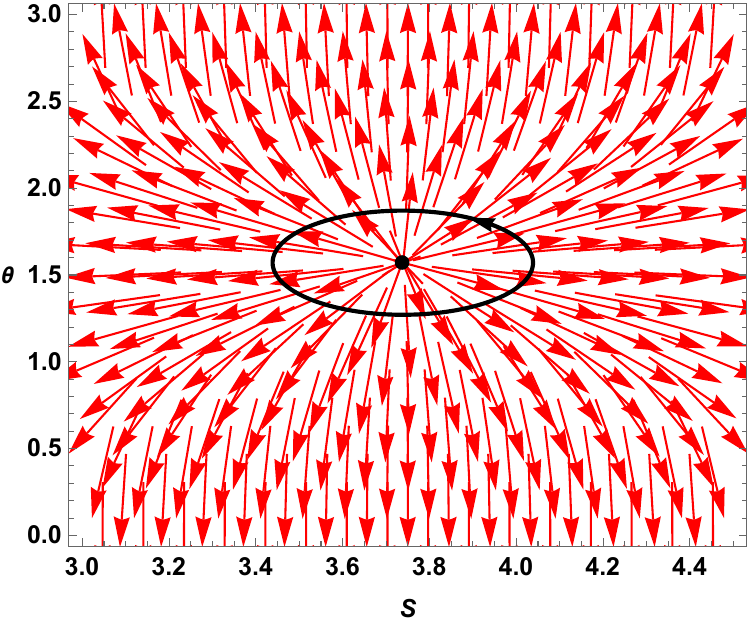}
			\caption{}
			\label{b8a}
		\end{subfigure}
		\begin{subfigure}{0.43\textwidth}
			\includegraphics[width=\linewidth]{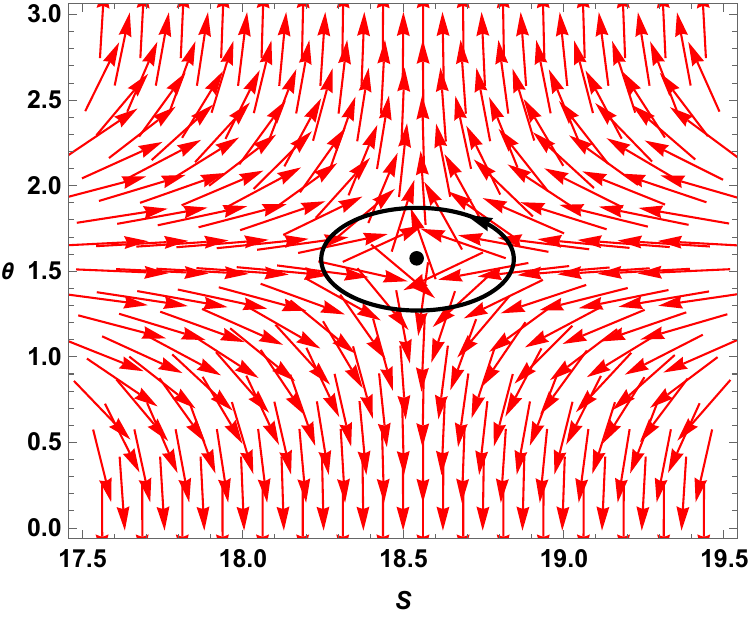}
			\caption{}
			\label{b8b}
		\end{subfigure}
		\begin{subfigure}{0.43\textwidth}
		\includegraphics[width=\linewidth]{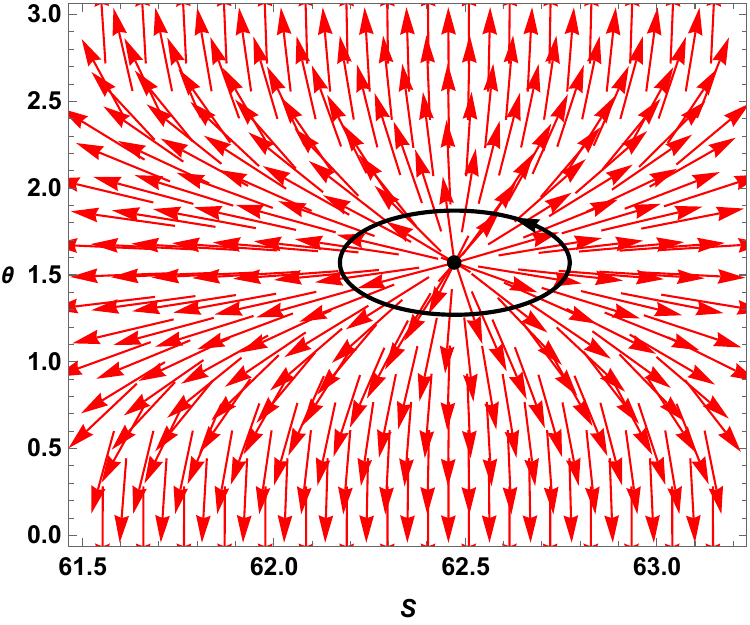}
		\caption{}
		\label{b8c}
	\end{subfigure}
	\begin{subfigure}{0.43\textwidth}
	\includegraphics[width=\linewidth]{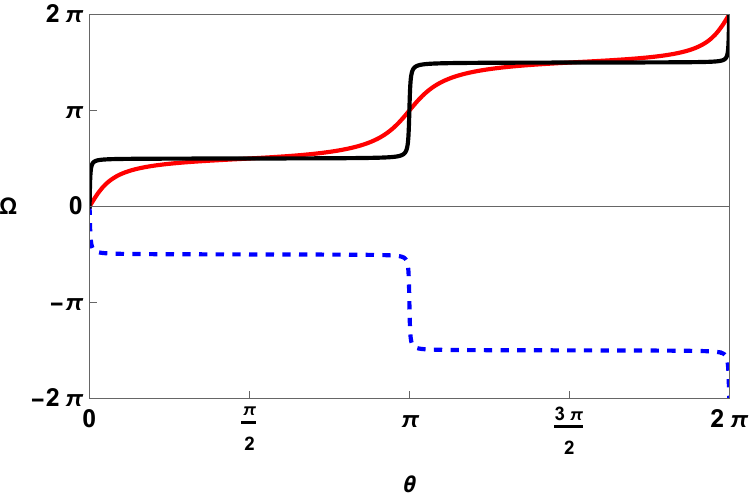}
		\caption{}
		\label{b8d}
	\end{subfigure}
	\caption{Thermodynamic topology of $D=5$ dimensional charged AdS BHs in the canonical ensemble. }
		\label{b8}
	\end{figure}

Figs.~\ref{b8a}, \ref{b8b}, and \ref{b8c} show the vector plots of the normalized vector field. The winding number for both the SBH and LBH is calculated to be $+1$, indicating stability, while the IBH has a winding number of $-1$, signifying its instability. By summing the winding numbers, the total topological charge for this particular defect curve is found to be $1 - 1 + 1 = 1$. 
In the scenario where the three phases merge into a single BH phase, we obtain a stable {BH} branch with a topological charge of 1. Therefore, in the canonical ensemble, the topological charge of the BH in GB statistics is 1. However, in the Barrow entropy framework, the topological charge can either be $1$ or $0$.
	We move our attention to grand canonical ensemble. The expression for $\tau$ in grand canonical ensemble is given by :
	\begin{equation}
	\tau_G=\frac{\mathcal{X}}{\mathcal{Y}} ,
	\end{equation}
	where, 
 \small
	\begin{align}
	\mathcal{X}=6^{2/3} \pi ^{4/3} (\Delta +2) L^2 S \gamma ^n \left(\frac{\left(\frac{3}{2 \pi }\right)^{\frac{4}{3 n}} \xi ^2 \left(S^{\frac{2}{\Delta +2}}\right)^{\frac{4}{3 n}}}{\gamma }+1\right)^n ,
	\end{align}
 \normalsize
 and
	\begin{align}
	\mathcal{Y}=& 6 \sqrt[3]{2} \gamma ^n \left(S^{\frac{2}{\Delta +2}}\right)^{4/3} \left(\frac{\left(\frac{3}{2 \pi }\right)^{\frac{4}{3 n}} \xi ^2 \left(S^{\frac{2}{\Delta +2}}\right)^{\frac{4}{3 n}}}{\gamma }+1\right)^n-L^2 \biggl(S^{\frac{2}{\Delta +2}}\biggr)^{2/3} \biggl(4 \sqrt[3]{3} \pi ^{2/3} \biggl(\phi ^2-1\biggr)
    \nonumber\\ &\,\times\, \gamma ^n \biggl(\frac{\biggl(\frac{3}{2 \pi }\biggr)^{\frac{4}{3 n}} \xi ^2 \biggl(S^{\frac{2}{\Delta +2}}\biggr)^{\frac{4}{3 n}}}{\gamma }+1\biggr)^n+\sqrt[3]{2} \biggl(S^{\frac{2}{\Delta +2}}\biggr)^{2/3}\biggr).
	\end{align}
	In this ensemble, three distinct topological classes are identified based on the values of $\Delta$, $L$, $q$, and $\gamma$. To illustrate, we examine three specific scenarios where these topological classes emerge.\\
	
\begin{figure}[!htp]	
		\centering
		\begin{subfigure}{0.43\textwidth}
			\includegraphics[width=\linewidth]{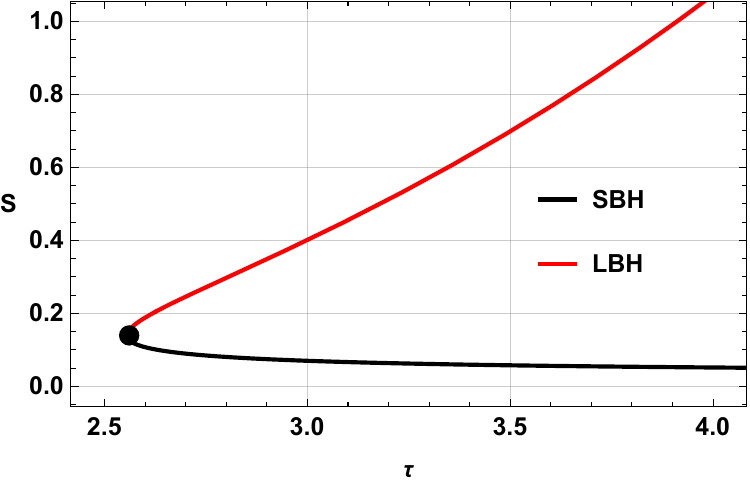}
			\caption{}
			\label{b9a}
		\end{subfigure}
		\begin{subfigure}{0.43\textwidth}
			\includegraphics[width=\linewidth]{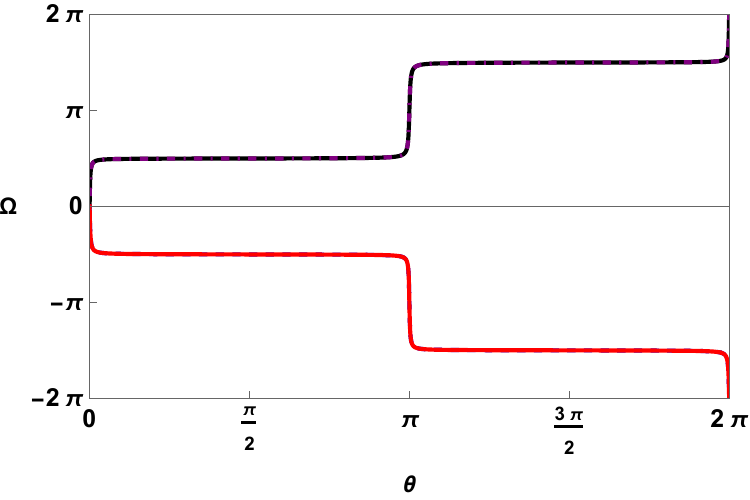}
			\caption{}
			\label{b9b}
		\end{subfigure}
		\caption{Topological charge $W=0$}
		\label{b9}
	\end{figure}

\begin{figure}[!htp]	
		\centering
		\begin{subfigure}{0.43\textwidth}
			\includegraphics[width=\linewidth]{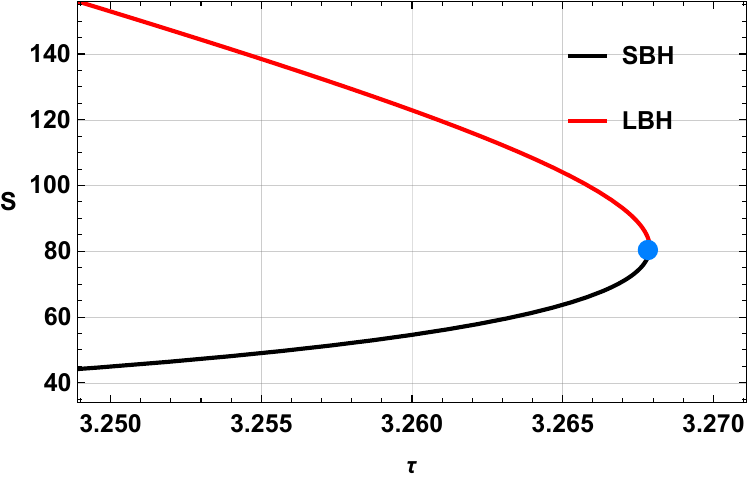}
			\caption{}
			\label{b10a}
		\end{subfigure}
		\begin{subfigure}{0.43\textwidth}
			\includegraphics[width=\linewidth]{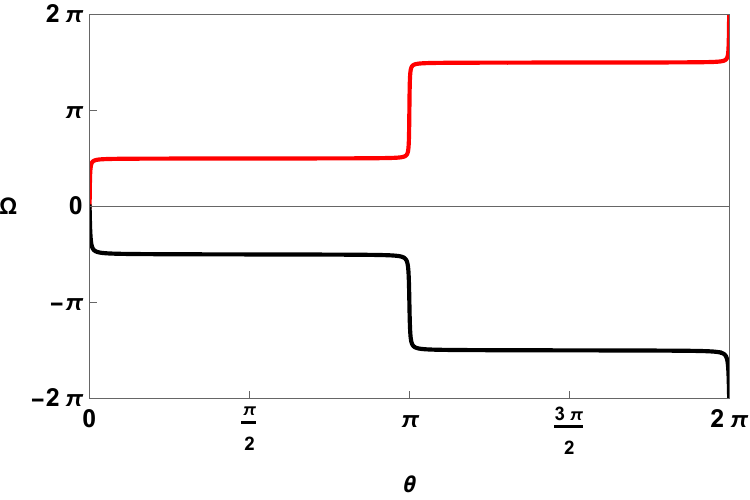}
			\caption{}
			\label{b10b}
		\end{subfigure}
		\caption{Topological class W=0 with a generation point  }
		\label{b10}
	\end{figure}

\begin{figure}[!htp]	
		\centering
		\begin{subfigure}{0.43\textwidth}
			\includegraphics[width=\linewidth]{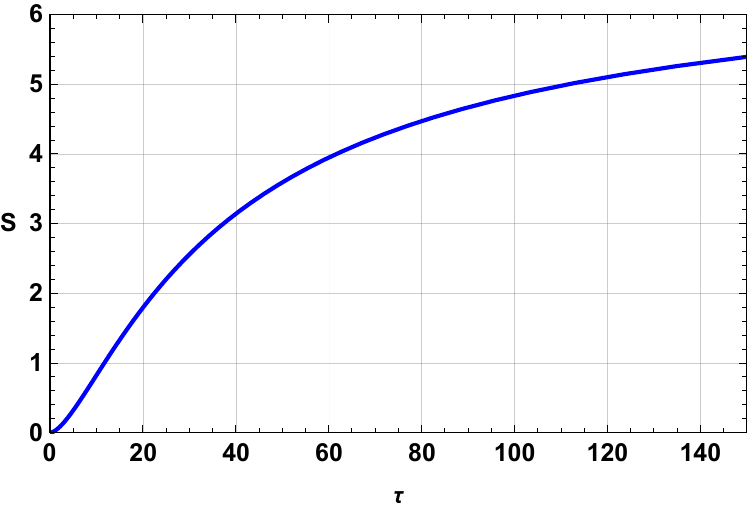}
			\caption{}
			\label{b11a}
		\end{subfigure}
		\begin{subfigure}{0.43\textwidth}
			\includegraphics[width=\linewidth]{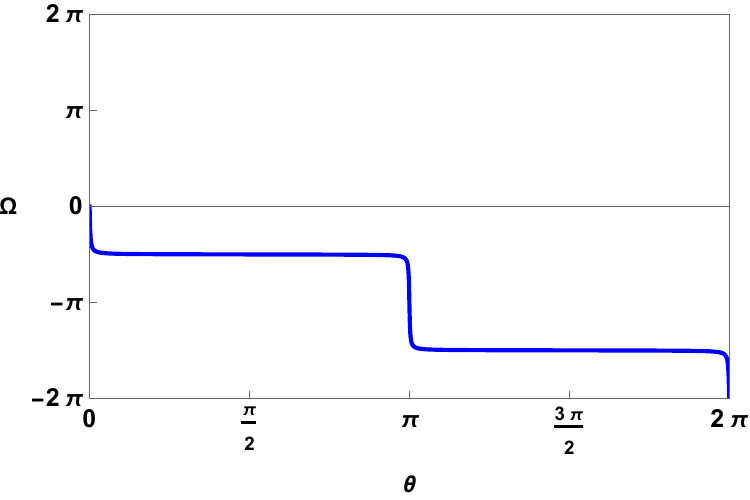}
			\caption{}
			\label{b11b}
		\end{subfigure}
		\caption{Topological class W=1}
		\label{b11}
	\end{figure}

For the first scenario, the thermodynamic parameter values are altered to $\phi = 0.5$, $\gamma = 0.0095$, $\xi = 1$, $\Delta = 1.5$, and $L = 1$. Here, two BH branches are observed, as shown in Fig. \ref{b9a}. The black solid line corresponds to the SBH branch, while the red solid line represents the LBH branch. The winding number calculations, depicted in Fig. \ref{b9b}, focus on the zero points for $\tau = 3$. Two zero points are identified at $S =0.07005$ for the SBH and $S = 0.4010$ for the LBH. 
The SBH exhibits a winding number of $+1$, indicating stability, whereas the LBH has a winding number of $-1$, suggesting instability. The total topological charge for this defect curve is the sum of these winding numbers, yielding $-1 + 1 = 0$. Additionally, an annihilation point is observed at $S =0.139431$ in this case.
	An intriguing scenario arises with parameters $\phi = 0.2$, $\gamma = 0.1$, $\xi = 1$, $\Delta = 0.5$, and $L = 1$.In this case, the local topology undergoes a significant change, with the SBHs becoming unstable and the LBHs gaining stability. Despite this shift in local topology, the global topology remains unchanged. The winding number calculation for $\tau = 3.26$ is illustrated in Fig. \ref{b10b}, where the winding number of the SBH is found to be $-1$, signifying its instability, while the LBH has a winding number of $+1$, indicating stability. A blue dot in the figure marks a generation point at $S=81.13101$, representing the transition from an unstable to a stable phase.The overall topological class is still found to be $0.$\\
	Next, we consider $\phi = 0.5$, $\gamma = 0.01$, $\xi = 1$, $\Delta = 0.5$, and $L = 1$, we find a single BH branch, but now with a winding number of $+1$ as illustrated in  Fig. \ref{b11a} and  Fig. \ref{b11b}

\section{Thermodynamics of $D=4$ charged AdS Black Holes }\label{ThermalBH4D}
\begin{figure*}[!htp]	
		\centering
		\begin{subfigure}{0.43\textwidth}
			\includegraphics[width=\linewidth]{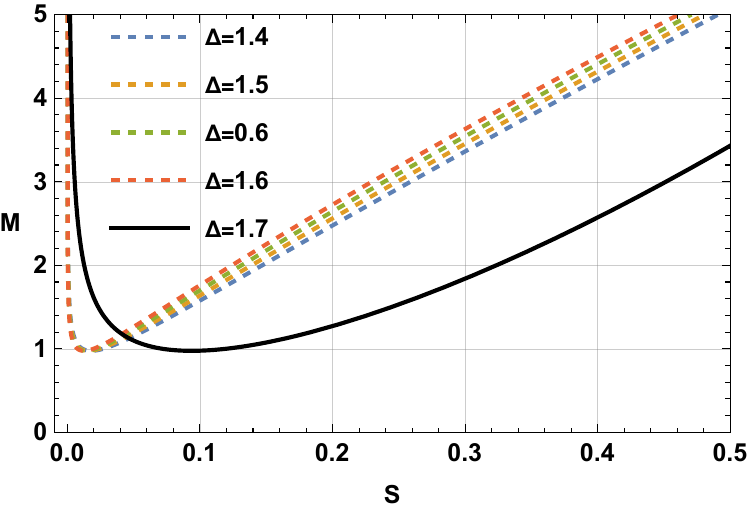}
			\caption{}
			\label{b1a}
		\end{subfigure}
		\begin{subfigure}{0.43\textwidth}
			\includegraphics[width=\linewidth]{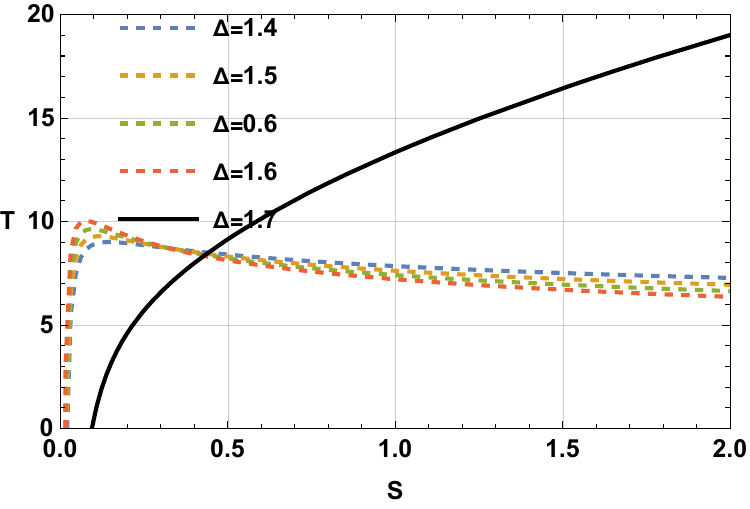}
			\caption{}
			\label{b1b}
		\end{subfigure}
		\begin{subfigure}{0.43\textwidth}
			\includegraphics[width=\linewidth]{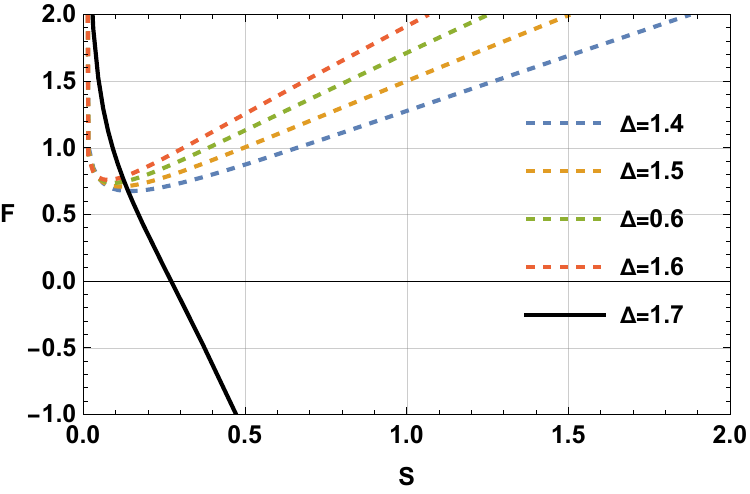}
			\caption{}
			\label{b1c}
		\end{subfigure}
		\begin{subfigure}{0.43\textwidth}
			\includegraphics[width=\linewidth]{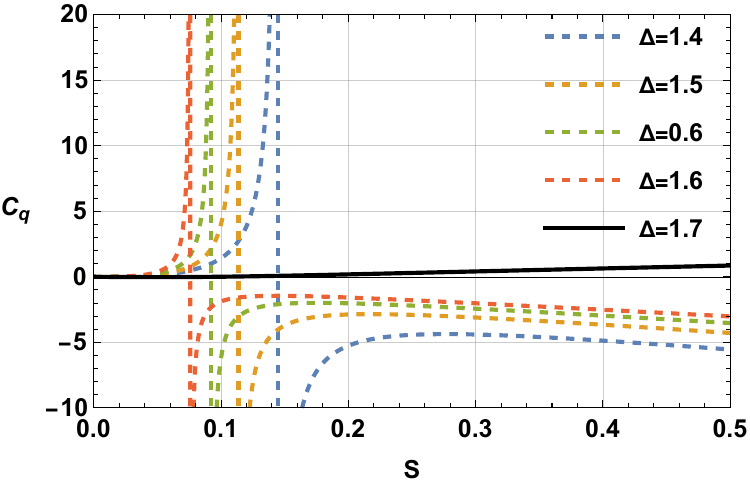}
			\caption{}
			\label{b1d}
		\end{subfigure}
		\begin{subfigure}{0.43\textwidth}
			\includegraphics[width=\linewidth]{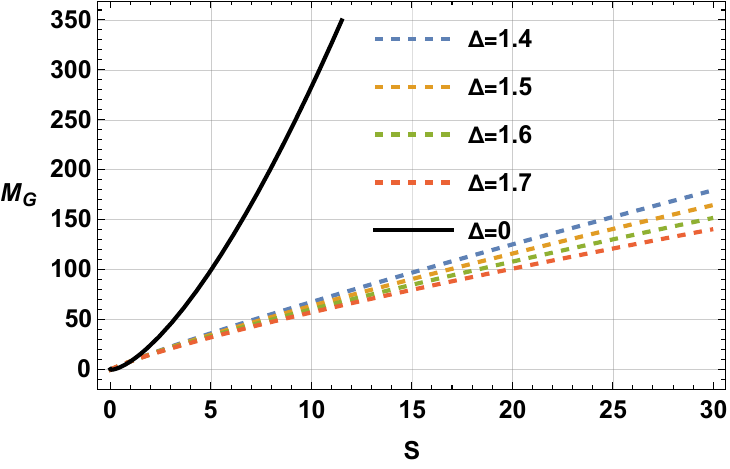}
			\caption{}
			\label{b1e}
		\end{subfigure}
		\begin{subfigure}{0.43\textwidth}
			\includegraphics[width=\linewidth]{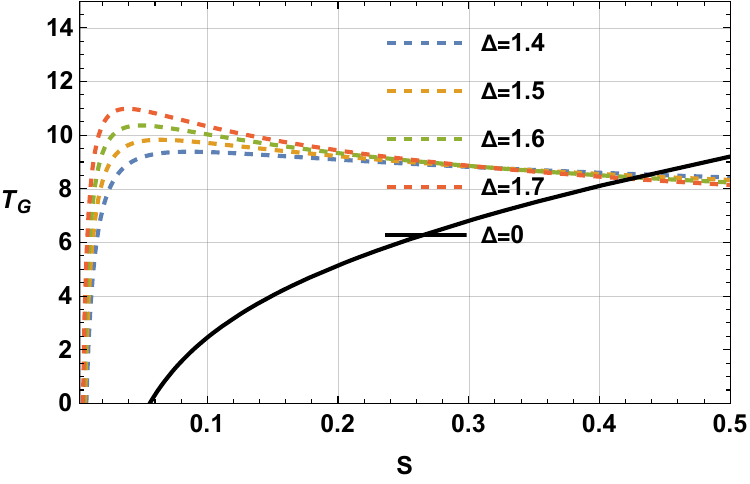}
			\caption{}
			\label{b1f}
		\end{subfigure}
		\begin{subfigure}{0.43\textwidth}
			\includegraphics[width=\linewidth]{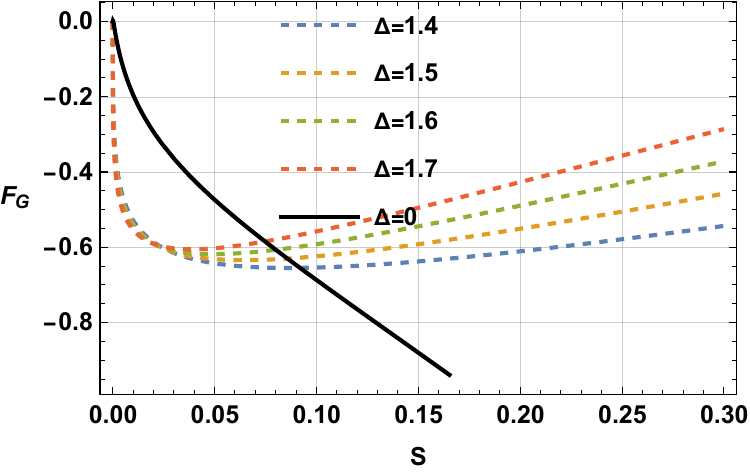}
			\caption{}
			\label{b1g}
		\end{subfigure}
		\begin{subfigure}{0.43\textwidth}
			\includegraphics[width=\linewidth]{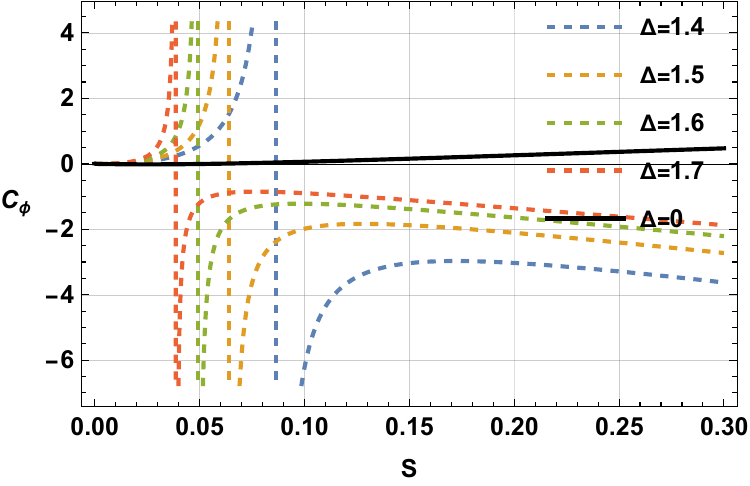}
			\caption{}
			\label{b1h}
		\end{subfigure}
		
		\caption{Thermodynamic property of $D=4$ dimensional  charged AdS BHs immersed in polytropic DE in the framework of Barrow entropy
		}
		\label{b1}
	\end{figure*}

The expression for mass is obtained as :
	\begin{align}
		&M=-\frac{\gamma ^{-n} \left(S^{\frac{2}{\Delta +2}}\right)^{3/2} \, _2F_1\left(n,n;n+1;-\frac{\pi ^{-\frac{3}{2 n}} \left(S^{\frac{2}{\Delta +2}}\right)^{\frac{3}{2 n}} \xi ^2}{\gamma }\right)}{6 \pi ^{3/2}} 
  +\frac{\left(S^{\frac{2}{\Delta +2}}\right)^{3/2}}{2 \pi ^{3/2} L^2}+\frac{\sqrt{\pi } q^2}{2 \sqrt{S^{\frac{2}{\Delta +2}}}}+\frac{\sqrt{S^{\frac{2}{\Delta +2}}}}{2 \sqrt{\pi }}.
		\label{massB4}
	\end{align}
	At $\Delta=0$, the mass term effectively corresponds to the mass in GB statistics.
	The temperature is now calculated again for Barrow entropy as
	\begin{align}
		T=\frac{d M}{d S}=& \frac{3 S^{\frac{2}{\Delta +2}-1} \sqrt{S^{\frac{2}{\Delta +2}}}}{2 \pi ^{3/2} (\Delta +2) L^2}-\frac{\gamma ^{-n} S^{\frac{2}{\Delta +2}-1} \sqrt{S^{\frac{2}{\Delta +2}}} \left(\frac{\pi ^{-\frac{3}{2 n}} \xi ^2 \left(S^{\frac{2}{\Delta +2}}\right)^{\frac{3}{2 n}}}{\gamma }+1\right)^{-n}}{2 \pi ^{3/2} (\Delta +2)}
        \nonumber\\
        &-\frac{\sqrt{\pi } q^2 S^{-\frac{2}{\Delta +2}-1} \sqrt{S^{\frac{2}{\Delta +2}}}}{2 (\Delta +2)}+\frac{\sqrt{S^{\frac{2}{\Delta +2}}}}{2 \sqrt{\pi } (\Delta +2) S}.
		\label{tempB4}
	\end{align}
	Next we calculate the heat capacity $C_q$ and free energy  at constant charge $q$ using the formulas 
	\begin{equation}
	C_q=\frac{dM}{dT}=\frac{dM}{dS}\frac{dS}{dT} ,
	\end{equation}
	\begin{equation}
	F=M-TS .
	\end{equation}
	The effect of  $\Delta$  on the BH properties is more evident for a higher range of values of parameter $\Delta$ and Barrow entropy $S$.   Fig.\ref{b1} shows the thermodynamic properties of the BH in the framework of Barrow entropy, where the effect of $\Delta$ is shown on each parameter while keeping $q=0.5,\Xi=1,\gamma=0.095,L=0.1$ fixed. Fig.\ref{b1a},  shows the plot between mass M and the Barrow entropy $S$. The black solid line that shows the $M$ vs $S$ plot, when  $\Delta$ is set to zero which is the reference line for comparison between GB and Barrow entropy framework. As the parameter $\Delta$ increases from zero, the plots of $M$ versus $S$ begin to diverge from the solid black line. These deviations become more pronounced for larger values of $\Delta$, indicating that the effect on the BH mass becomes significant at lower entropy values when $\Delta$ is larger, compared to when $\Delta$ is smaller. Similarly, temperature $T$ is plotted against $S$ in Fig.\ref{b1b}. where we see two BH phases for higher values of $\Delta$. The BH branch is also visible in free energy $F$ against the Barrow entropy $S.$  The thermal stability of the BH branch is depicted by Fig.\ref{b1d} where we have plotted $C_q$ against the Barrow entropy $S$. It is seen that for all values of $q$ and $L$, the heat capacity is positive for SBH branch, which means that SBH branch is thermally stable and the opposite is spotted for a LBH branch. Here it is seen that there exist Davies type phase transition as at some point, the specific heat diverges.\\
Next we discuss the thermodynamic properties of these BH in grand canonical ensemble within the Barrow entropy framework, where the  potential $\phi$ is kept fixed instead of its conjugate parameter $q.$ $\phi$ can be calculated as :
	\begin{equation}
	\phi =\frac{\partial M}{\partial q}=\frac{\sqrt{\pi } q}{\sqrt{S^{\frac{2}{\Delta +2}}}} ,
	\label{phiGC}
	\end{equation}
	solving Eq. \eqref{phiGC}, we can obtain expression of $q$ in terms of $\phi$. Next $q$ is substituted in the equation given below and the grand canonical mass $M_G$ is obtained
 \small 
\begin{align}
	M_G=&M-q \phi
    =\frac{\left(S^{\frac{2}{\Delta +2}}\right)^{3/2}}{2 \pi ^{3/2} L^2}-\frac{\phi ^2 \sqrt{S^{\frac{2}{\Delta +2}}}}{2 \sqrt{\pi }}+\frac{\sqrt{S^{\frac{2}{\Delta +2}}}}{2 \sqrt{\pi }}  
    -\frac{\gamma ^{-n} \left(S^{\frac{2}{\Delta +2}}\right)^{3/2} \, _2F_1\left(n,n;n+1;-\frac{\pi ^{-\frac{3}{2 n}} \left(S^{\frac{2}{\Delta +2}}\right)^{\frac{3}{2 n}} \xi ^2}{\gamma }\right)}{6 \pi ^{3/2}},
\end{align}
 \normalsize
	following the same steps as explained above, the other thermodynamic quantities are obtained. It is seen that they follow the same thermodynamic behaviour as that of canonical ensemble. The grand canonical mass and temperature is plotted in Fig.\ref{b1e} and Fig.\ref{b1f} respectively while keeping $\phi=0.5,\Xi=1,\gamma=0.0095,L=0.1$ constant. The free energy and heat capacity plot in GC ensemble found to be showing similar behaviour to that in canonical ensemble which depicts that in GC ensemble also, the BH shows Davies-type phase transition within the Barrow entropy framework.\\
	
\begin{figure}[!htp]	
		\centering
		\begin{subfigure}{0.43\textwidth}
			\includegraphics[width=\linewidth]{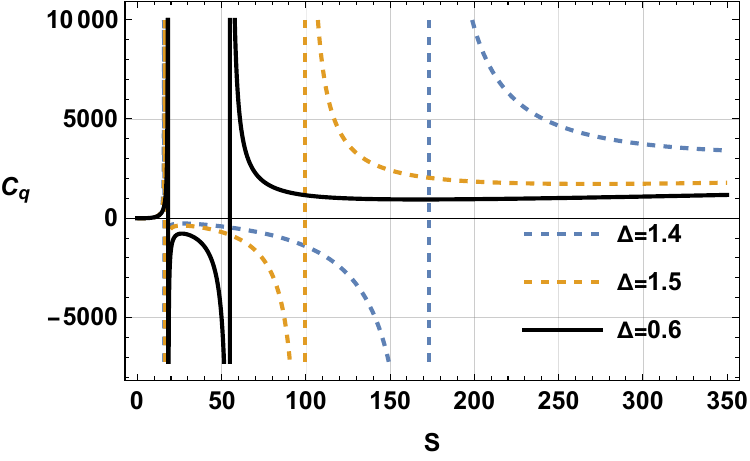}
			\caption{}
			\label{c1}
		\end{subfigure}
		\hspace{0.5cm}
		\begin{subfigure}{0.43\textwidth}
			\includegraphics[width=\linewidth]{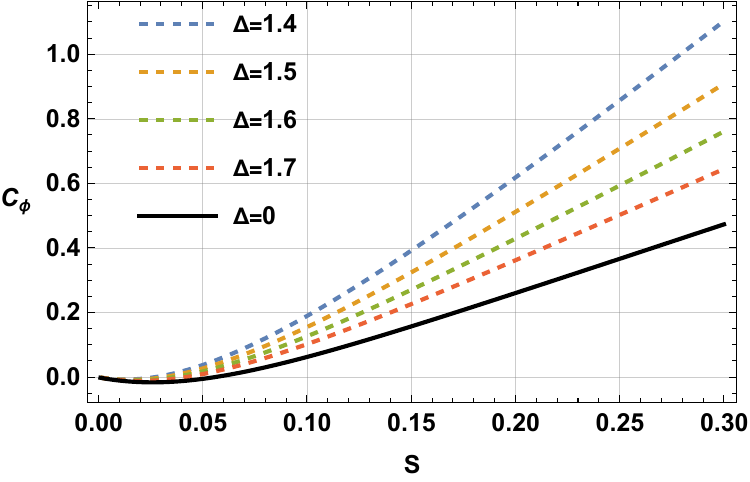}
			\caption{}
			\label{c2}
		\end{subfigure}
		\caption{Comparison of heat capacity curve in both ensemble within GB statistics framework}
		\label{c11}
		\end{figure}

	To study the thermodynamic properties of these BHs in GB statistics, we consider smaller value of the parameter $\Delta$. The phase transition properties are most evident in the heat capacity vs entropy curve, hence we have decided to plot only the heat capacity curve. From Fig.\ref{c1}, it is evident that in canonical ensemble,  these 4D BHs exhibit Van der Waals-like phase transitions within GB statistics framework depending upon the values of $q$ and $L$. On the other hand, in Barrow statistics, they exhibit Davies type phase transition. For grand canonical ensemble, we do not see any phase transition in GB statistics. The Fig.\ref{c2} depicts there exist only one stable BH branch for all values of $\phi$ and $L.$
  
	\subsection{Thermodynamic topology of $4D$  Black Holes}
 
\begin{figure*}[!htp]
		\centering
		\begin{subfigure}{0.43\textwidth}
			\includegraphics[width=\linewidth]{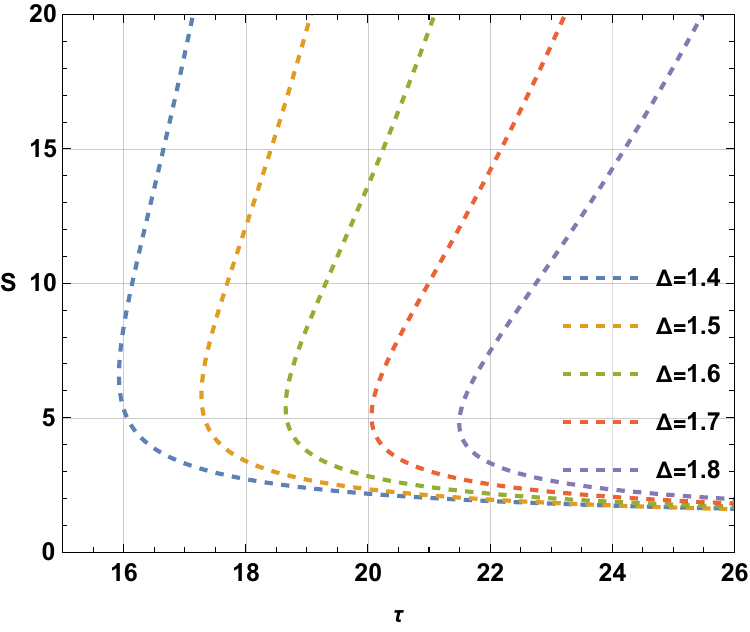}
			\caption{}
			\label{b2a}
		\end{subfigure}
		\begin{subfigure}{0.43\textwidth}
			\includegraphics[width=\linewidth]{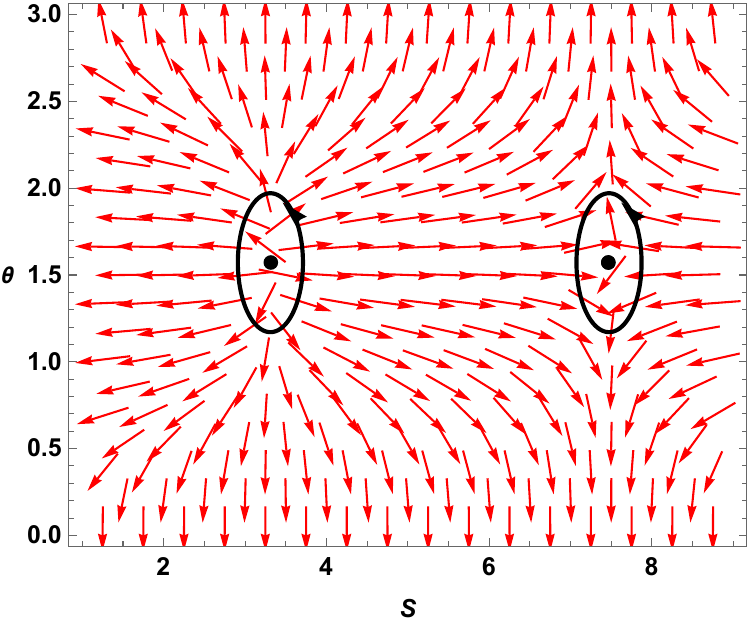}
			\caption{}
			\label{b2b}
		\end{subfigure} \hspace{0.6cm} 
		\begin{subfigure}{0.43\textwidth}
			\includegraphics[width=\linewidth]{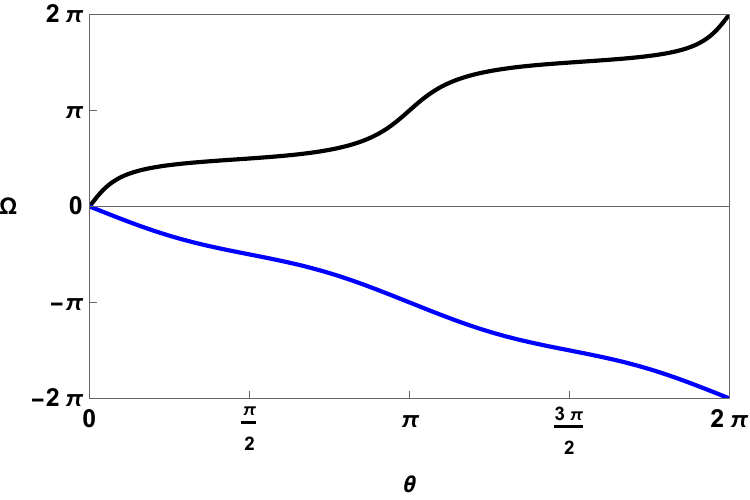}
			\caption{}
			\label{b2c}
		\end{subfigure}
		\begin{subfigure}{0.43\textwidth}
			\includegraphics[width=\linewidth]{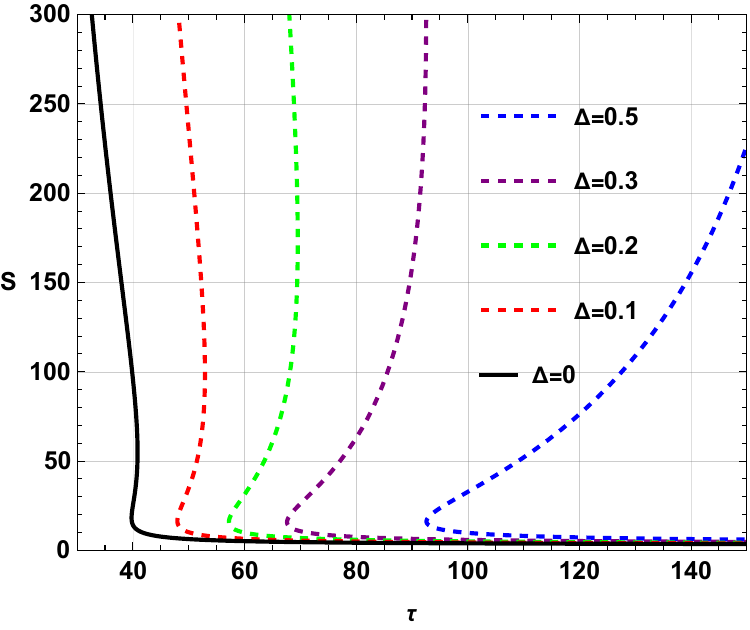}
			\caption{}
			\label{b2d}
		\end{subfigure}
		\begin{subfigure}{0.43\textwidth}
			\includegraphics[width=\linewidth]{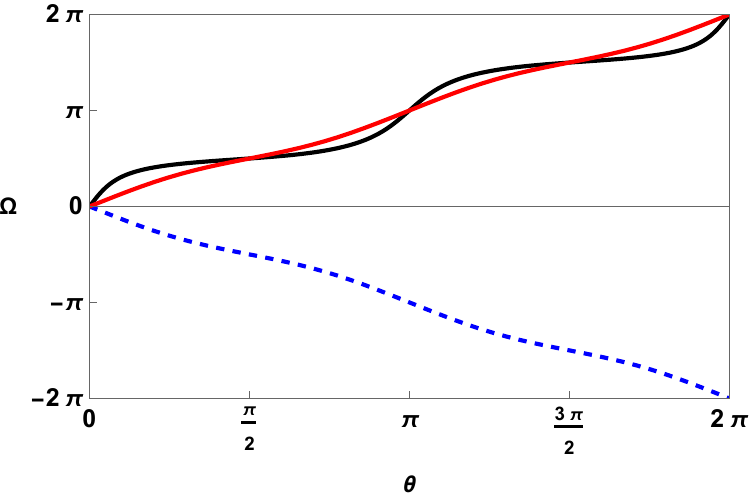}
			\caption{}
			\label{b2e}\
		\end{subfigure} \hspace{0.6cm} 
		\caption{Topological classes of 4D charged AdS BHs immersed in polytropic DE using Barrow entropy.The upper panel corresponds to topological class of BHs in  canonical ensemble in Barrow entropy framework and the lower panel shows the same in GB statistics framework}
		\label{b2}
	\end{figure*}

The expression for $\tau$ is obtained as :
	\begin{equation}
		\tau =\frac{\mathcal{A}}{\mathcal{B}} ,
	\end{equation}
 where
  \begin{small}
 \begin{equation}
     \mathcal{A}=6 \pi ^2 (\Delta +2) L^2 \gamma ^n S^{\frac{2 (\Delta +4)}{\Delta +2}} \Bigg(\frac{\left(\frac{\pi }{2}\right)^{2/n} \xi ^2 \left(S^{\frac{2}{\Delta +2}}\right)^{2/n}}{\gamma }+1\Bigg)^n ,
 \end{equation}
 and
 \begin{align}
     &\mathcal{B}=3 \pi ^4 \gamma ^n S^{\frac{\Delta +10}{\Delta +2}} \bigg(\frac{\left(\frac{\pi }{2}\right)^{2/n} \xi ^2 \left(S^{\frac{2}{\Delta +2}}\right)^{2/n}}{\gamma }+1\bigg)^n-L^2 \Bigg(\pi ^4 S^{\frac{\Delta +10}{\Delta +2}}+48 q^2 S \Bigg(\left(\frac{\pi }{2}\right)^{2/n} \xi ^2 \left(S^{\frac{2}{\Delta +2}}\right)^{2/n}+\gamma\Bigg)^n\Bigg).
 \end{align}
 \end{small}
	The defect curve ( $\tau$ vs $S$) is plotted for Barrow entropy case in  Fig.\ref{b2b}. As it is seen from this figure, there are two branches in the defect curve for different values of $\Delta$ while we have kept $q=0.5,L=1,\Xi=1,\gamma=0.0095 $  constant. Next, to calculate the topological charge, we choose any value of $\tau$ from the defect curve and draw the normalized vector plot around the selected zero point as shown in Fig.\ref{b2b}.From all the curves in  Fig.\ref{b2b} we choose the purple dashed curve where $\Delta=1.8.$ For $\tau=22$, the zero points are located at $S_{0}=3.31139$ and $S_{0}=7.47261$ represented by black dots in Fig.\ref{b2b}. Hence we draw the normalized vector field in the range $1\leq S \leq 9$ with the component $\phi^S$ and $\phi^\theta$.	

For the calculation of topological charge,  we construct a contour around each zero point using the Eq.(\ref{contour}) as  represents by the black contours in Fig.\ref{b2b}:
	\begin{equation}
		\begin{cases}
			S=0.3 \cos\nu +S_{0}, \\ 
			\\ 
			\theta =0.3 \sin\nu +\frac{\pi }{2}.%
		\end{cases}%
		\label{contour1}
	\end{equation}%
	We find the deflection from  Eq.(\ref{deflection}) where we solve the contour integration by simply substituting $S$ and $\theta$ by this parametrized contour values in Eq.(\ref{contour1}).  From deflection, winding number can be  calculated using  Eq.(\ref{winding}).
	Fig. \ref{b2c} displays the contour plots, which illustrate the value of deflection at $\theta=2\pi$. The winding number for $S_0 = 3.31139$  is found to +1 as shown by the black line and that for $S_0=7.47261$  is calculated to be $-1$ illustrated by the blue line. A positive winding number implies a stable BH branch. Hence, the SBH is stable and the LBH is unstable.Since the phase transition is taking place from winding number $+1$ to $-1$, hence we identify the critical points as annihilation points,\\
	
	Now we move on to GB statistics frame work. As it is evident in Fig.\ref{b2d}, the number of branch for lower value of $\Delta$ is found to be either one or three. From our previous analysis, it is well expected that the topological charge must be +1. In Fig.\ref{b2e}, we confirmed it by plotting the parametrized contours for $\Delta=40,q=0.5, L=10,\Xi=1,\gamma=0.0095 $ and $\tau=40$  The black and red solid line in Fig.\ref{b2e}, represents the winding number calculation for SBH and LBH respectively which is found to be $+1$. The blue dashed line depicts the winding number for IBH which is $-1.$. Adding all the winding numbers for each branch, we found the topological charge of $+1$ as expected.
	Hence, $4D$ charged BHs in canonical ensemble immersed in polytropic DE within the framework of Barrow entropy has topological charge of $0$ but the same is found to be $+1$ within GB statistics framework. The topological charge does not change with change in value of all the thermodynamic parameters apart from the Barrow entropy parameter $\Delta$\\

 In grand canonical ensemble, the expression for $\tau$  is given by :
	\begin{equation}
	\tau_G=\frac{\mathcal{A}_1}{\mathcal{B}_1} ,
	\end{equation}
 where,
 \begin{align}
   \mathcal{A}_1=  2 \pi ^{3/2} (\Delta +2) L^2 S \gamma ^n \left(\frac{\pi ^{-\frac{3}{2 n}} \xi ^2 \left(S^{\frac{2}{\Delta +2}}\right)^{\frac{3}{2 n}}}{\gamma }+1\right)^n,
 \end{align}
 and 
 \begin{align}
     \mathcal{B}_1= &\sqrt{S^{\frac{2}{\Delta +2}}} \biggl( 3 \gamma ^n S^{\frac{2}{\Delta +2}} 
\biggl( \frac{\pi ^{-\frac{3}{2 n}} \xi ^2 \biggl(S^{\frac{2}{\Delta +2}}\biggr)^{\frac{3}{2 n}}}{\gamma } + 1 \biggr)^n 
\nonumber \\
&\,- L^2 \biggl( \pi  \biggl(\phi ^2 - 1\biggr) \gamma ^n 
\biggl( \frac{\pi ^{-\frac{3}{2 n}} \xi ^2 \biggl(S^{\frac{2}{\Delta +2}}\biggr)^{\frac{3}{2 n}}}{\gamma } + 1 \biggr)^n 
+ S^{\frac{2}{\Delta +2}} \biggr) \biggr).
 \end{align}
 
\begin{figure*}[!htp]
		\centering
		\begin{subfigure}{0.43\textwidth}
			\includegraphics[width=\linewidth]{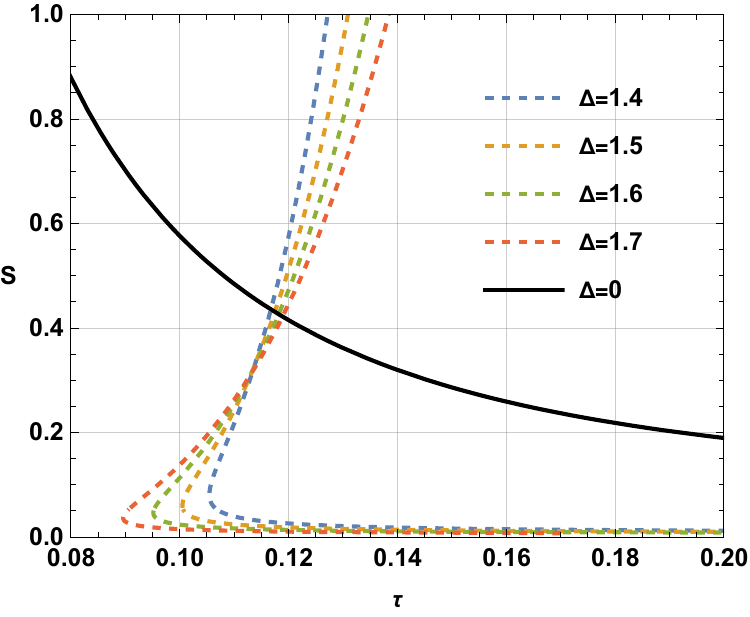}
			\caption{}
			\label{g4a}
		\end{subfigure}
		\hspace{0.5cm}
		\begin{subfigure}{0.43\textwidth}
			\includegraphics[width=\linewidth]{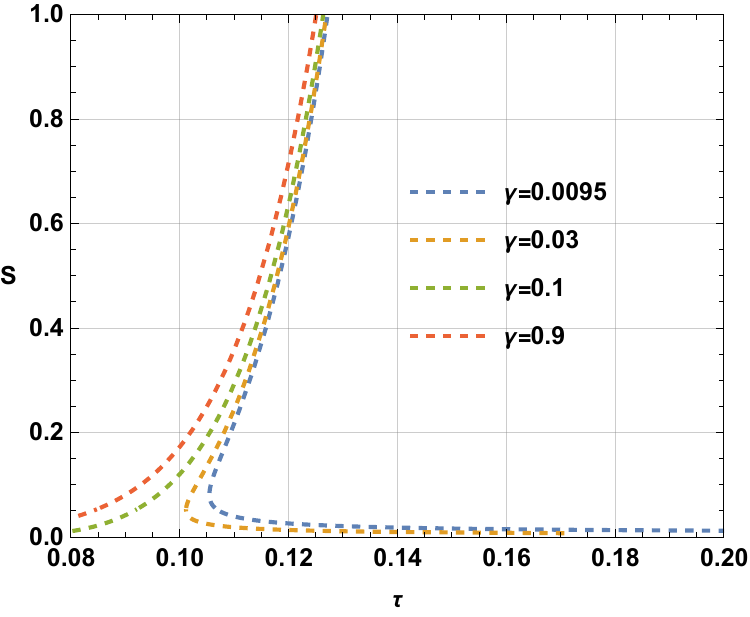}
			\caption{}
			\label{g4b}
		\end{subfigure}
			\hspace{0.5cm}
		\begin{subfigure}{0.43\textwidth}
			\includegraphics[width=\linewidth]{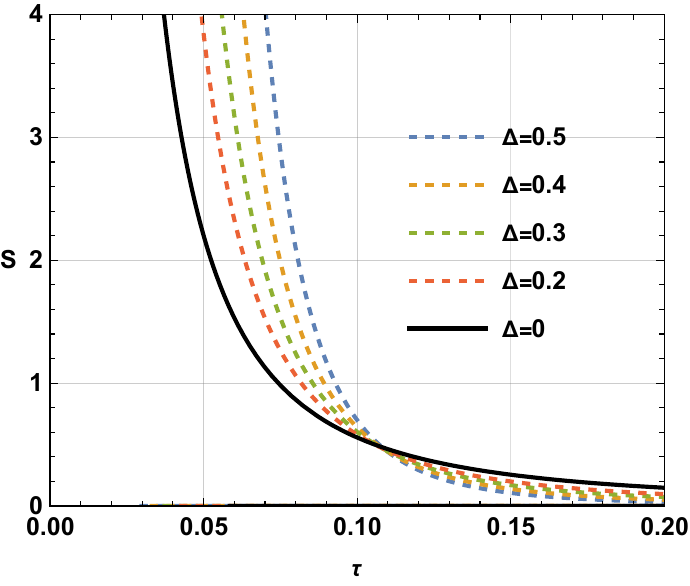}
			\caption{}
			\label{g4c}
		\end{subfigure} 
		\caption{Topological classes of 4D charged AdS BHs immersed in polytropic DE in grand canonical ensemble.}
		\label{g4}
	\end{figure*}

	In Fig.\ref{g4a}, the defect curve for 4D BHs in the grand canonical ensemble is shown. The dashed lines represent the $\tau$ vs $S$ curve within the Barrow entropy framework, while the black solid line corresponds to the same curve in GB statistics. As depicted in Fig.\ref{g4a}, the topological charge in the Barrow entropy framework is found to be $W = 0$, with the SBH stable and the LBH unstable, where we observe annihilation points. In contrast, the topological charge in the GB statistics framework is $W = 1$. Furthermore, within the Barrow entropy framework, the topological charge changes to $W = -1$ depending on the parameter $\gamma$. In Fig.\ref{g4b}, BHs in the single branch belong to the topological class $W = -1$, while BHs with two branches belong to the topological class $W = 0$, featuring an annihilation point. Meanwhile, Fig.\ref{g4c} shows that the topological class in GB statistics remains consistently $W = +1$, independent of any thermodynamic parameters. Notably, no Van der Waals-like phase transition is observed in the grand canonical ensemble.
 
\section{Sparsity of Black Hole radiation}\label{sparsityBHrad}
As an interesting further characteristic of BHs and the Hawking radiation flux, we now proceed to the study of sparsity, defined as the average time difference between the emission of successive quanta. Relative to the black body, Hawking radiation is considerably sparser, as evidenced, e.g., in ~\cite{Spars}. This is a key feature that distinguishes the two distinct types of radiation system. 

For $D$-dimensional BHs, sparsity is governed by the parameter,
\begin{equation}
\label{sparseq}
    \eta =\frac{\mathcal{H}}{\Tilde{g} }\left(\frac{\lambda_t^{D-2} }{\mathcal{A}_{eff}}\right),
\end{equation}
which corresponds to the well-known expression for $D=4$~\cite{Spars}. For this purpose, $\mathcal{H}$ is a dimensionless constant, $\Tilde{g}$ is the spin degeneracy factor of the emitted quanta, $\lambda_t=2\pi/T$ represents their thermal wavelength, and $\mathcal{A}_{eff}=\sigma_{\text{capt}}\mathcal{A}_{BH}$ refers to the effective area of the BH where $\sigma_{\text{capt}}$ stands for the capture cross-section of the photon by the Schwarzschild BH \cite{Ahmedov:2021ohg}. In the most straightforward case of Schwarzschild BHs of dimension $(1+3)$, taking into account the emission of massless bosons, the constant is $\eta=64\pi^3/27\simeq73.49$. For the sake of comparison, we note that $\eta\ll1$ for black bodies.  

Accordingly, with a view to uncovering on how the polytropic structure affects the sparsity characteristic relevant to higher-dimensional charged AdS BHs, consider, responsive to this, the corresponding high-dimensional effective area of BH involved in the capture cross-section as follows \cite{Ahmedov:2021ohg}
\begin{equation}\label{areaspars1}
\mathcal{A}_{eff}=\mathcal{A}_{BH}\frac{2^{4/(D-3)}(D-3)(D-1)^{(D-1)(D-3)}}{\left[2^{3/(D-3)}(D-1)-2^{D/(D-3)} \right]^2}\pi,
\end{equation}
where the BH area in higher-dimensions can be- represented by
\begin{equation}\label{areaspars2}
    \mathcal{A}_{BH}=\frac{2\pi^{(D-1)/2}}{\Gamma\left(\frac{D-1}{2}\right)}r_h^{D-2}.
\end{equation}

Corrections imposed on Eq.~\eqref{sparseq} by generalized entropies and / or uncertainty relations were investigated in~\cite{Cor1,Cor2,Cor3,Luciano:2023fyr,Sekhmani:2024udl,Sekhmani:2024kfj,Sekhmani:2024dhc}. Furthermore, the computation of sparsity in $D$-dimensional generic Tangherlini BHs has been carried out in~\cite{Schuster:2019xvp}. The aim herein is to explore how Eq.~\eqref{sparseq} arises for charged AdS BHs with a surrounding polytropic structure in high-dimensional spacetimes. In this respect, we point out that the straightforward substitution of the modified Hawking temperature which is obtained by using \eqref{rbar} in \eqref{Tbar}, such that
\begin{align}
     T&=\frac{1}{\pi }\Biggl\{4^{\frac{1}{2-D}-1} \bigg(\frac{S^{\frac{2}{\Delta +2}}}{\omega_{D-2} }\bigg)^{\frac{1}{2-D}} \Bigg(4^{\frac{2}{D-2}} \bigg(\frac{S^{\frac{2}{\Delta +2}}}{\omega_{D-2} }\bigg)^{\frac{2}{D-2}} 
     \Bigg(-\frac{2}{D-2}\bigg(\gamma +\xi ^2 \bigg(4^{\frac{1}{D-2}}
   \bigg(\frac{S^{\frac{2}{\Delta +2}}}{\omega_{D-2} }\bigg)^{\frac{1}{D-2}}\bigg)^{\frac{D-1}{n}}\bigg)^{-n}
   \nonumber\\
   &-\frac{1}{L^2}+\frac{48 \pi ^2 Q^2 \bigg(4^{\frac{1}{D-2}} \bigg(\frac{S^{\frac{2}{\Delta +2}}}{\omega_{D-2}}\bigg)^{\frac{1}{D-2}}\bigg)^{4-2 D}}{\omega_{D-2}
   ^2}\Bigg) 
   +D \Bigg(-\frac{16 \pi ^2 Q^2 \bigg(4^{\frac{1}{D-2}} \bigg(\frac{S^{\frac{2}{\Delta +2}}}{\omega_{D-2}}\bigg)^{\frac{1}{D-2}}\bigg)^{6-2
   D}}{\omega_{D-2} ^2}+1\Bigg)\nonumber\\&+4^{\frac{2}{D-2}}\frac{D}{L^2} \bigg(\frac{S^{\frac{2}{\Delta +2}}}{\omega_{D-2}}\bigg)^{\frac{2}{D-2}}-3\Bigg)\Biggr\},
    \end{align}
into Eq.~\eqref{sparseq} yields
     \begin{align}
        \eta&=(D-1)^{(3-D) (D-1)} \pi ^{\frac{1}{2} (D (D+2)-9)}\left(\frac{\mathcal{C}}{D-3}\right)\Gamma \left(\frac{D-1}{2}\right)\nonumber\times \frac{2^{3 D-\frac{4}{D-3}-5} \left(2^{\frac{D}{D-3}}-8^{\frac{1}{D-3}} (D-1)\right)^2  }{\left( \left(\pi ^{\frac{1}{2}-\frac{D}{2}} \Gamma
   \left(\frac{D-1}{2}\right) S^{\frac{2}{\Delta +2}}\right)^{\frac{1}{D-2}}\right)^{D-2}},
     \end{align}
     
     where we have defined
     \begin{small}
         \begin{align}
&\mathcal{C}=\Bigg(\frac{\mathcal{C}_1}{ \mathcal{C}_2-8  Q^2 \Gamma \bigg(\frac{D-1}{2}\bigg)^{\frac{2(D-1)}{D-2}} \pi ^{\frac{-3}{D-2}}  \bigg(2^{\frac{\Delta }{2}+1} S\bigg)^{\frac{12}{(\Delta +2)(D-2)}}}\Bigg)^{D-2}\nonumber\\
         &\mathcal{C}_1=(D-2) (\Delta +2) S \left(\pi ^{\frac{1-D}{2}} \Gamma \left(\frac{D-1}{2}\right) S^{\frac{2}{\Delta
   +2}}\right)^{\frac{3}{D-2}}\times\Gamma \left(\frac{D-1}{2}\right) \bigg(2^{\frac{1}{D-2}} \bigg(\pi ^{\frac{1-D}{2}} \Gamma \left(\frac{D-1}{2}\right) S^{\frac{2}{\Delta +2}}\bigg)^{\frac{1}{D-2}}\bigg)^D,\nonumber\\
   &\mathcal{C}_2=\Biggl\{\frac{(D-2) (D-1) \bigg(\pi ^{\frac{1-D}{2}} \Gamma \left(\frac{D-1}{2}\right) \left(2^{\frac{\Delta }{2}+1} S\right)^{\frac{2}{\Delta
   +2}}\bigg)^{\frac{2}{D-2}}}{L^2}
   D^2-2 \bigg(\pi ^{\frac{1-D}{2}} \Gamma \left(\frac{D-1}{2}\right) \left(2^{\frac{\Delta }{2}+1} S\right)^{\frac{2}{\Delta +2}}\bigg)^{\frac{2}{D-2}} \nonumber\\
   &\times\bigg(\gamma +\xi ^2 \bigg(2^{\frac{1}{D-2}} \bigg(\pi
   ^{\frac{1-D}{2}} \Gamma \left(\frac{D-1}{2}\right) S^{\frac{2}{\Delta +2}}\bigg)^{\frac{1}{D-2}}\bigg)^{\frac{D-1}{n}}\bigg)^{-n}
   -5 D+6\Biggr\}\pi ^D 
   \nonumber\\
   &\times \bigg(2^{\frac{1}{D-2}} \bigg(\pi ^{\frac{1-D}{2}} \Gamma
   \left(\frac{D-1}{2}\right) S^{\frac{2}{\Delta +2}}\bigg)^{\frac{1}{D-2}}\bigg)^{2 D}.
     \end{align}
     \end{small}
     
\begin{figure*}[!htp]	
		\centering
		\begin{subfigure}{0.43\textwidth}
			\includegraphics[width=\linewidth]{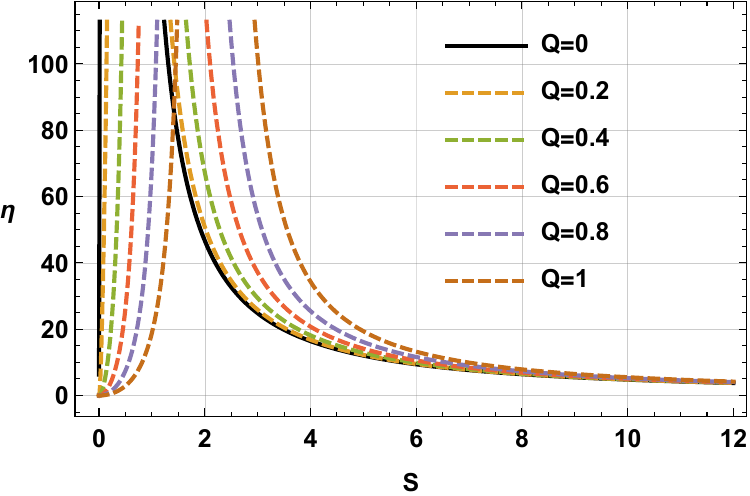}
			\caption{$\xi=1, \gamma=0.095, \Delta=0.4$}
			\label{sp1}
		\end{subfigure}
		\begin{subfigure}{0.43\textwidth}
			\includegraphics[width=\linewidth]{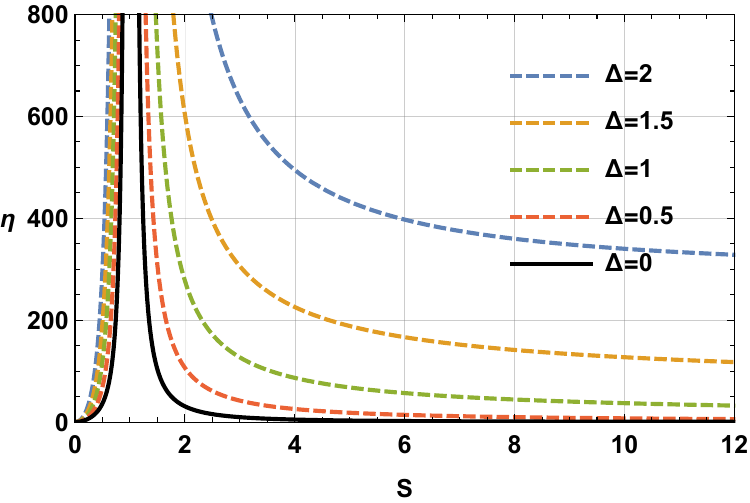}
			\caption{$\xi=1, \gamma=0.095, Q=0.5$}
			\label{sp2}
		\end{subfigure}
		\begin{subfigure}{0.43\textwidth}
		\includegraphics[width=\linewidth]{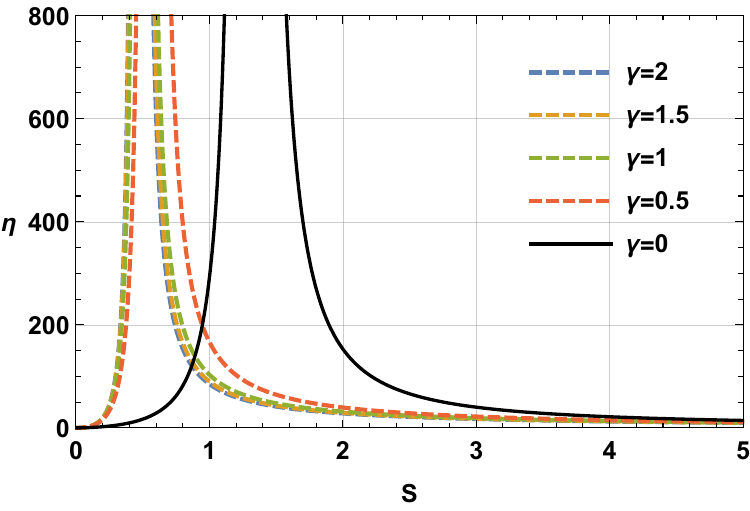}
		\caption{$\xi=1, \Delta=0.4, Q=0.5$}
		\label{sp3}
	\end{subfigure}
	\caption{Sparity effects for the fixed-parameter polytropic structure of the BH system in $4D$ spacetime.
		}
		\label{SPP4}
	\end{figure*}

\begin{figure*}[!htp]	
		\centering
		\begin{subfigure}{0.43\textwidth}
			\includegraphics[width=\linewidth]{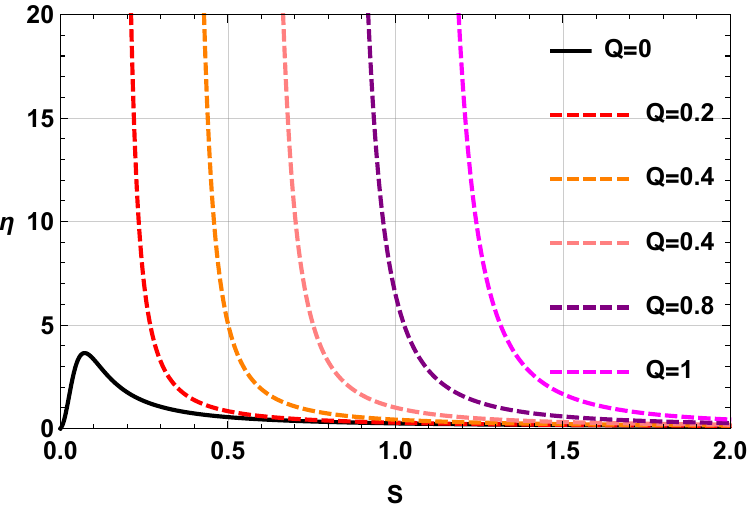}
			\caption{$\xi=1, \gamma=0.095, \Delta=0.4$}
			\label{sp4}
		\end{subfigure}
		\begin{subfigure}{0.43\textwidth}
			\includegraphics[width=\linewidth]{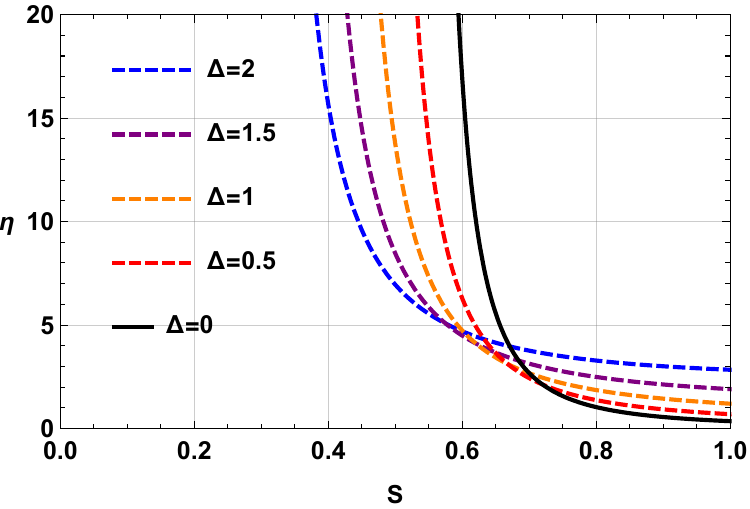}
			\caption{$\xi=1, \gamma=0.095, Q=0.5$}
			\label{sp5}
		\end{subfigure}
		\begin{subfigure}{0.43\textwidth}
		\includegraphics[width=\linewidth]{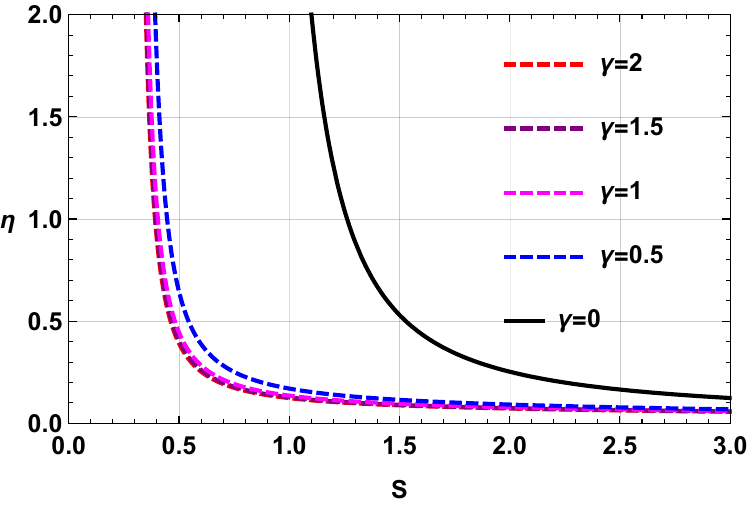}
		\caption{$\xi=1, \Delta=0.4, Q=0.5$}
		\label{sp6}
	\end{subfigure}
	\caption{Sparity effects for the fixed-parameter polytropic structure of the BH system in  $5D$ spacetime.
		}
		\label{SPP5}
	\end{figure*}

Examination of the plot in Figs.~\ref{SPP4} and \ref{SPP5} offers some interesting information about the dimensions of $4D/5D$ spacetime. Unlike Schwarzschild's BHs, the sparsity fluctuates ( i.e. decreases) as $S$ increases, i.e. as the size of the BHs increases. For $S$ sufficiently large, $\eta\ll1$ is achieved: in this regime, the radiation emitted by AdS BHs with a surrounding polytropic structure appears to be almost classical and completely consistent with the spectrum of a black body. Alternatively, we observe that for a given $S$, the larger the size of the BH, the sparser the radiation, and vice versa. 


On the one hand, this is caused by the surround polytropic structure ($\gamma$ decreases), which slows down the emission of quanta and therefore increases the density of the radiation as $Q$ and $\Delta$ increases (see Figs.~\ref{sp1}-~\ref{sp1} and Figs.~\ref{sp4}-~\ref{sp5}), while remaining observed as $D$ decreases $(5D\rightarrow 4D)$ from a point of view regarding Figs.~\ref{SPP4} and  \ref{SPP5}. For the sake of comparison, we note that in~\cite{Schuster:2019xvp} - where no polytropic structure is being considered - the density is lossy in high dimensions. Broadly speaking, this outcome is valid and serves to highlight the impact of the polytropic structure on the fluctuation of the Hawking radiation sparsity.

 \section{Thermal geometry}\label{Sectthermalgeometry}
 
In investigating in a depth-way the relevant heat capacity phase transition points, an alternative approach revealing this feature can be taken into account, such as geometrical thermodynamics (GT). In particular, the characteristic of the heat capacity phase transition points are correlated to divergent points of the thermodynamic Ricci curvature scalar. Applying the GT tools offer on the other hand a practical way to gain a better understanding about thermodynamic phase transitions from a geometric point of view, especially from the Ricci scalar curvature of the thermodynamic metric. In the realm of these endeavors, a wide number of GT tools are regarded efficiently to predict the phase transition point of heat capacity such as Weinhold, Ruppeiner and Quevedo \cite{Weinhold:1975xej, Ruppeiner:1995zz, Quevedo:2006xk, Quevedo:2008xn}. However, it has been demonstrated that the phase transition points provided by this family of metrics are not consistent with those provided by the canonical ensemble approach. For that reason, the application of the so-called Hendi, Panahiyan, Eslam Panah and Momennia (HPEM) metric is therefore efficace to set out a presice axact position of the heat capacity phase transition points \cite{Hendi:2015rja,Hendi:2016pvx,Hendi:2015hoa,Hendi:2015xya}. The associated metric for HPEM is given by
\newpage
\begin{figure}[!htp]	
		\centering
		\begin{subfigure}{0.43\textwidth}
			\includegraphics[width=\linewidth]{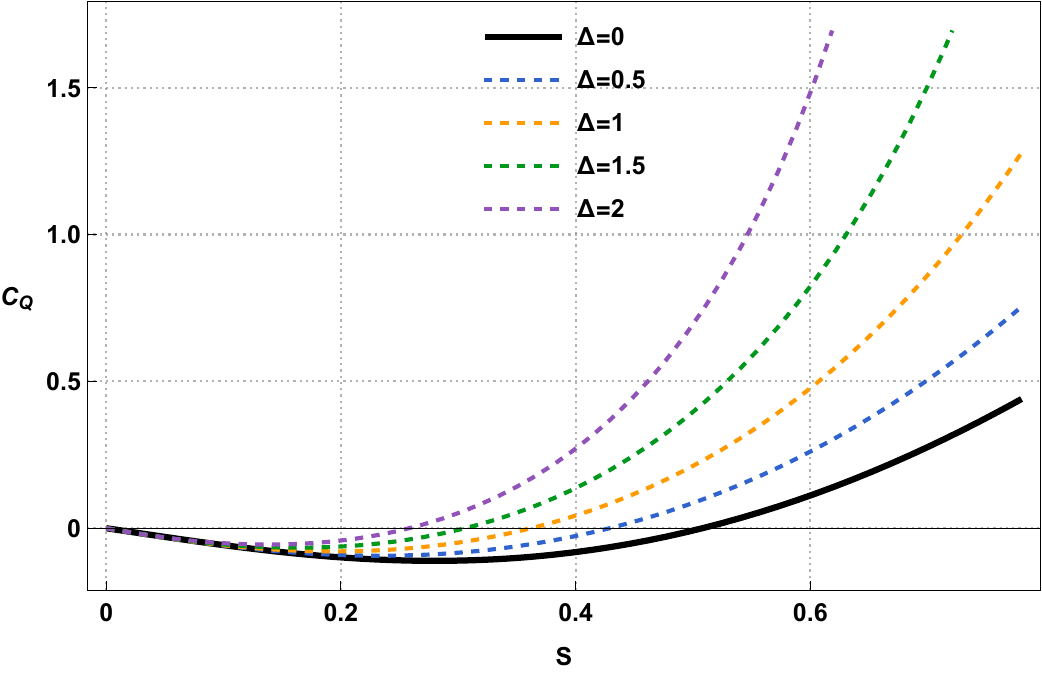}
			\caption{}
			\label{ccc1}
		\end{subfigure}
		\hspace{0.5cm}
		\begin{subfigure}{0.43\textwidth}
			\includegraphics[width=\linewidth]{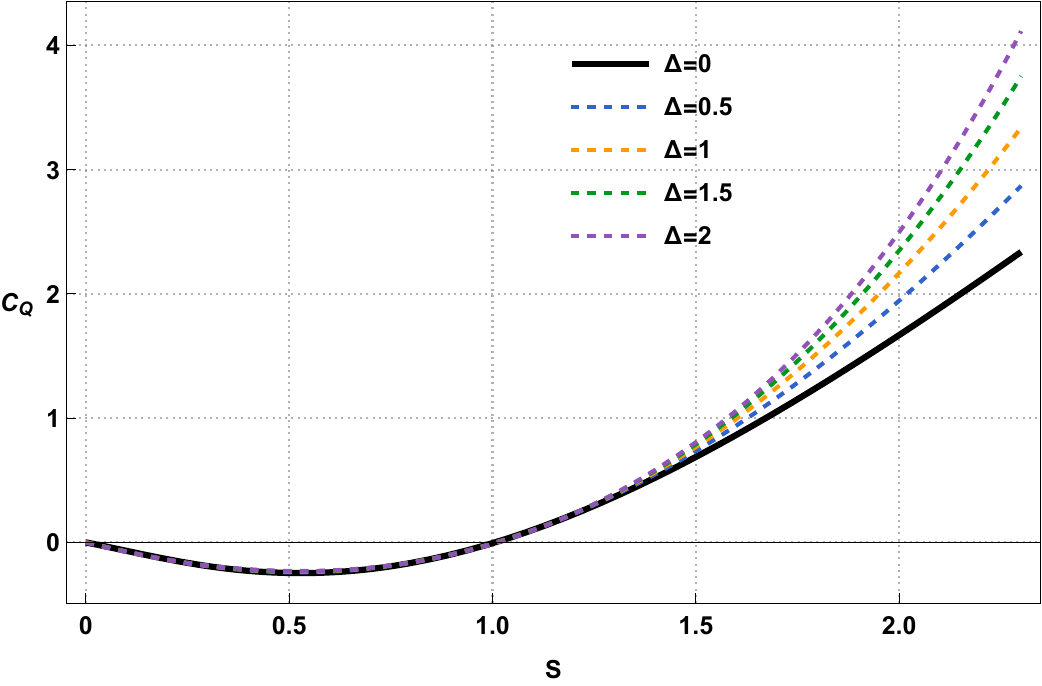}
			\caption{}
			\label{ccc2}
		\end{subfigure}
		\caption{Heat capacity $C_Q$ variation against $S$ ($C_Q$ roots, i.e., physical limitation point) for various values of the Barrow parameter $\Delta$ with setting $\gamma=0.095$, $L=1$, $Q=0.5$, $\xi=1$, and $n=2$.}
		\label{CQ1}
		\end{figure}

  \begin{figure}[!htp]	
		\centering
		\begin{subfigure}{0.43\textwidth}
			\includegraphics[width=\linewidth]{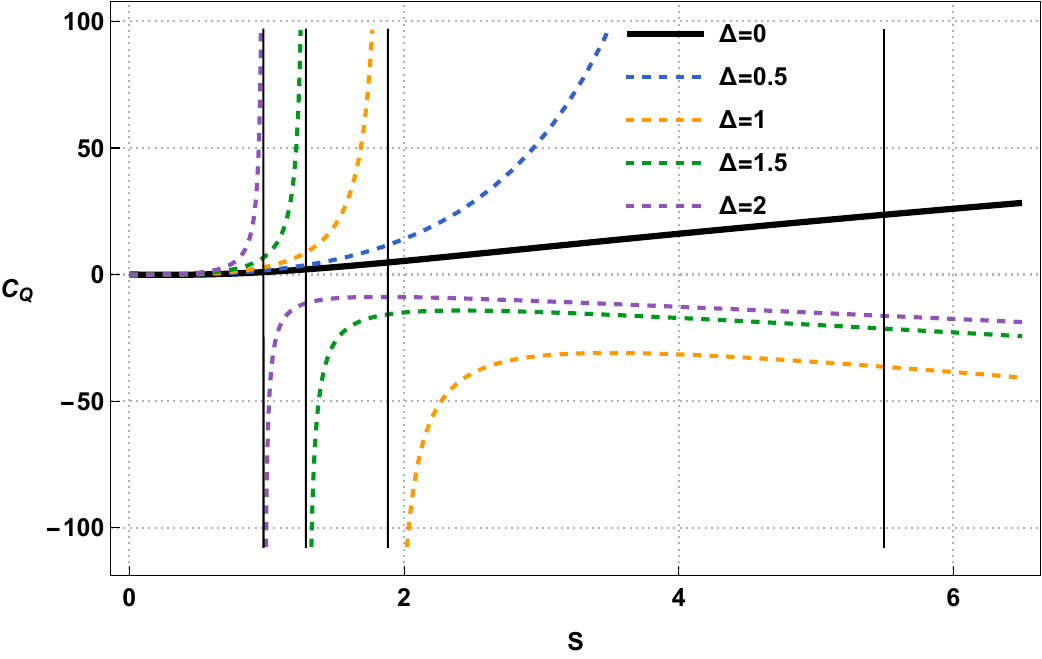}
			\caption{}
			\label{p1}
		\end{subfigure}
		\hspace{0.5cm}
		\begin{subfigure}{0.43\textwidth}
			\includegraphics[width=\linewidth]{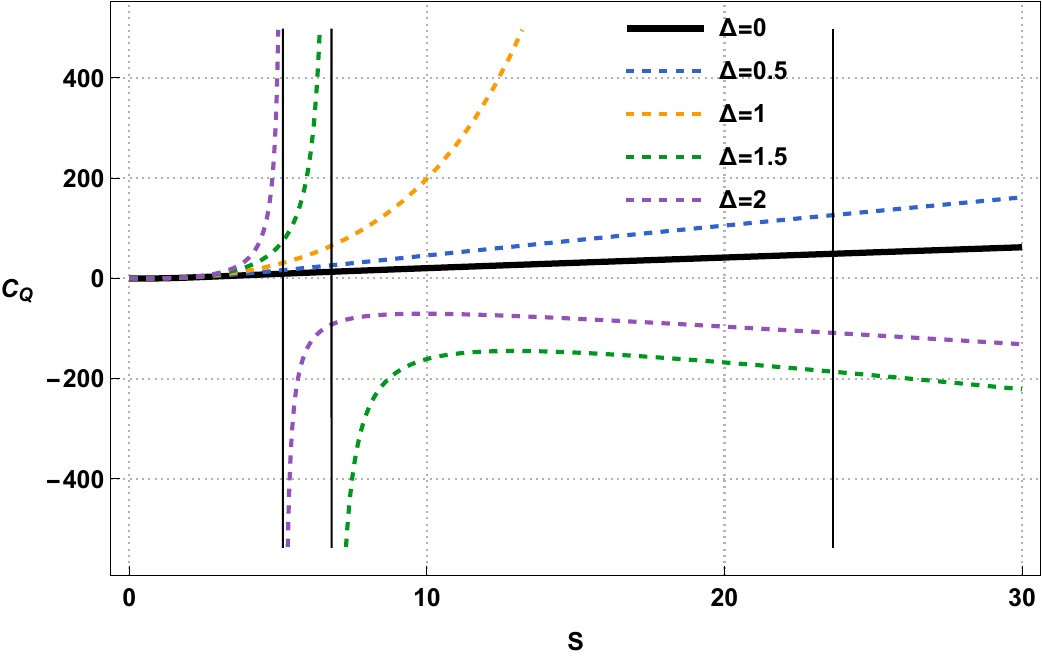}
			\caption{}
			\label{p2}
		\end{subfigure}
		\caption{Heat capacity $C_Q$ variation against $S$ ($C_Q^{-1}$ roots, i.e., phase transition critical point) for various values of the Barrow parameter $\Delta$ with setting $\gamma=0.095$, $L=1$, $Q=0.5$, $\xi=1$, and $n=2$.}
		\label{CQ2}
		\end{figure}
  
\begin{figure}[!htp]	
		\centering
		\begin{subfigure}{0.43\textwidth}
			\includegraphics[width=\linewidth]{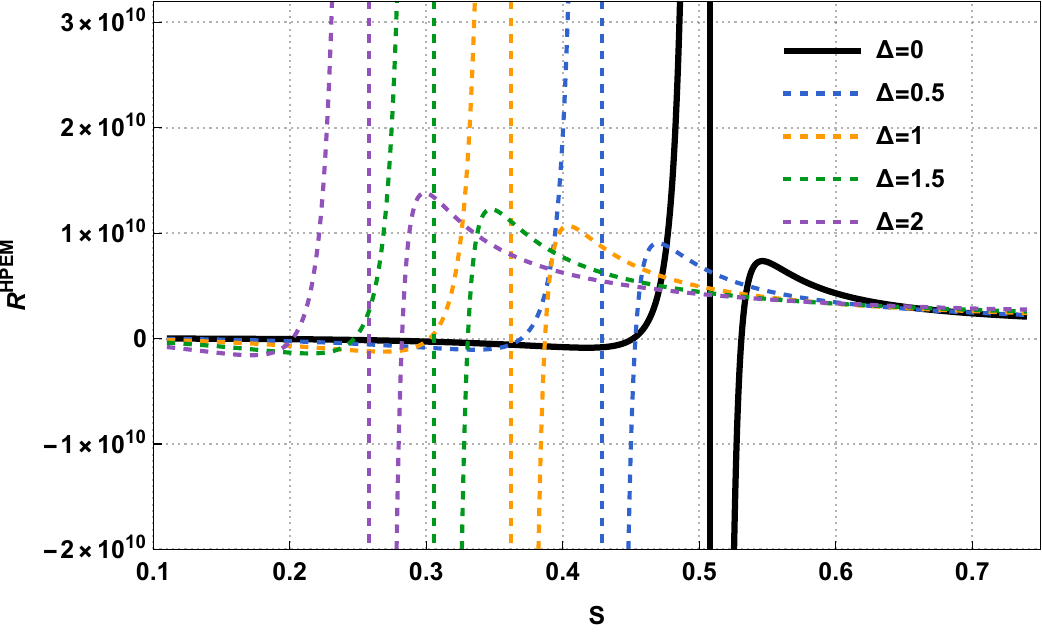}
			\caption{}
			\label{ha}
		\end{subfigure}
		\hspace{0.5cm}
		\begin{subfigure}{0.43\textwidth}
			\includegraphics[width=\linewidth]{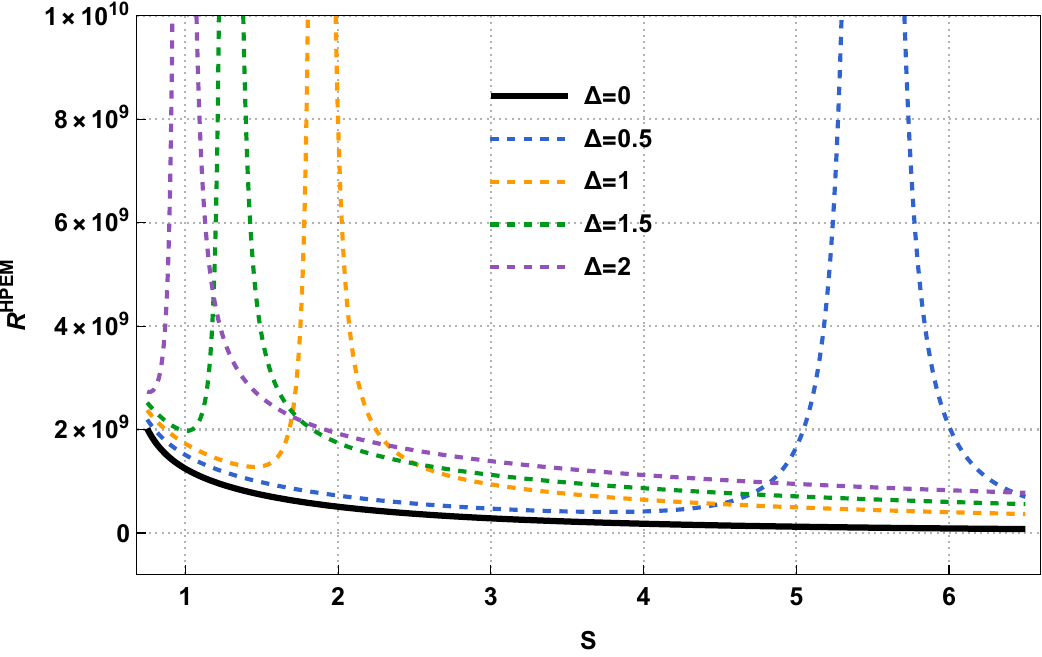}
			\caption{}
			\label{hb}
		\end{subfigure}
		\caption{Ricci scalar of HPEM’s metric $R^{\text{HPEM}}$ against $S$ for various values of the Barrow parameter $\Delta$ in  $5D$ spacetime.}
		\label{HPEM5D}
		\end{figure}

    \begin{equation}
    \mathrm{d}S^2_{HPEM}=\frac{S\,M_S}{\left(\prod_{i=2}^n\frac{\partial^2M}{\partial \xi_i^2}\right)^3}\left(-M_{SS}\mathrm{d}S^2+\sum_{i=2}^n\left(\frac{\partial^2M}{\partial \chi_i^2}\right)\mathrm{d}\chi_i^2\right),
\end{equation}
where, $\chi(\chi\neq S)$, $M_S=\frac{\partial M}{\partial S}$, and $M_{SS}=\frac{\partial^2 M}{\partial S^2}$ are, respectively, extensive parameters. Accordingly, the background of metrics expressed in view of the first law of BH thermodynamics as follows
 \begin{align}
\mathrm{d}S^2_{HPEM}&=\frac{S\,M_S}{\left(\frac{\partial^2M}{\partial L^2}\frac{\partial^2M}{\partial Q^2}\frac{\partial^2M}{\partial \gamma^2}\right)^3}\bigg(-M_{SS}\mathrm{d}S^2+M_{LL}\mathrm{d}L^2 
+M_{QQ}\mathrm{d}Q^2+M_{\gamma\gamma}\mathrm{d}\gamma^2\bigg).
\end{align}  
Likewise, in the realm of mass representation, a diagonal matrix form relevant to HPEM geometry can be established according to the following expression :
\begin{equation}
    g^{HPEM}=\frac{S\,M_S}{\left(\frac{\partial^2M}{\partial L^2}\frac{\partial^2M}{\partial Q^2}\frac{\partial^2M}{\partial \gamma^2}\right)^3}\begin{pmatrix}
-M_{SS} &  &  & \\
 & M_{LL} &  & \\
 &  & M_{Q Q} & \\
 &  &  & M_{\gamma \gamma}
\end{pmatrix}.
\end{equation}
\begin{figure}[!htp]	
		\centering
		\begin{subfigure}{0.43\textwidth}
			\includegraphics[width=\linewidth]{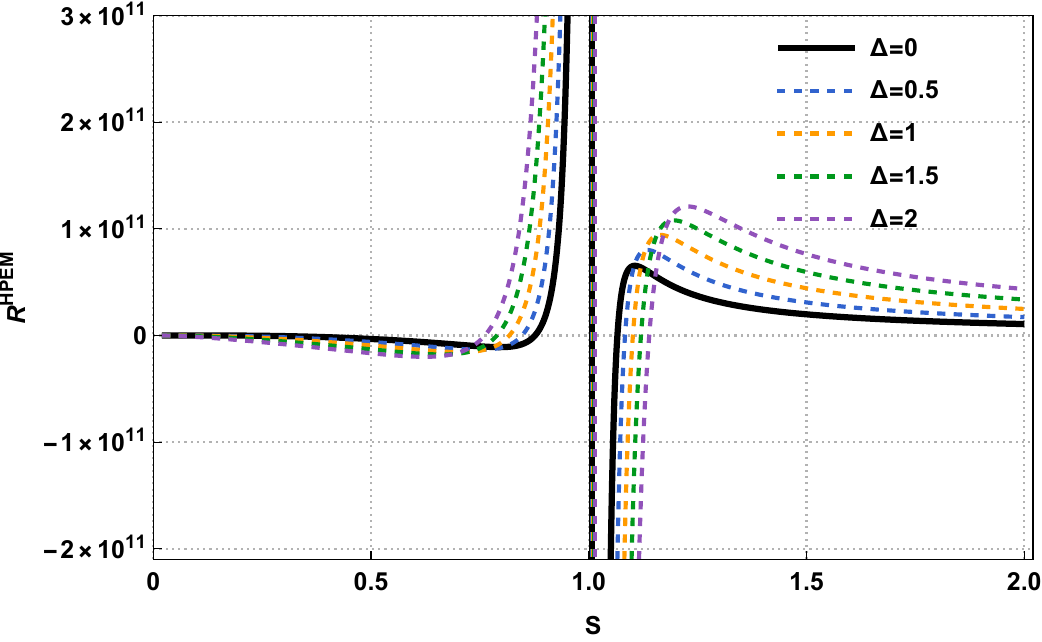}
			\caption{}
			\label{h1}
		\end{subfigure}
		\hspace{0.5cm}
		\begin{subfigure}{0.43\textwidth}
			\includegraphics[width=\linewidth]{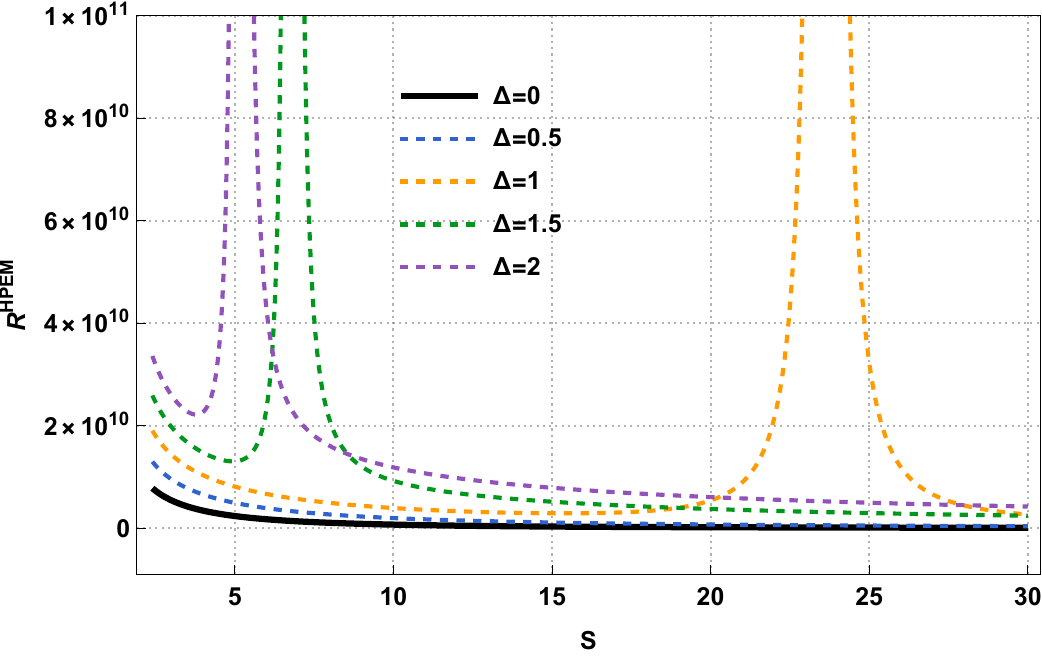}
			\caption{}
			\label{h2}
		\end{subfigure}
		\caption{Ricci scalar of HPEM’s metric $R^{\text{HPEM}}$ against $S$ for various values of the Barrow parameter $\Delta$ in  $4D$ spacetime.}
		\label{HPEM4D}
		\end{figure}
In this formalism, the associated Ricci scalar curvature employed to carry out these measurements has the following denominator
    \begin{equation}
   \text{denom}(R^{HPEM}) = 
     2M_{SS}^2 S^3 M_{S}^3,
\end{equation}
where solving $\text{denom}(R^{HPEM})=0$ gives us a precise insight into the number and location of each specific heat capacity phase transition point.

In order to draw out information and gain a detailed insight into the behaviour of the HPEM $R^{HPEM}$ scalar curvature, Figs.~\ref{HPEM5D} and \ref{HPEM4D} depict an intrinsic analysis. For this purpose, a mandatory step is to take into account the behavior of the heat capacity at constant $Q$ (see Figs.~\ref{CQ1} and \ref{CQ2}) to establish such a one-to-one correspondence with the features of $R^{HPEM}$. Manifestly, the HPEM scalar curvature for a given $\Delta$ parameter spectrum involves, whether in the case of $5D$ or $4D$, two divergent points (one with a change of sign and the second refers to a positive divergent point). Practically speaking, these two divergent points are consistently linked to the heat capacity phase transition points. To be concrete, the heat capacity phase transition point entails what is called the physical limitation point by solving the set, $\left(\frac{\partial M}{\partial S}\right)=0$ \cite{Davies:1977bgr,EslamPanah:2018ums}, while the other specific point refers to the phase transition critical point emerging by solving $\left(\frac{\partial^2 M}{\partial S^2}\right)=0$ \cite{Davies:1977bgr,EslamPanah:2018ums,Tzikas:2018cvs}. Therefore, it is straightforward that the set of physical limitation points in $5D$ (see Fig. \ref{ccc1}) is perfectly predicted by the HPEM scalar curvature, as shown in Fig. \ref{ha}. However, similar findings are explored in $4D$ (see Figs.~\ref{ccc2} and \ref{h1}). On the other hand, with respect to the exploration of the phase transition critical points relevant for the heat capacity (divergent point), the HPEM scalar curvature predicts these exactly by means of a positive divergent point. For the $5D$ case, the phase transition critical points (Fig. \ref{p1}) are correlated to the divergent point of $R^{HPEM}$ (see Fig. \ref{hb}), while for the $4D$ case, Fig. \ref{p2} coherently maps to Fig. \ref{h2}. Furthermore, it is of interest to note that the impact carried out by the Barrow parameter $\Delta$ in a decreasing sense reduced the set of phase transition critical points (divergent points of the heat capacity), which are predicted in terms of positive divergent points pertinent to the HPEM scalar curvature $R^{HPEM}$. This is graphically shown in the case of $5D$ where for $\Delta<0.5$, the system features not any phase transition critical point; in turn, in the case of $4D$, the HPEM scalar curvature $R^{HPEM}$ remains a feature with three positive divergent points satisfying the observation condition such that $\Delta\geq 1$.

 \begin{table*}[!htp]
    \centering
    \caption{Topological Classes in Different BH Dimensions and Ensembles}
    \scalebox{0.85}{\begin{tabular}{|c|c|c|c|}
        \hline
        \textbf{ Dimension} & \textbf{Ensemble} & \textbf{Barrow Entropy} & \textbf{Gibbs-Boltzmann (GB)} \\
        \hline
        \multirow{2}{*}{5D} & Canonical & $W=1$, $W=0$ & $W=1$ \\
        \cline{2-4}
        & Grand Canonical & $W=0$ (annihilation), $W=0$ (generation) & $W=0$ (generation), $W=1$ \\
        \hline
        \multirow{2}{*}{4D} & Canonical & $W=1$, $W=0$ (annihilation), \\ & & Van der Waals-like phase transition & $W=1$ \\
        \cline{2-4}
        & Grand Canonical & $W=0$, $W=-1$, $W=1$ & $W=1$ \\
        \hline
    \end{tabular}}
    \label{tabb1}
\end{table*}

\section{Conclusion}\label{conc}

The sections \ref{sec2} to \ref{secther} were for the polytropic solution and thermodynamics basis needed for the higher dimension AdS BH detailled thermodynamic study. These section provide us the material on polytropic gases pressure and density solutions useful for the thermodynamic topology of the Barrow entropy in higher dimensions and its possible generalizations. The possibilities of possible DE states for the background have been treated for the main fluid models allowing AdS BHs, whether quintessence, chaplygin gas and in general polytropic fluids as showed in Table \ref{Tab1}. In Figs. \ref{fig3} and \ref{figb}, we present in addition the strong energy condition evolution and the metric function ${  A(r)}$ with the radius coordinate in terms of fixed parameter to better simulate the parameter impacts on the AdS BH models by the thermodynamic topology approach. The ${  A(r)}$ evolution indicates the thermodynamic effects on the metric, but also on curvature and Einstein FEs. This is essential for higher dimensions BH solutions.

The section \ref{secther} concerns the thermodynamic variables themselves evolution for polytropic gases and typically AdS BHs expected thermodynamic quantities evolutions. This implies the BH temperature profile, entropy, conserved charges, mass evolution, the heat capacity and so on... All these variables are essential for polytropic Barrow entropy parameters determination for $4$ and $5$ dimensional AdS BHs solution and spacetime topology. From these ingredients, we can study all possible higher-dimensional AdS BH solutions without any constraints.

 In Section \ref{topology}, we conducted a detailed analysis of the thermodynamic topology of these BHs within the Barrow entropy framework, with a particular focus on both 4-dimensional and 5-dimensional BH solutions.Our analysis was performed in two distinct ensembles: the canonical ensemble and the grand canonical ensemble.\\

Starting with the 5-dimensional BHs, where the behavior became more intricate, we evaluated their thermodynamic quantities and analyzed their topological structure. In the canonical ensemble, we identified topological classes characterized by the charges \(W=1\) and \(W=0\) within the Barrow entropy framework, while only the topological class \(W=1\) appeared in the Gibbs-Boltzmann (GB) statistics framework. Next, in the grand canonical ensemble, we observed a topological class \(W=0\) with an annihilation point in the Barrow entropy framework, alongside another topological class \(W=0\) with a generation point. Under the GB statistics framework, we found the topological classes \(W=0\) and \(W=1\), with \(W=0\) featuring a generation point.\\

Extending our study to 4-dimensional BHs, we found two distinct topological classes with charges \(W=1\) and \(W=0\) in the canonical ensemble, depending on the values of the thermodynamic parameters within the Barrow entropy framework. The topological charge \(W=0\) also contained annihilation points. In the GB statistics framework, only the topological class \(W=1\) was present. Furthermore, we observed a Van der Waals-like phase transition in this ensemble. In contrast, the grand canonical ensemble revealed three topological classes with charges \(W=0\), \(W=-1\), and \(W=1\) within the Barrow entropy framework, while only the topological class \(W=1\) was found in the GB statistics framework, showcasing an even richer structure. Notably, the topological charge in both ensembles was found to depend on the thermodynamic and ensemble parameters.
A key feature of both BH solutions was the significant influence of the Barrow entropy parameter $\Delta$. For values of $\Delta$ close to zero, the thermodynamic topology closely resembled the behavior predicted by Gibbs-Boltzmann (GB) statistics. However, as $\Delta$ deviated from zero, the system exhibited substantial deviations from GB statistics, both globally and locally, under Barrow statistics. The results are summarized in the Table \ref{tabb1} for highlighting and comparing 4D and 5D AdS BH main results. These $4$ and $5$ dimensional AdS BH solutions are the necessary cases for generalization to the $D>5$ AdS BH solution classes with respect to the Barrow statistic and still by using Canonical and Grand Canonical statistic approaches. Once again, we have to take into account the main differences between these Canonical and Grand Canonical solutions.

In sect \ref{sparsityBHrad}, we found that the inter-quanta average time difference is impacted by the higher-dimensional AdS BH for the Hawking Radiation (an highly sparser case). For polytropic structures, the AdS charge BH area is corrected according to Eqs. \eqref{areaspars1} and \eqref{areaspars2}. The Hawking radiation temperature is then corrected in these case. The sparsity effect also implies several corrections for polytropic fluid parameters and ultimately the spacetime structure around an AdS BH versus a Schwarzschild BH under this background. The 4 and 5 dimensions AdS BHs solutions clearly show the impacts on the main polytropic variables and parameters when we look at the Fig. \ref{SPP4} and \ref{SPP5}. We may obtain similar effect for $D>5$ AdS BHs cases solutions.

In sect \ref{Sectthermalgeometry},  we finally determined the thermodynamic-dependent spacetime structures for polytropic AdS BHs. These structures directly clarify the effects of Barrow thermal radiation on the higher-dimensional spacetime structure $D>3$. Yes, thermodynamics and Barrow entropy directly influence the metric representation of spacetime. We plotted the effects for the 4 and 5-dimensional AdS BHs studied in detail in this paper to make this clear. We could do the same for the cases of BHs with $D>5$. This was at the same time the ultimate goal of this paper.

We did the detailed study of polytropic Barrow Entropy system and topology for higher dimension AdS BHs. There are naturally several possible extensions of this study such as Murnaghan fluids and more complex fluids. We can ask us how the current paper 4D and 5D BH solutions could be generalized to more complex fluid and thermodynamic topological structures. In these cases, it would be good to know how thermodynamic topologies will behave for more complex structures. How will the 4 and 5-dimensional AdS BHs solutions change, both for the temperature, mass, entropy profiles and for the invariants involved. Would additional dimensional solutions $D>5$ be needed that are more suited to more complex fluids? How will sparsity be possible for such cases? This will have to be investigated tactfully. Beyond GR, one could in some future study these types of topological thermodynamic phenomena and structures for alternative theories, such as in teleparallel $F(T)$ gravity \cite{metricaffinebarrow,teleparabarrow1,teleparabarrow2}. In such a case, one would base the study on the evolution of the spacetime torsion tensor and scalar. Another implication can be the Tsallis entropy arising within teleparallel frameworks \cite{teleparabarrow2}. One could also repeat such a challenge for symmetric teleparallel $F(Q)$ gravity and other intermediate theories. Even though the challenge is very big, it is good and reasonable to believe in these possibilities.

	
	
	
	\section*{Acknowledgements}
	
	SKM is thankful for continuous support and encouragement from the administration of the University of Nizwa for this research work.
	B. Hazarika would like to thank DST-INSPIRE. {We would like to thanks the referee for the paper's appreciation.}
	
	\section*{Declarations}
	
	\begin{itemize}
		\item SKM research works are supported by the TRC Project (Grant No. BFP/RGP/CBS/24/203).
		\item B. Hazarika research works are supported by the Ministry of Science and Technology fellowship program, Govt. of India for awarding the DST/INSPIRE Fellowship[IF220255] for financial support. 	
	\end{itemize}


\end{document}